\documentclass[%
 aip,
 amsmath,amssymb,
 reprint,%
]{revtex4-1}

\usepackage{graphicx}
\usepackage{dcolumn}
\usepackage{bm}

\usepackage[utf8]{inputenc}
\usepackage[T1]{fontenc}
\usepackage{mathptmx}
\usepackage{etoolbox}

\usepackage{color}

\makeatletter
\def\@email#1#2{%
 \endgroup
 \patchcmd{\titleblock@produce}
  {\frontmatter@RRAPformat}
  {\frontmatter@RRAPformat{\produce@RRAP{*#1\href{mailto:#2}{#2}}}\frontmatter@RRAPformat}
  {}{}
}%
\makeatother
\begin{document}

\preprint{AIP/123-QED}

\title[Local momentum balance in electromagnetic gyrokinetic systems]{Local momentum balance in electromagnetic gyrokinetic systems}
\author{H. Sugama}
\thanks{Corresponding author: sugama.hideo@nifs.ac.jp}
\affiliation{
National Institute for Fusion Science, 
Toki 509-5292, Japan
}
\affiliation{
Department of Advanced Energy, University of Tokyo, 
Kashiwa 277-8561, Japan
}

\date{\today}

\begin{abstract}
The Eulerian variational formulation is presented to obtain governing equations of the electromagnetic turbulent gyrokinetic system. A local momentum balance in the system is derived from the invariance of the Lagrangian of the system under an arbitrary spatial coordinate transformation by extending the previous work [H. Sugama {\it et al}., Phys.\ Plasmas {\bf 28}, 022312 (2021)]. Polarization and magnetization due to finite gyroradii and electromagnetic microturbulence are correctly described by the gyrokinetic Poisson equation and Amp\`{e}re's law which are derived from the variational principle. Also shown is how the momentum balance is influenced by including collisions and external sources. Momentum transport due to collisions and turbulence is represented by a symmetric pressure tensor which originates in a variational derivative of the Lagrangian with respect to the metric tensor. The relations of the axisymmetry and quasi-axisymmetry of the toroidal background magnetic field to a conservation form of the local momentum balance equation are clarified. In addition, an ensemble-averaged total momentum balance equation is shown to take the conservation form even in the background field with no symmetry when a constraint condition representing the macroscopic Amp\`{e}re's law is imposed on the background field. Using the WKB representation,  the ensemble-averaged pressure tensor due to the microturbulence is expressed in detail and it is verified to reproduce the toroidal momentum transport derived in previous works for axisymmetric systems. The local momentum balance equation and the pressure tensor obtained in this work present a useful reference for elaborate gyrokinetic simulation studies of momentum transport processes.
\end{abstract}

\maketitle

\section{INTRODUCTION}

Gyrokinetics~\cite{Antonsen,CTB,F-C,B&H,Sugama2000,Schekochihin,Krommes} 
is a powerful theoretical framework based on which a large number of analytical and numerical  studies on microinstabilities and turbulent processes in magnetized plasmas~\cite{Horton} have been done.
The original (or classical) gyrokinetic theory~\cite{Antonsen,CTB,F-C,Sugama1996} 
adopts the WKB approximation (or ballooning representation)~\cite{WKB} 
and treats the perturbed parts of particle distribution functions and electromagnetic fields with gyroradius-scale perpendicular wavelengths. 
This type of gyrokinetic theory is widely employed as the basic model for local flux-tube gyrokinetic simulations~\cite{Dimits,GENE,GYRO,GKV,GKW} 
to evaluate turbulent particle and heat fluxes. 
The other type of (or modern) gyrokinetic theory uses the Lie transformation method~\cite{Littlejohn1982}  
to obtain gyrocenter coordinates which obey the Lagrangian and/or Hamiltonian dynamics derived from the variational formulation.~\cite{B&H,Sugama2000} 
The modern theory guarantees favorable conservation 
properties~\cite{Scott,Brizard2011,Parra_PPCF2011,Qin2014,Sugama2017,Fan}
of gyrokinetic equations for total distribution functions 
(including both background and fluctuation parts), 
which are generally used for long-time global gyrokinetic 
simulations.~\cite{GTC,Idomura2017,XGC,Wang2009,GYSELA,ORB5,ELMFIRE,Gkeyll} 
It is also noted here that classical gyrokinetic equations are shown to be consistently derived from  modern ones by properly taking account of different phase-space coordinate systems used in the two type of theories.~\cite{Sugama2022}

Over the years, momentum transport processes have been attracting much attention because they determine profiles of plasma flows
such as background plasma rotations and ${\bf E}\times{\bf B}$ zonal flows, 
which are regarded as important factors for stabilizing or regulating instabilities and improving plasma 
confinement.~\cite{Burrell}
Also, there are a lot of activities in designing advanced magnetic configurations such as toroidal systems with quasi-symmetry,~\cite{QS1,QS2,QS3,QS4} in which reduction of neoclassical transport and increase of plasma flows are expected. 
In the present paper, local momentum balance equations which describe the momentum transport processes in electromagnetic gyrokinetic turbulence are derived by extending 
the previous work~\cite{Sugama2021} 
on the momentum balance in electrostatic turbulence based on the 
Eulerian variational formulation,~\cite{Newcomb} 
 which is also called the Euler-Poincar\'{e} reduction 
procedure.~\cite{B&T,Hirvijoki2020,Cendra,Marsden,Squire,Sugama2018}

In conventional studies, momentum balance equations are obtained 
by taking the first-order velocity-space moment of 
a kinetic equation or from a variation in an action integral of the Lagrangian 
under infinitesimal translation or rotation. 
In the former derivation, it is unclear how the momentum balance in the direction perpendicular to the background magnetic field can be obtained from the gyrokinetic equation in the gyrophase-averaged form.  In the latter, Noether's theorem can be applied to connect the symmetry condition of the system directly with the canonical momentum conservation equation,~\cite{Sugama2013,Brizard2021}
 in which, however, local momentum transport is represented by the asymmetric canonical pressure tensor because of the vector potential included in the canonical momentum. 
In this work as well as in the previous works,~\cite{Sugama2018,Sugama2021} 
 the invariance of the Lagrangian under arbitrary infinitesimal transformations of general spatial coordinates is used to derive the local momentum balance equation which contains the symmetric pressure tensor obtained by taking the variational derivatives of the Lagrangian with respect to the $3 \times 3$ metric tensor components. 
This is analogous to the derivation of energy-momentum conservation laws from the invariance of an action integral under arbitrary transformations of spatiotemporal coordinates in the theory of general relativity.~\cite{Landau} 
The relations of the symmetry and quasi-symmetry properties of the background magnetic field to the momentum balance equation are investigated with the help of the symmetric pressure tensor. 
In addition, the effects of collisions and/or external momentum sources can be easily included in the local momentum balance equation, by which both collisional~\cite{Hinton1976,H&S,Helander} 
and turbulent transport processes are described. 

In extending the previous study for electrostatic turbulence~\cite{Sugama2021}
to the case of electromagnetic turbulence, 
one need to consider the average and fluctuating parts of the magnetic field and 
accordingly those parts of the magnetic potential. 
Then, as shown in Ref.~\cite{Sugama2022} and this paper, 
the variational derivative of the gyrokinetic Lagrangian with respect to the 
fluctuating part of the vector potential is used to correctly represent 
both the average and fluctuating parts of the local particle flux and 
the current density which appears in Amp\`{e}re's law. 
On the other hand, the variational derivative of the gyrokinetic Lagrangian 
with respect to the average part of the vector potential 
also takes a form similar to the particle flux and 
it appears in the momentum balance equation 
derived using the variational technique in the present study. 
Comparison between the above-mentioned two types of particle fluxes 
shows that their average parts coincides with each other to the leading order 
in the small gyroradius expansion although their fluctuating parts do not. 
It is also shown in the present work 
that, using the ensemble average of the latter particle flux 
to self-consistently determine the average part of the magnetic field from Amp\`{e}re's law, 
the conservation form of the ensemble-averaged local 
momentum balance equation can be derived. 

Currently, large-scale gyrokinetic simulations such as global ones solving phenomena from a device size to an ion gyroradius scale and cross-scale flux-tube simulations treating interactions between both ion and electron gyroradius scales are actively conducted. 
Huge simulations including all scales from the machine size to the electron gyroradius remain a challenging future task, for which global simulations need to treat full finite ion gyroradius effects at least as done in flux-tube simulations. 
In principle, the gyrokinetic model and the momentum balance equation presented in the present work contain all scales ranging from macroscopic equilibrium gradient lengths to microscopic turbulence wavelengths of the order of the electron gyroradius. 
The macroscopic behaviors of the momentum transport processes are described by 
the ensemble-averaged momentum balance equation, which is shown to take the conservation form under a condition to adjust the background field to the macroscopic Amp\`{e}re's law. 
Furthermore, the WKB representation is used to explicitly express the full gyroradius effects of the 
electromagnetic turbulence on the symmetric pressure tensor, 
the {\it ij}th component of which represents 
the turbulent transport of the $i$th momentum component in the $j$th direction. 
These expressions can be applied to evaluation of the local momentum transport by the flux-tube simulations.

The rest of this paper is organized as follows.
In Sec.~II, 
equations of the gyrocenter motion in turbulent electromagnetic fields are derived   
as the Euler-Lagrange equations from the Lagrangian given as a function of the gyrocenter coordinates. 
In Sec.~III, the Lagrangian for the whole gyrokinetic 
system consisting of particles of all species and 
electromagnetic fields is presented to derive gyrokinetic Vlasov equations 
for gyrocenter distribution functions and the 
gyrokinetic Poisson and Amp\`{e}re equations for electrostatic and vector 
potentials based on the Eulerian (or Euler-Poincar\'{e}) variational formulation. 
Then, the gyrokinetic and field parts of the Lagrangian are all represented 
in terms of general spatial coordinates in Sec.~IV and the invariance of the 
Lagrangian under an arbitrary infinitesimal transformation of spatial coordinates is used 
to derive the momentum balance equations for a single-particle-species system 
and for a system including all particle species and electromagnetic fields in 
Sec.~V. 
In Sec.~VI, 
axisymmetric, non-axisymmetric, and quasi-axisymmetric toroidal systems 
are investigated from the viewpoint of momentum balance, 
and Sec.~VII presents the ensemble-averaged momentum balance equation, 
which is shown to take the conservation form when the background field 
is determined by the condition representing the macroscopic Amp\`{e}re's law.
The ensemble-averaged pressure tensor caused by the electromagnetic turbulence 
is expressed in detail using the WKB representation in Sec.~VIII. 
Finally, conclusions are given in Sec.~IX. 
In Appendix~A, the potential field included in the gyrocenter 
Hamiltonian is represented by gyroradius expansion around the gyrocenter, 
which is used in Appendix~B to expand the electromagnetic interaction part 
of Lagrangian density in terms of the electrostatic and vector potentials 
and their derivatives. 
In the same way as in Appendix~B, 
charge and current densities are expanded 
in Appendices~C and D, respectively, where the polarization and 
magnetization parts are identified. 
Energy balance equations in electromagnetic gyrokinetic 
turbulence are presented in Appendix~E.

\section{EQUATIONS OF GYROCENTER MOTION IN TURBULENT ELECTROMAGNETIC FIELDS}

The Lagrangian for describing the gyrocenter motion of the charged particle is given by~\cite{B&H,Sugama2000,Sugama2022}
\begin{equation}
\label{LGYa}
L_{GYa} ({\bf Z}, \dot{\bf Z}, t)  
 \equiv 
 \frac{e_a}{c} {\bf A}_a^* ({\bf X}, U, t) \cdot \dot{\bf X}
+ \frac{m_a c}{e_a} \mu \dot{\vartheta}  
- H_{GYa} ({\bf Z}, t)  
,
\end{equation}
where the modified vector potential ${\bf A}_a^*$ is defined by
$
{\bf A}_a^* ({\bf X}, U, t)
 \equiv  
{\bf A} ({\bf X}, t)
+ (m_a c/e_a) U {\bf b} ({\bf X}, t)
$,
the subscript $a$ represents the particle species with mass $m_a$ and
charge $e_a$, 
and 
$\dot{} \equiv d/dt$ represents the time derivative 
along the motion of the particle in phase space. 
The gyrocenter phase-space coordinates ${\bf X}$, $U$, 
$\mu \equiv m v_\perp^2/(2B)$, 
and $\vartheta$ 
denote the gyrocenter position, 
the velocity component parallel to the magnetic field, 
the magnetic moment, and the gyrophase angle, respectively,  
The vector potential and the unit vector parallel to the background magnetic field ${\bf B}$ 
are written by ${\bf A}$ and ${\bf b} \equiv {\bf B}/B$, respectively. 
Here, it is supposed that ${\bf A}$ can weakly depend on time $t$ and accordingly 
the background magnetic field ${\bf B} = \nabla \times {\bf A}$ 
is allowed to slowly vary in time. 
Thus we can treat the inductive electric field 
${\bf E}_T \equiv 
- c^{-1} \partial {\bf A}/\partial t$ which drives the ohmic current in tokamaks. 

The gyrocenter Hamiltonian 
which appears on the right-hand side of Eq.~(\ref{LGYa}) is given by 
\begin{equation}
\label{HGYa}
H_{GYa} ({\bf Z}, t)   \equiv \frac{1}{2} m_a U^2 + \mu B ({\bf X}, t)+ e_a \Psi_a ({\bf Z}, t) 
,
\end{equation}
and the potential $\Psi_a$ including effects of 
the turbulent electromagnetic fields is defined by~\cite{Sugama2022}      
\begin{eqnarray}
\label{Psi_a}
& & 
\hspace*{-5mm} e_a \Psi_a  ({\bf Z}, t) 
 \equiv 
e_a \langle
\psi_a ({\bf Z}, t) 
 \rangle_\vartheta
- \frac{e_a}{c}  {\bf v}_{Ba}  \cdot 
\langle \widehat{\bf A} 
({\bf X} + \boldsymbol{\rho}_a, t) 
\rangle_\vartheta
\nonumber \\
& & 
\mbox{}
\hspace*{10mm} 
+ \frac{e_a^2}{2 m_a c^2} 
\langle
|\widehat{\bf A}
({\bf X} + \boldsymbol{\rho}_a, t) |^2
\rangle_\vartheta
- \frac{e_a^2}{2 B} 
\frac{\partial}{\partial \mu} 
\langle
(\widetilde{\psi}_a)^2
\rangle_\vartheta
,
\hspace*{4mm}
\end{eqnarray}
where 
$\boldsymbol{\rho}_a$ is
the gyroradius vector given by  
$\boldsymbol{\rho}_a \equiv {\bf b} \times {\bf v} / \Omega_a$, 
 $\Omega_a \equiv e_a B/(m_a c)$ is the gyrofrequency, 
and 
\begin{equation}
\label{spsia}
\psi_a ({\bf Z}, t) 
\equiv
\phi ({\bf X} + \boldsymbol{\rho}_a, t) 
- \frac{\bf v}{c} \cdot 
\widehat{\bf A}
({\bf X} + \boldsymbol{\rho}_a, t) 
. 
\end{equation}
The gyrophase average $\langle Q \rangle_\vartheta$ 
and the gyrophase-dependent part 
%
$\widetilde{Q}$ 
of an arbitrary function $Q$ of the gyrocenter 
phase-space coordinates 
${\bf Z} \equiv ({\bf X}, U, \mu, \vartheta)$ are represented by  
\begin{equation} 
\label{2-3}
\langle Q \rangle_\vartheta 
\equiv
\frac{1}{2\pi}
\oint  Q \, d\vartheta
\hspace*{3mm}
\mbox{and}
\hspace*{3mm}
\widetilde{Q}
\equiv
Q -
\langle Q \rangle_\vartheta
,
\end{equation}
respectively. 
The particle's velocity 
${\bf v}$ is written as 
\begin{equation}
\label{vdef}
{\bf v}
\equiv
U {\bf b} ({\bf X}, t)
-
W
[ \sin \vartheta \; 
{\bf e}_1 ({\bf X}, t) + \cos \vartheta \; 
{\bf e}_2 ({\bf X}, t) ]
,
\end{equation}
where 
$({\bf e}_1, {\bf e}_2, {\bf b})$ are unit vectors which form a right-handed orthogonal triad and are regarded as functions of $({\bf X}, t)$. 
The magnetic moment is given by 
$
\mu \equiv m_a W^2 / 2B
$, 
and 
\begin{equation}
\label{vBa}
{\bf v}_{Ba}
\equiv
\frac{c}{e_a B} {\bf b} \times 
\left( 
m_a U^2 {\bf b} \cdot \nabla {\bf b} 
+ \mu \nabla B 
\right)
\end{equation}
is the first-order drift velocity consisting of 
curvature drift and $\nabla B$ drift. 
Second-order terms retained in Eq.~(\ref{Psi_a}) are necessary 
for correctly deriving the gyrokinetic Poisson and 
Amp\`{e}re equations from variational derivatives 
with $\phi$ and $\widehat{\bf A}$, respectively, as 
shown in Sec.~III. 
Especially, it is shown in Ref.~\cite{Sugama2022} that 
the second-order term 
$- (e_a/c) {\bf v}_{Ba}\cdot \langle \widehat{\bf A} 
\rangle_\vartheta$, which is often neglected in 
conventional studies, should be kept for the variational 
derivative to  
obtain the gyrokinetic Amp\`{e}re's law 
including both equilibrium and turbulent parts 
accurately. 
In Appendix~A, 
$\Psi_a$ is expanded in the gyroradius and 
decomposed into several parts which have different 
dependences on electrostatic and magnetic 
fluctuations. 
It is noted in Ref.~\cite{Sugama2011,Parra2012,Calvo2015} 
that basic equations including terms of higher-order, 
which are not considered here, 
are required for accurately describing 
the flux-surface-averaged momentum balance along the symmetry direction in up-down symmetric tokamaks and stellarator-symmetric quasisymmetric stellarators 
where the low-flow ordering is assumed as in the present work. 

Using the gyrocenter Lagrangian in Eq.~(\ref{LGYa}), 
the Euler-Lagrangian equations are given by
\begin{equation}
\label{ELeq1}
\frac{d}{dt}
\left(
\frac{\partial L_{GYa}}{\partial \dot{\bf Z}}
\right)
-
\frac{\partial L_{GYa}}{\partial {\bf Z}}
=0
. 
\end{equation}
from which the gyrocenter motion equations are obtained as 
\begin{equation}
\label{dZdt}
\frac{d {\bf Z}}{dt}
=
\{ {\bf Z}, H_{GYa} \}
+
\{ {\bf Z}, {\bf X} \}
\cdot \frac{e_a}{c} 
\frac{\partial {\bf A}_a^*}{\partial t} 
. 
\end{equation}
with the Poisson brackets defined by 
\begin{eqnarray}
& & 
\{ {\bf X}, {\bf X} \}
= \frac{c}{e_a B_{a \parallel}^*} 
{\bf b} \times {\bf I}
, 
\hspace*{3mm}
\{ {\bf X}, U \}
= \frac{{\bf B}_a^*}{m_a B_{a\parallel}^*} 
, 
\nonumber \\
& & 
\{  {\bf X}, \vartheta \} =  0
,
\hspace*{3mm}
\{ U , \vartheta \} = 0, 
\hspace*{3mm}
\{  \vartheta , \mu \} =  \frac{e_a}{m_a c}
.
\end{eqnarray}
Equations~(\ref{dZdt}) are rewritten as 
\begin{eqnarray}
\label{dXdt}
\frac{d{\bf X}}{dt}
& = & 
\frac{1}{B^*_{a\parallel}}
\biggl[
\biggl( U + \frac{e_a}{m_a} 
\frac{\partial \Psi_a}{\partial U}
\bigg) {\bf B}_a^*
\nonumber \\
& & \mbox{}
\hspace*{5mm}
+
c {\bf b} \times \left( 
\frac{\mu}{e_a}\nabla B
+ \nabla \Psi_a
+ \frac{1}{c} \frac{\partial {\bf A}_a^*}{\partial t}
\right)
\biggl]
,
\hspace*{5mm}
\end{eqnarray}
\begin{equation}
\label{dUdt}
\frac{d U}{dt}
=
-\frac{{\bf B}_a^*}{m_a B^*_{a\parallel}}
\cdot
\left(
\mu \nabla B
+
e_a \nabla \Psi_a
+ \frac{e_a}{c} \frac{\partial {\bf A}_a^*}{\partial t}
\right)
,
\end{equation}
\begin{equation}
\label{dmdt}
\frac{d\mu}{dt}
=
0
,
\end{equation}
and 
\begin{eqnarray}
\label{dthdt}
\frac{d\vartheta}{dt}
=
\Omega_a
+
\frac{e_a^2}{m_a c} 
\frac{\partial \Psi_a}{\partial \mu}
,
\end{eqnarray}
where ${\bf B}_a^*$ and $B^*_{a\parallel}$ are defined by
\begin{equation}
\label{Ba}
{\bf B}_a^*
\equiv
\nabla \times {\bf A}_a^*
\hspace*{3mm}
\mbox{and}
\hspace*{3mm}
B^*_{a\parallel}
\equiv
{\bf B}_a^* \cdot {\bf b}
,
\end{equation}
respectively. 
The gyrocenter motion given by Eqs.~(\ref{dXdt})--(\ref{dthdt}) 
satisfies Liouville's theorem, which is expressed as
\begin{equation}
\label{Liouville}
\frac{\partial D_a({\bf Z}, t)}{\partial t}
+ \frac{\partial}{\partial {\bf Z}}
\cdot \left(
 D_a ({\bf Z}, t)
\frac{d {\bf Z}}{d t}
\right)
=
0
, 
\end{equation}
where the Jacobian $D_a({\bf Z}, t)$ is given by
$
D_a({\bf Z}, t)
=
B^*_{a\parallel}/m_a
$.

\section{GYROKINETIC VLASOV-POISSON-AMP\`{E}RE SYSTEM}

The action integral for the gyrokinetic Vlasov-Poisson-Amp\`{e}re system is given by 
\begin{equation}
\label{IGKF}
I  \equiv  \int_{t_1}^{t_2} dt \; L_{GKF} 
\equiv  \int_{t_1}^{t_2} dt \; ( L_{GK} + L_F ) 
,
\end{equation}
where the gyrokinetic Lagrangian $L_{GK}$ is defined by the phase-space integral of 
the gyrocenter distribution function $F_a$ multiplied by the gyrocenter 
Lagrangian $L_{GYa}$ [see Eq.~(\ref{LGYa})] as 
\begin{equation}
\label{LGK}
L_{GK}
\equiv 
\sum_a L_{GKa}
\equiv 
\sum_a
\int d^6 Z
\; 
F_a ({\bf Z}, t) 
L_{GYa}  ({\bf Z}, {\bf u}_{aZ} ({\bf Z}, t) , t)  
.
\end{equation}
Here, based on the Eulerian picture, 
the temporal change rates of the gyrocenter coordinates are regarded as 
functions of $({\bf Z}, t)$ and they are represented by 
\begin{equation}
\label{uaZ}
{\bf u}_{aZ} ({\bf Z}, t) = [{\bf u}_{aX} ({\bf Z}, t), 
u_{aU} ({\bf Z}, t), u_{a\mu} ({\bf Z}, t), u_{a\vartheta} ({\bf Z}, t)], 
\end{equation}
which are used in Eq.(\ref{LGK}) to evaluate 
$L_{GYa}  ({\bf Z}, {\bf u}_{aZ} ({\bf Z}, t) , t)$. 
Then, the distribution function $F_a$ satisfies 
\begin{equation}
\label{Vlasov}
 \frac{\partial F_a ({\bf Z}, t)}{\partial t} 
+ \frac{\partial}{\partial {\bf Z}}
\cdot
\bigl[ 
F_a ({\bf Z}, t) \; {\bf u}_{aZ} ({\bf Z}, t) 
\bigr]
= 0
, 
\end{equation}
where the functional form of ${\bf u}_{aZ} ({\bf Z}, t)$ is 
determined later by the Euler-Poincar\'{e} 
variational principle. 

Here, the Darwin approximation is made to 
remove electromagnetic waves propagating at light speed, 
and  the Lagrangian $L_F$ is defined by~\cite{Sugama2000}
%
\begin{eqnarray}
\label{LF}
L_F 
& \equiv & 
\frac{1}{8 \pi} 
\int_V d^3 x 
\left[
|{\bf E}_L ({\bf x}, t) |^2
- 
|{\bf B} ({\bf x}, t) + \widehat{\bf B} ({\bf x}, t)|^2
\right.
\nonumber \\ 
& & \mbox{}
\hspace*{17mm}
\left. 
+ 
\frac{2}{c} \lambda ({\bf x}, t) 
\nabla \cdot \widehat{\bf A} ({\bf x}, t) 
\right]
. 
\end{eqnarray}
%
where $V$ represents the spatial domain of the system, 
$\lambda$ plays the role of a Lagrange undetermined multiplier to 
derive the Coulomb gauge condition
\begin{equation}
\label{Coulomb}
\nabla \cdot \widehat{\bf A} =0
,
\end{equation}
from the variational condition $\delta I/ \delta \lambda = 0$ 
(or $\delta L_{GKF}/ \delta \lambda = \delta L_F/ \delta \lambda = 0$), 
and ${\bf E}_L$ is the longitudinal (or irrotational) part 
of the electric field written in terms of electrostatic potential 
$\phi$ as
\begin{equation}
\label{EL}
{\bf E}_L \equiv - \nabla \phi
.
\end{equation}
%


Now, the trajectories of particles in the phase space as well as 
the electrostatic potential and perturbed vector potential are virtually varied to derive 
the governing equations of the collisionless electromagnetic gyrokinetic turbulent system 
from the Eulerian variational principle. 
Following the same procedure as in Refs.~\cite{Sugama2018,Sugama2022}, 
the infinitesimal variation in the phase-space trajectory  
is represented in the Eulerian picture by
\begin{equation}
\delta {\bf Z}_{aE} ({\bf Z}, t)
= [\delta {\bf X}_{aE} ({\bf Z}, t) , 
      \delta  U_{a\parallel E} ({\bf Z}, t), 
     \delta \mu_{aE} ({\bf Z}, t), 
     \delta \zeta_{aE} ({\bf Z}, t)]. 
\end{equation}
   Then, 
the variations in the functional forms of 
${\bf u}_{aZ} =  ({\bf u}_{aX}, u_{aU}, u_{a \mu}, u_{a \vartheta}) $ 
and $F_a ({\bf Z}, t)$ are written in the Eulerian picture as 
\begin{eqnarray}
\label{duZ}
\hspace*{-3mm}
\delta {\bf u}_{aZ}({\bf Z}, t) 
& = & 
\left( \frac{\partial }{\partial t}
+ {\bf u}_{aZ} ({\bf Z}, t) \cdot \frac{\partial }{\partial {\bf Z}}
\right) \delta {\bf Z}_{aE} ({\bf Z}, t) 
\nonumber \\ 
& & \mbox{}
- 
\delta {\bf Z}_{aE}({\bf Z}, t)  \cdot \frac{\partial }{\partial {\bf Z}}
 \; {\bf u}_{aZ}({\bf Z}, t) 
, 
\end{eqnarray}
and
\begin{eqnarray}
\label{dF}
\delta F_a ({\bf Z}, t) 
& = & 
- \frac{\partial}{\partial {\bf Z}} \cdot
[ F_a ({\bf Z}, t) \delta {\bf Z}_{aE} ({\bf Z}, t) ]
, 
\end{eqnarray}
respectively. 
The variations in the electrostatic potential and 
the perturbed vector potential are denoted 
by $\delta \phi$ and $\delta \widehat{\bf A}$, respectively. 
Using Eqs.~(\ref{IGKF}), (\ref{LGK}), (\ref{duZ}), and (\ref{dF}),  
the variation in the action integral  $I_{GKF}$ is given by 
\begin{eqnarray}
\label{dIGKF}
\delta I_{GKF}
& = & 
\sum_a
\int_{t_1}^{t_2}  dt 
\int d^6 Z
\; F_a
\nonumber \\ & & 
\times
\left[
\left( \frac{\partial L_{GYa}}{\partial {\bf Z}} \right)_u
-
\left(\frac{d}{d t}\right)_a
\left( \frac{\partial L_{GYa}}{\partial {\bf u}_{aZ}} \right)
\right]\cdot \delta {\bf Z}_{aE}
\nonumber \\
& & 
\mbox{} 
+ 
\int_{t_1}^{t_2} dt \int_V d^3 x   
\left[
\frac{\delta L_{GKF} }{\delta\phi }
\delta \phi 
+
\frac{\delta L_{GKF} }{\delta \widehat{\bf A}}
\cdot
\delta \widehat{\bf A}
\right]
\nonumber \\
& & 
\mbox{} 
+ \mbox{B.T.}
.
\end{eqnarray}
Here, B.T.\ represents boundary terms that appear due to partial integrals 
and $(d/dt)_a$ denotes the time derivative along the phase-space trajectory 
defined by 
\begin{equation}
\label{ddt}
\left( 
\frac{d}{dt}
\right)_a
 \equiv  
\frac{\partial}{\partial t } 
+  {\bf u}_{aZ} \cdot \frac{\partial}{\partial {\bf Z} } 
, 
\end{equation}
from which one obtains  
$(d/dt)_a {\bf Z} = {\bf u}_{aZ}$. 
The variational derivative  $\delta {\cal F} [f] / \delta f$ 
of any functional ${\cal F} [f]$ of a function $f$ 
in three-dimensional space is defined as a function in the space 
which satisfies  
\begin{equation}
\label{funder1}
\int d^3 x 
\frac{\delta {\cal F} [f] }{\delta f} ({\bf x}) \varphi({\bf x}) 
=
\lim_{\epsilon \rightarrow 0} 
 \frac{{\cal F} [f + \epsilon \varphi] - {\cal F} [f]}{\epsilon}
,
\end{equation}
from which one can also write 
\begin{equation}
\label{funder2}
\frac{\delta {\cal F} [f ({\bf y})]}{\delta f ({\bf x})} 
\equiv 
\left. 
\frac{d}{d\epsilon} {\cal F} [f ({\bf y}) + 
\epsilon \delta^3 ({\bf y} - {\bf x})]
\right|_{\epsilon=0} 
,
\end{equation}
where 
$\delta^3 ({\bf y} - {\bf x}) \equiv
\delta (y^1-x^1) \delta (y^2-x^2) \delta (y^3-x^3)$. 
From Eq.~(\ref{funder2}), one has 
\begin{equation}
\label{dphdph}
\frac{ \delta \phi ({\bf X} + \boldsymbol{\rho}_a ) }{
\delta \phi ({\bf x}) }
= 
\delta 
( {\bf X} + \boldsymbol{\rho}_a - {\bf x} )
.
\end{equation}

Now, it is required that $\delta I_{GKF} = 0$ holds for any variations $\delta {\bf Z}_{aE}$, 
$\delta\phi $, and $\delta \widehat{\bf A}$ which vanish on the boundaries of the integral region.
Then, it is found from Eq.~(\ref{dIGKF}) that 
\begin{equation}
\label{ELeq2}
\left(\frac{d}{d t}\right)_a \left( \frac{\partial L_{GYa}}{\partial {\bf u}_{aZ}} \right)
-
\left( \frac{\partial L_{GYa}}{\partial {\bf Z}} \right)_u
= 0
,
\end{equation}
$\delta L_{GKF}/\delta \phi=0$, and $\delta L_{GKF}/\delta \widehat{\bf A} = 0$ 
need to be satisfied. 
Here, since Eq.~(\ref{ELeq2}) is equivalent to Eq.~(\ref{ELeq1}), one finds that 
${\bf u}_{aZ}({\bf Z}, t)$ should be given by the right-hand side of 
Eq.~(\ref{dZdt}). 
Thus, the gyrokinetic Vlasov equation is given by Eq.~(\ref{Vlasov}) with 
Eq.~(\ref{dZdt}) substituted into ${\bf u}_{aZ}({\bf Z}, t)$. 
 
The gyrokinetic Poisson equation is derived from $\delta L_{GKF}/\delta \phi=0$ 
and written as 
\begin{equation}
\label{GKP}
\nabla \cdot {\bf E}_L
= 
4 \pi \rho_c
,
\end{equation}
where  the charge density $\rho_c$ is
given by 
\begin{eqnarray}
\label{rhoc}
\rho_c 
& = & 
- \frac{\delta L_{GK}}{\delta \phi} 
= 
- \sum_a \frac{\delta L_{GKa}}{\delta \phi} 
\nonumber \\
& = & 
\sum_a e_a 
\int d^6 Z
\; 
\delta^3 ({\bf X} + 
\boldsymbol{\rho}_a - {\bf x} )
\biggl(
F_a 
+
\frac{e_a \widetilde{\psi}_a}{B} 
\frac{\partial F_a}{\partial \mu}
\biggr)
,
\hspace*{3mm}
\end{eqnarray}
and 
$
\delta L_F/\delta \phi
=
(1/4\pi)
\nabla \cdot {\bf E}_L
$
is used. 

The gyrokinetic  Amp\`{e}re's law is derived from 
$\delta L_{GKF}/\delta \widehat{\bf A} = 0$ as 
\begin{equation}
\label{GKA1}
\nabla \times ( {\bf B} + \widehat{\bf B} ) 
=
\frac{4\pi}{c} {\bf j} -  \frac{1}{c} \nabla \lambda
, 
\end{equation}
where the electric current density ${\bf j}$ is given by
\begin{eqnarray}
\label{j}
{\bf j} 
& =  & 
c \frac{\delta L_{GK}}{\delta \widehat{\bf A}} 
= 
c \sum_a  \frac{\delta L_{GKa}}{\delta \widehat{\bf A}} 
\nonumber \\
& = & 
\sum_a e_a
\int d^6 Z
\; \delta^3 ({\bf X} + 
\boldsymbol{\rho}_{a} - {\bf x} )
\nonumber \\
& & 
\times \biggl[
F_a ({\bf Z}, t) \biggl( {\bf v}  - \frac{e_a}{m_a c} \widehat{\bf A} 
+
{\bf v}_{Ba} \biggr)
+
\frac{e_a \widetilde{\psi}_a}{B} 
\frac{\partial F_a}{\partial \mu} {\bf v}
\biggr]
,
\hspace*{8mm}
\end{eqnarray}
and 
$
\delta L_F/\delta \widehat{\bf A}
=
- (1/4\pi)
\nabla \times ( {\bf B} + \widehat{\bf B} ) 
- (1 / 4 \pi c) \nabla \lambda
$
is used. 
It is noted here that 
an arbitrary vector field ${\bf a}$ is written as  
$
{\bf a} = {\bf a}_L + {\bf a}_T
$
where the longitudinal  (or irrotational) 
and transverse (or solenoidal) parts of ${\bf a}$ 
are given by
$
{\bf a}_L ({\bf x}) 
= - (4\pi)^{-1} \nabla \int d^3 x' \,
(\nabla' \cdot {\bf a} ({\bf x}') ) 
/ |{\bf x} - {\bf x}'|
$
and
$
{\bf a}_T ({\bf x}) 
= (4\pi)^{-1} \nabla \times 
( \nabla \times
\int d^3 x' \,
{\bf a} ({\bf x}') 
/ |{\bf x} - {\bf x}'| )
$, respectively.~\cite{Jackson}  
Then, the longitudinal part of Eq.~(\ref{GKA1}) 
gives
\begin{equation}
\label{Lpart}
\nabla \lambda
=
4\pi {\bf j}_L
.
\end{equation}
From the transverse part of Eq.~(\ref{GKA1}), 
the gyrokinetic  Amp\`{e}re's law is written as  
\begin{equation}
\label{GKA2}
\nabla \times ( {\bf B} + \widehat{\bf B} ) 
=
\frac{4\pi}{c} {\bf j}_T
. 
\end{equation}
In Eqs.~(\ref{Lpart}) and (\ref{GKA2}), ${\bf j}_L$ and ${\bf j}_T$ 
represent the longitudinal and transverse parts of 
${\bf j}$, respectively.

\section{REPRESENTATION IN GENERAL SPATIAL COORDINATES}

In this section,
general spatial coordinates are used to represent
the Lagrangian of the electromagnetic turbulent gyrokinetic system defined in 
Sec.~III. 
The Lagrangian is given as the integral of the Lagrangian density with respect 
to the general spatial coordinates, and it is invariant under an arbitrary spatial 
coordinate transformation. 

\subsection{The Lagrangian density represented in general spatial coordinates}

  The action integral $I_{GKF}$ in Eq.~(\ref{IGKF}) is written here  as 
\begin{equation}
\label{IGKF2}
I_{GKF}  \equiv  \int_{t_1}^{t_2} dt \; L_{GKF} 
\equiv \int_{t_1}^{t_2}  dt \int_V d^3 x\; {\cal L}_{GKF} 
,
\end{equation}
where the Lagrangian density ${\cal L}_{GKF}$ is given by 
\begin{eqnarray}
\label{LGKF}
{\cal L}_{GKF} 
& \equiv & 
{\cal L}_{GK} + {\cal L}_{F} 
\nonumber \\ 
{\cal L}_{GK} & \equiv & 
\sum_a
\int d^3 v
\; F_a (x, v, t) L_{GYa}  (x, v, t) 
\nonumber \\ 
{\cal L}_{F} 
& \equiv & 
\frac{\sqrt{g(x)}}{8 \pi}
\Bigl[ 
 g^{ij}(x) 
(E_L)_i (x, t) (E_L)_j (x, t) 
\nonumber \\ & & \mbox{}
-  g_{ij}(x)  \bigl\{
B^i (x,t) + \widehat{B}^i (x,t)
\bigr\}
 \bigl\{
B^j (x,t) + \widehat{B}^j (x,t)
\bigr\}
\nonumber \\ & & \mbox{}
+
\frac{2}{c}
\lambda (x, t) 
g^{ij}(x) \nabla_i \widehat{A}_j (x,t)
\Bigr]
.
\end{eqnarray}
Here, $x \equiv (x^i)_{i=1,2,3}$, $v \equiv (U, \mu, \vartheta)$, 
$d^3 x \equiv dx^1 dx^2 dx^3$, and $d^3 v \equiv d U d\mu d\vartheta$ are used, 
and $\nabla_j$ represents a covariant derivative.  
%
In the equation for ${\cal L}_{GK}$ shown in Eq.~(\ref{LGKF}), 
$x \equiv (x^i)_{i=1,2,3}$  represent the coordinates not of the 
position of the particle but that of the gyrocenter (denoted by ${\bf X}$ in Sec.~II). 
It should be emphasized that in this section,  
$x \equiv (x^i)_{i=1,2,3}$ are general spatial coordinates  which 
can be either Cartesian or any other curved coordinates. 
However, 
the spatial position vector ${\bf r} = {\bf r} (x)$ 
is  assumed to be a function of only the spatial coordinates 
$x \equiv (x^i)_{i=1,2,3}$ and it is independent of time $t$. 
The gyrocenter distribution function in 
the $(x, v)$-space 
is denoted by $F_a$, and 
the number of particles of species $a$ in 
the phase-space volume element 
$d^3 x d^3 v \equiv dx^1 dx^2 dx^3 d U d\mu d\vartheta$ 
at time $t$ 
is given by $F_a (x, v, t) d^3 x d^3 v$. 
This paper employs the summation convention that 
the same symbol used for a pair of upper and lower indices  
within a term 
[such as seen in Eq.~(\ref{LGKF}) as well as 
in the equations shown below]  
indicates summation over the range $\{ 1, 2, 3 \}$ 
of the symbol index.
The contravariant metric tensor components $g^{ij}$
in the general spatial coordinates $x \equiv (x^i)$ are
related to the covariant components $g_{ij}$ by 
$
g^{ik} g_{kj} = \delta^i_j
$,
where $\delta^i_j$ represents the Kronecker delta. 
The determinant of the covariant metric tensor matrix 
is denoted by  
$
g (x) \equiv  \det [ g_{ij} (x) ]
$.
As the spatial position vector ${\bf r}$ is
a function of only the spatial coordinates 
$x \equiv (x^i)$, 
$g_{ij}(x)$,  $g^{ij}(x)$, and $g(x)$ are 
all independent of time $t$. 

The gyrocenter Lagrangian $L_{GYa}$, 
which enters  ${\cal L}_{GK}$ in Eq.~(\ref{LGKF}), 
is represented in the Eulerian picture by 
\begin{eqnarray}
\label{LGYa2}
L_{GYa} 
& \equiv & 
\left( \frac{e_a}{c} A_j (x, t)+ m_a U b^i (x, t) g_{ij}(x) \right) 
u_{ax}^j (x, v, t)
\nonumber \\ & & \mbox{}
+ \frac{m_a c}{e_a} \mu u_{a\vartheta} (x, v, t) 
- H_{GYa} (x, U, \mu, t)
,
\end{eqnarray}
where 
$b^i \equiv B^i/B$ is the $i$th contravariant component 
of the unit vector parallel to the background magnetic field and
the background field strength is given by 
$
B (x, t)
\equiv
\sqrt{ g_{ij}(x) B^i (x, t) B^j (x, t)}
$.
   The contravariant components of the background and perturbed magnetic fields 
are expressed in terms of the covariant components of the vector potentials as 
\begin{equation}
\label{BB}
B^i (x, t)
\equiv 
\frac{\epsilon^{ijk}}{\sqrt{g(x)}}\frac{\partial A_k(x, t)}{\partial x^j}
, 
\hspace*{5mm}
\widehat{B}^i (x, t)
\equiv 
\frac{\epsilon^{ijk}}{\sqrt{g(x)}}\frac{\partial \widehat{A}_k(x, t)}{\partial x^j}
, 
\end{equation}
where the Levi-Civita symbol is denoted by  
\begin{eqnarray}
\label{eijk}
& & 
\epsilon^{ijk} 
 \equiv   \epsilon_{ijk}
\nonumber \\
&  & 
\equiv
\left\{
\begin{array}{cl}
1 
&
\mbox{($(i, j, k) = (1,2,3), (2,3,1), (3,1,2)$)} 
\\
-1
& 
\mbox{($(i, j, k) = (1,3,2), (2,1,3), (3,2,1)$)} 
\\
0 
& 
\mbox{(otherwise)}. 
\end{array}
\right. 
\end{eqnarray}
The gyrocenter Hamiltonian is written here as  
\begin{equation}
\label{HGYa2}
 H_{GYa} (x, U, \mu, t) 
\equiv \frac{1}{2} m_a U^2 + \mu B (x, t)+ e_a \Psi_a (x, U, \mu, t)
,
\end{equation}
and the fluctuation part is given by    
\begin{eqnarray}
\label{Psia0}
\Psi_a (x, \mu, t)
& \equiv  & 
\phi (x, t) + \Psi_{E1a} (x, \mu, t) + \Psi_{\widehat{A}1a} (x, U, \mu, t) 
\nonumber \\ & & \mbox{}
+ \Psi_{E2a} (x, \mu, t)+  \Psi_{E\widehat{A}a} (x, U, \mu, t) 
\nonumber \\ & & \mbox{}
+ \Psi_{\widehat{A}2a} (x, U, \mu, t)
, 
\end{eqnarray}
where 
\begin{eqnarray}
\label{PsiE1a0}
& & \Psi_{E1a} (x, \mu, t) 
=
\sum_{n=1}^\infty  
\frac{\alpha_a^{j_1 \cdots j_n}(x, \mu, t)}{n!}
\nabla_{j_1} \cdots \nabla_{j_n} \phi (x, t)
\nonumber \\
& &  
\hspace*{5mm}
=
- \sum_{n=1}^\infty  
\frac{\alpha_a^{j_1 \cdots j_n}(x, \mu, t)}{n!}
\nabla_{j_1} \cdots \nabla_{j_{n-1}}  (E_L)_{j_n} (x, t)
, 
\hspace*{5mm}
\end{eqnarray}
and 
\begin{eqnarray}
\label{PsiA1a0}
& & 
\hspace*{-5mm}
\Psi_{\widehat{A}1a} (x, U, \mu, t)
=
 - \frac{1}{c}
\sum_{n=0}^\infty
\frac{1}{n!}
\bigl[
\alpha_a^{j_1 \cdots j_n} \;  ( U b^i + v_{Ba}^i )
+ \Omega_a 
\nonumber \\
&  & \mbox{} 
\hspace*{10mm}
\times \sqrt{g} \varepsilon_{k l m} \; 
\alpha_a^{j_1 \cdots j_n k} \; b^l \; g^{i m}
\bigr] \; 
\nabla_{j_1} \cdots \nabla_{j_n}  \widehat{A}_i (x, t)
.
\end{eqnarray}
Here, $\alpha_a^{j_1 \cdots j_n}(x, \mu, t)$  
is defined  by  Eqs.~(\ref{alpha})--(\ref{eta}) 
in Appendix~A with
using 
$
h^{ij}
\equiv 
g^{ij} - b^i b^j
$.
The second-order parts 
$\Psi_{E2a} (x, \mu, t)$, 
$\Psi_{E\widehat{A}a} (x, U, \mu, t)$, 
and 
$\Psi_{\widehat{A}2a} (x, U, \mu, t)$
on the right-hand side of Eq.~(\ref{Psia0}) are obtained from 
Eqs.~(\ref{PsiE2a2}), (\ref{PsiEAa2}), and (\ref{PsiA2a2}), respectively, 
with the partial derivative $\partial_j$ replaced by the covariant derivative $\nabla_j$.

The temporal change rates of the gyrocenter coordinates 
in Eq.~(\ref{uaZ})
are denoted here by 
$u_{ax}^i(x, v, t)$, 
$u_{aU}(x, v, t)$, 
$u_{a\mu}(x, v, t)$, and 
$u_{a\vartheta}(x, v, t)$. 
Then, Eq.~(\ref{Vlasov}) for the distribution function
$F_a (x, v, t)$ is written as 
\begin{eqnarray}
\label{Vlasov2}
& & 
\hspace*{-2mm}
\frac{\partial F_a}{\partial t} 
+ \frac{\partial}{\partial x^j}
( F_a u_{ax}^j  )
+ \frac{\partial}{\partial U}
( F_a u_{a U}  )
+ \frac{\partial}{\partial \mu}
( F_a u_{a \mu} )
+ \frac{\partial}{\partial \vartheta}
( F_a u_{a \vartheta}  )
\nonumber \\ 
& & 
 \mbox{}
= 0
. 
\end{eqnarray}

The gyrocenter Hamiltonian $H_{GYa}$ given in Eq.~(\ref{HGYa2}) takes a functional form, 
\begin{eqnarray}
\label{HGYaf}
H_{GYa} 
& = & 
H_{GYa} 
 \bigl[ v,   
\partial_j A_i (x,t), \partial_{jk} A_i (x,t), 
\{ \partial_J \phi (x,t) \},  
\nonumber \\ & & \mbox{}
\hspace*{8mm}
\{ \partial_J \widehat{A}_i (x,t) \}, 
 \{ \partial_J g_{ij} (x) \}   \bigr]
, 
\end{eqnarray}
which depends on the velocity space coordinates
(except for $\vartheta$) as well as the 
general spatial coordinates $x \equiv (x^i)_{i=1,2,3}$ 
through the field variables 
$[ \partial_j A_i (x,t), \partial_{jk} A_i (x,t),  \{ \partial_J \phi (x,t) \},  
\{ \partial_J \widehat{A}_i (x,t) \}, 
\{ \partial_J g_{ij} (x) \} ]$. 
Here, 
the notation
$J \equiv (j_1, j_2, \cdots, j_n )$ 
($n = 0, 1, 2, \cdots ; j_1, j_2, \cdots, j_n = 1, 2, 3$) 
is used 
to write 
\begin{equation}
\label{pJQ}
\partial_J {\cal F} \equiv 
\left\{
\begin{array}{lc}
 {\cal F}
&
(n=0)
\\
\partial_{j_1 j_2 \cdots j_n} {\cal F}
\;
\equiv 
\;
\partial^n {\cal F}/
\partial x^{j_1} \partial x^{j_2} \cdots \partial x^{j_n}
&
(n \geq 1)
\end{array}
\right.
\end{equation}
where 
 ${\cal F}$ is an arbitrary function of $x = (x^i)_{i=1,2,3}$. 
Then, 
$
\{ \partial_J \phi \} 
 \equiv 
\{ \phi, \partial_j \phi, \partial_{jk} \phi, 
\partial_{jkl} \phi, \cdots \}
$, 
and the definitions of other compact notations
$\{ \partial_J \widehat{A}_i \}$, and 
 $\{ \partial_J g_{ij} (x) \}$ in Eq.~(\ref{HGYaf}) 
are understood in the same way.
One obtains $(E_L)_i \equiv - \partial_i \phi$ from Eq.~(\ref{EL}) which 
is used to replace 
$\{ \partial_J \phi \}$ with  
$
\{ \phi, \{ \partial_J (E_L)_i \}  \}
$
where
$
\{ \partial_J (E_L)_i \} \equiv   
\{ (E_L)_i,  \partial_j (E_L)_i, \partial_{jk} (E_L)_i, 
\partial_{jkl} (E_L)_i, \cdots   \}
$.
Note that high-order spatial derivative terms due to finite gyroradii 
enter the gyrocenter Hamiltonian $H_{GYa}$ 
as seen in Eqs.~(\ref{PsiE1a0}) 
and (\ref{PsiA1a0}) where the covariant 
derivatives contain the spatial derivatives of $g_{ij}$ through the 
Christoffel symbols [see Eq.~(A4) in Ref.~\cite{Sugama2021}]. 

It is found from Eqs.~(\ref{LGYa2}) and  (\ref{HGYaf}) 
that the functional form of the gyrocenter Lagrangian $L_{GYa}$ 
is written as 
\begin{eqnarray}
\label{LGYaf}
L_{GYa} 
& = & 
L_{GYa} 
 \bigl[ v,   u_{ax}^i (x, v, t),  u_{a\vartheta}(x, v, t),  
A_i (x, t), \partial_j A_i (x,t), 
\nonumber \\ & & 
\mbox{} 
\partial_{jk} A_i (x,t), 
\{ \partial_J \phi (x,t) \},  \{ \partial_J \widehat{A}_i (x,t) \},  
 \{ \partial_J g_{ij} (x) \}   \bigr]
,
\hspace*{6mm}
\end{eqnarray}
where
$u_{ax}^i (x, v, t)$, $u_{a\vartheta}(x, v, t)$, $\phi(x,t)$, and 
$\widehat{A}_i (x,t)$ are 
the functions, the governing equations of which are derived 
from the variation principle in Sec.~IV.C  
while the dependence of $L_{GYa}$ on $A_i (x, t)$, $\partial_j A_i (x, t)$, 
$\partial_{jk} A_i (x, t)$,
and $\partial_J g_{ij} (x, t)$ is also explicitly shown 
because their variations need to be taken into account to evaluate 
the variation of $L_{GYa}$ in Sec.~IV where 
the local momentum balance is derived using   
the general spatial coordinate transformation which causes 
the variations in the functional forms of both  
$(u_{ax}^i, u_{a\vartheta}, \phi, \widehat{A}_i )$ and 
$(A_i, g_{ij})$. 

\subsection{The Lagrangian density associated with the electromagnetic interaction}

It is found from Eqs.~(\ref{LGKF}), (\ref{LGYa2}), and (\ref{HGYa2}) that 
the part of the Lagrangian density including $\Psi_a$ is given by 
\begin{eqnarray}
\label{LPsi}
& & 
\hspace*{-6mm} 
{\cal L}_\Psi 
 \equiv  
\sum_a 
{\cal L}_{\Psi a} 
\equiv 
- \sum_a
\int d^3 v \, F_a  
\, e_a \Psi_a 
\nonumber \\
& & 
=
- \rho_c^{(g)} \phi  + {\cal L}_ {E 1}   
+ {\cal L}_ {\widehat{A}1}  
+ {\cal L}_{E 2} + {\cal L}_ {E\widehat{A}}  
+ {\cal L}_ {\widehat{A}2}  
, 
\end{eqnarray}
which determines the electromagnetic interaction of particles. 
Here, the gyrocenter charge density $\rho_c^{(g)}$ is defined by 
\begin{equation}
\label{rhog}
\rho_c^{(g)} \equiv 
\sum_a e_a N_a^{(g)} 
\equiv
\sum_a e_a
\int d^3 v \, F_a 
, 
\end{equation}
where $N_a^{(g)}$ represents the gyrocenter density. 
The terms 
${\cal L}_ {E 1}$,  ${\cal L}_ {\widehat{A}1}$,  
${\cal L}_{E 2}, {\cal L}_ {E\widehat{A}}$, and 
${\cal L}_ {\widehat{A}2}$  
on the right-hand side of Eq.~(\ref{LPsi}), 
are defined by 
\begin{equation}
\label{LE1}
{\cal L}_ {E 1} 
 \equiv   
\sum_a {\cal L}_ {E 1 a} 
=
\sum_{k=1}^\infty 
Q_0^{j_1 \cdots j_{2k}} 
\nabla_{j_1} \cdots \nabla_{j_{2k-1}} (E_L)_{j_{2k}}
,
\end{equation}
\begin{eqnarray}
\label{LA1}
 {\cal L}_{\widehat{A}1} 
& \equiv & 
\sum_a {\cal L}_ {\widehat{A}1a}
=
\sum_{n=1}^\infty 
R_0^{j_1 \cdots j_{n}} 
\nabla_{j_1} \cdots  \nabla_{j_{n-1}}
\widehat{A}_{j_{n}}
,
\end{eqnarray}
\begin{eqnarray}
\label{LE2}
{\cal L}_{E2} 
&  \equiv   & 
\sum_a {\cal L}_{E 2a}
= 
\frac{1}{2} \sum_{m=1}^\infty  \sum_{n=1}^\infty  
\chi_{E}^{i_1 \cdots i_m ;  j_1 \cdots j_n}
\nabla_{i_1} \cdots  \nabla_{i_{m-1}} (E_L)_{i_m}
\nonumber \\ 
& & 
\hspace*{10mm}
\mbox{} 
\times
\nabla_{j_1} \cdots  \nabla_{j_{n-1}} (E_L)_{j_n}
,
\end{eqnarray}
\begin{eqnarray}
\label{LEA}
 {\cal L}_{E\widehat{A}} 
& \equiv & 
\sum_a
{\cal L}_ {E\widehat{A}a}
=
\sum_{m=1}^\infty \sum_{n=1}^\infty 
\chi_{E\widehat{A}}^{j_1 \cdots j_m ; k_1, \cdots k_n} 
\nabla_{j_1} \cdots  \nabla_{j_{m-1}}
 (E_L)_{j_m}
\nonumber \\ 
&  & \mbox{} 
\hspace*{10mm}
\times
\nabla_{k_1} \cdots  \nabla_{k_{n-1}}
\widehat{A}_{k_n},
\end{eqnarray}
and 
\begin{eqnarray}
\label{LA2}
 {\cal L}_{\widehat{A}2} 
& \equiv & 
\sum_a {\cal L}_ {\widehat{A}2a}  
=
\frac{1}{2}
\sum_{m=1}^\infty \sum_{n=1}^\infty 
\chi_{\widehat{A}}^{j_1 \cdots j_m ; k_1, \cdots k_n} 
\nabla_{j_1} \cdots  \nabla_{j_{m-1}}
\widehat{A}_{j_m}
\nonumber \\ 
&  & \mbox{} 
\hspace*{10mm}
\times
\nabla_{k_1} \cdots  \nabla_{k_{n-1}}
\widehat{A}_{k_n},
\end{eqnarray}
respectively.
Here, $Q_0^{j_1 \cdots j_{2k}}$, $R_0^{j_1 \cdots j_{n}}$, 
$\chi_{E}^{i_1 \cdots i_m ;  j_1 \cdots j_n}$,  
$\chi_{E\widehat{A}}^{j_1 \cdots j_m ; k_1, \cdots k_n}$, 
and
$ {\cal L}_{\widehat{A}2}$ included in 
Eqs.~(\ref{LE1})--(\ref{LA2}) are given 
by 
\begin{eqnarray}
\label{Q0}
& &
\left[
Q_0^{j_1 \cdots j_{2k}}, R_0^{j_1 \cdots j_{n}}
\right]
\equiv 
\sum_a 
\left[
Q_{0a}^{j_1 \cdots j_{2k}}, R_{0a}^{j_1 \cdots j_{n}}
\right]
,
\nonumber \\
& & 
\left[
\chi_{E}^{i_1 \cdots i_m ;  j_1 \cdots j_n},  
\chi_{E\widehat{A}}^{j_1 \cdots j_m ; k_1, \cdots k_n},
\chi_{\widehat{A}}^{j_1 \cdots j_m ; k_1, \cdots k_n} 
\right]
\nonumber \\
& &
\hspace*{2mm}
 \equiv
\sum_a 
\left[
\chi_{Ea}^{i_1 \cdots i_m ;  j_1 \cdots j_n},  
\chi_{E\widehat{A}a}^{j_1 \cdots j_m ; k_1, \cdots k_n}, 
\chi_{\widehat{A}a}^{j_1 \cdots j_m ; k_1, \cdots k_n} 
\right]
, 
\hspace*{3mm}
\end{eqnarray}
where $Q_{0a}^{j_1 \cdots j_{2k}}$, $R_{0a}^{j_1 \cdots j_{n}}$, 
$\chi_{Ea}^{i_1 \cdots i_m ;  j_1 \cdots j_n}$,  
$\chi_{E\widehat{A}a}^{j_1 \cdots j_m ; k_1, \cdots k_n}$, 
and
$\chi_{\widehat{A}a}^{j_1 \cdots j_m ; k_1, \cdots k_n}$ are defined by 
Eqs.~(\ref{B4}), (\ref{B6}),  and (\ref{B12}) 
in Appendix~B. 
As seen in Appendices~C and D, 
the charge and current densities in
the gyrokinetic Poisson and Amp\`{e}re equations are derived from 
${\cal L}_\psi$, of which the components  ${\cal L}_ {E 1}$,  ${\cal L}_ {\widehat{A}1}$,  
${\cal L}_{E 2}, {\cal L}_ {E\widehat{A}}$, and 
${\cal L}_ {\widehat{A}2}$ are associated with 
the polarization charge and the magnetization current. 

\subsection{Derivation of gyrokinetic Vlasov-Poisson-Amp\`{e}re equations 
in general spatial coordinates}

Here, general spatial coordinates $x = (x^i)_{i=1,2,3}$ are used for the Eulerian 
variational derivation of the gyrokinetic Vlasov-Poisson-Amp\`{e}re equations for the 
the electromagnetic gyrokinetic system. 
The virtual variations in the phase-space trajectory  
are now represented in the Eulerian picture by
$\delta x_{aE}^i (x, v, t)$, 
$\delta  U_{a E} (x, v, t)$, 
$\delta \mu_{aE} (x, v, t)$, 
and 
$\delta \vartheta_{aE} (x, v, t)$.
Then, Eqs.~(\ref{duZ}), (\ref{dF}), and (\ref{dIGKF}) are rewritten as 
\begin{eqnarray}
\label{duvmt}
& & 
\hspace*{-3mm}
[ \delta u_{ax}^i , \delta u_{a U}, 
\delta u_{a \mu}, \delta u_{a \vartheta} ]
\nonumber \\
&  &  =
\biggl( \frac{\partial }{\partial t}
+ u_{ax}^j \frac{\partial }{\partial x^j}
+ u_{a U}  \frac{\partial }{\partial U}
+ u_{a \mu}  \frac{\partial }{\partial \mu}
\nonumber \\ 
& & \mbox{}
\hspace*{8mm}
+ u_{a \vartheta} \frac{\partial }{\partial \vartheta}
\biggr) 
[\delta x_{aE}^i, \delta U_{a E}, 
 \delta \mu_{aE}, \delta \vartheta_{aE} ]
\nonumber \\ 
& & \mbox{}
- \biggl( 
\delta x_{aE}^j \frac{\partial }{\partial x^j}
+ \delta U_{a E} \frac{\partial }{\partial U}
+  \delta \mu_{aE} \frac{\partial }{\partial \mu}
\nonumber \\ 
& & \mbox{}
\hspace*{8mm}
+ \delta \vartheta_{aE} \frac{\partial }{\partial \vartheta}
\biggr) 
[ u_{ax}^i ,  u_{a U}, 
u_{a \mu}, u_{a \vartheta} ]
,
\end{eqnarray}
\begin{eqnarray}
\label{dFGK2}
\delta F_a
& = & 
- \frac{\partial}{\partial x^j}
( F_a \delta x_{aE}^j  )
- \frac{\partial}{\partial U_{a}}
( F_a \delta U_{a E})
- \frac{\partial}{\partial \mu}
( F_a \delta \mu_{aE})
\nonumber \\ & & \mbox{}
- \frac{\partial}{\partial \vartheta}
( F_a \delta \vartheta_{aE})
,
\end{eqnarray}
and 
\begin{eqnarray}
\label{dIGKF2}
& & 
\hspace*{-5mm}
\delta I_{GKF}
= 
\sum_a
\int_{t_1}^{t_2}  dt \int_V d^3 x
\int d^3 v
\; F_a
 \nonumber \\
& & 
\hspace*{1mm}
\mbox{} \times 
\left[
\left\{ 
\left( \frac{\partial L_{GYa}}{\partial x^i} \right)_u
-
\left(\frac{d}{d t}\right)_a
\left( \frac{\partial L_{GYa}}{\partial u_{ax}^i} \right)
\right\} \delta x_{aE}^i 
\right. 
\nonumber \\
& & 
\hspace*{2mm}
\mbox{} + 
\left( \frac{\partial L_{GYa}}{\partial U} \right)_u \delta U_{a E}
+ 
\left( \frac{\partial L_{GYa}}{\partial \mu } \right)_u \delta \mu_{aE}
\nonumber \\
& & 
\left. 
\hspace*{2mm}
+ 
  \left\{ 
\left( \frac{\partial L_{GYa}}{\partial \vartheta} \right)_u
-    \left(\frac{d}{d t}\right)_a 
\left( \frac{\partial L_{GYa}}{\partial u_{a\vartheta}} \right)
\right\} 
\delta \vartheta_{aE}
\right]
\nonumber \\
& & 
\hspace*{2mm} \mbox{} 
+ 
\int_{t_1}^{t_2} dt \int_V d^3 x   
\left( 
\frac{\delta L_{GK} }{\delta \phi} \delta \phi 
+
\frac{\delta L_{GK} }{\delta \widehat{A}_i} \delta  \widehat{A}_i
+
\frac{\delta L_{GK} }{\delta \lambda} \delta  \lambda
\right) 
\nonumber \\
& & 
\hspace*{2mm} \mbox{} 
+ \mbox{B.T.}
, 
\end{eqnarray}
respectively, 
where  
$( \partial L_{GYa}/\partial x^i )_u$, 
$( \partial L_{GYa}/\partial U )_u$, 
$( \partial L_{GYa}/\partial \mu )_u$, 
 and 
$( \partial L_{GYa}/\partial \vartheta )_u$ denote  
the derivatives of $L_{GYa}$ in $x^i$, $U$, $\mu$, 
and $\vartheta$, respectively, 
with $(u_{ax}^i, u_{a\vartheta})$ kept fixed in $L_{GYa}$, and 
the time derivative along the phase-space trajectory 
is represented by 
\begin{equation}
\label{ddt2}
\left( 
\frac{d}{dt}
\right)_a
 \equiv  
\frac{\partial}{\partial t } 
+  u_{ax}^k \frac{\partial}{\partial x^k} 
+ u_{aU} \frac{\partial}{\partial U} 
+  u_{a\mu} \frac{\partial}{\partial \mu} 
+  u_{a\vartheta} \frac{\partial}{\partial \vartheta} 
. 
\end{equation}
Using Eq.~(\ref{dIGKF}) and $\delta I_{GKF} = 0$, 
one first obtains 
\begin{equation}
\label{dpidt}
\left( 
\frac{d}{dt}
\right)_a 
p_{ai}
=
\left( \frac{\partial L_{GYa}}{\partial x^i} \right)_u
, 
\end{equation}
where 
$
p_{ai}
\equiv 
\partial L_{GYa}/\partial u_{ax}^i
= 
(e_a/c) A_i (x, t) + m_a U b_i (x, t)
\equiv (e_a/c) A^*_{ai} (x, U, t) 
$ 
represents the covariant vector component 
of the canonical momentum. 
Equation~(\ref{dpidt}) can be deformed to obtain 
\begin{equation}
\label{mupb}
m_a u_{aU} b_i   = 
 e_a \left( - \frac{\partial \Psi_a}{\partial x^i}
-   \frac{1}{c}
\frac{\partial  A^*_{ai}}{\partial t}
 + \frac{1}{c} \sqrt{g} \epsilon_{ijk} u_x^j 
B^{*k} \right)
- \mu \frac{\partial B}{\partial x^i}
, 
\end{equation}
where 
$
B^{*i}_a  \equiv
(\epsilon^{ijk} / \sqrt{g}) (\partial A^*_{ak}/ \partial x^j )
$.
Next, 
combining Eq.~(\ref{mupb}) with 
$
( \partial L_{GYa}/\partial U )_u
= m_a ( 
u_{ax}^i b_i  - U )
= 0 
$, 
$
(\partial L_{GYa}/\partial \mu )_u
= 
(m_a c/e_a) u_{a\vartheta} - B 
- e_a \partial \Psi_a / \partial \mu
= 0 
$, 
and 
$
( d / d t )_a
( \partial L_{GYa} / \partial u_{a\vartheta} )
= 
( m_a c/ e_a)  u_{a\mu}
=
( \partial L_{GYa} / \partial \vartheta )_u
=
0  
$, 
the gyrocenter motion equations are derived as 
\begin{eqnarray}
\label{uxieq}
\hspace*{-3mm}
u_{ax}^i
& = & 
\frac{1}{B^*_{a\parallel}}
\biggl[ \left( U + \frac{e_a}{m_a} \frac{\partial \Psi_a}{\partial U} \right) 
B^{*i}_a
\nonumber \\  & & \mbox{}
\hspace*{5mm}
+
c \frac{\epsilon^{ijk}}{\sqrt{g}} b_j
\biggl( 
 \frac{\mu}{e_a} \frac{\partial B}{\partial x^k}
+ \frac{\partial \Psi_a}{\partial x^k}
+ \frac{1}{c}
\frac{\partial  A^*_{ak}}{\partial t}
\biggr)
\biggr]
,
\end{eqnarray}
\begin{equation}
\label{uvpeq}
m_a u_{aU} 
 =  
- \frac{B^{*i}_a}{B^*_{a\parallel}}
 \left[
 \mu \frac{\partial B}{\partial x^i}
+
e_a   \left( \frac{\partial \Psi_a}{\partial x^i}
+ \frac{1}{c}
\frac{\partial  A^*_{ai}}{\partial t} 
\right)
\right]
,
\end{equation}
\begin{equation}
\label{umueq}
u_{a\mu}
= 
0
,
\end{equation}
   and 
\begin{equation}
\label{uvteq}
u_{a\vartheta} 
= 
\Omega_a
+ \frac{e_a^2}{m_a c} 
\frac{\partial \Psi_a}{\partial \mu}
, 
\end{equation}
 where 
$
B^*_{a\parallel}
\equiv
B^{*i}_a b_i
$.
Substituting Eqs.~(\ref{uxieq})--(\ref{uvteq}) into 
Eqs.~(\ref{Vlasov2}) and taking its average with respect to 
the gyrophase $\vartheta$, 
the gyrokinetic Vlasov equation is derived as 
\begin{eqnarray}
\label{GKE}
& & 
\hspace*{-1mm}
\frac{\partial \overline{F}_a}{\partial t} 
+ \frac{\partial}{\partial x^i}
\biggl[ \overline{F}_a \frac{1}{B^*_{a\parallel}}
\biggl\{ 
\left( U + \frac{e_a}{m_a} \frac{\partial \Psi_a}{\partial U} \right)
 B^{*i}_a
+
c \frac{\epsilon^{ijk}}{\sqrt{g}} b_j
\nonumber \\ 
& & \mbox{}
\hspace*{34mm}
\times
\biggl( 
 \frac{\mu}{e_a} \frac{\partial B}{\partial x^k}
+ \frac{\partial \Psi_a}{\partial x^k}
+ \frac{1}{c}
\frac{\partial  A^*_{ak}}{\partial t}
\biggr)
\biggr\}
 \biggr]
\nonumber \\ 
& & 
\hspace*{6mm}
\mbox{}
+ \frac{\partial}{\partial U}
\biggl[ 
\overline{F}_a 
\frac{B^{*i}_a}{m_a B^*_{a\parallel}}
 \biggl\{
- e_a   \biggl( \frac{\partial \Psi_a}{\partial x^i}
+ \frac{1}{c}
\frac{\partial  A^*_{ai}}{\partial t} 
\biggr)
- \mu \frac{\partial B}{\partial x^i}
\biggr\}
 \biggr]
\nonumber \\ 
& & \mbox{}
\hspace*{2mm}
 = 0
, 
\end{eqnarray}
where 
$
\overline{F}_a
\equiv 
\langle F_a \rangle_\vartheta
\equiv
\oint F_a d\vartheta / 2\pi 
$
is the gyrophase-averaged 
distribution function.

The remaining conditions for $\delta I_{GKF} = 0$ are given by 
$\delta L_{GK}/\delta \phi = \delta L_{GK}/\delta \widehat{A}_i 
= \delta L_{GK}/\delta \lambda= 0$. 
The Coulomb gauge condition is obtained as 
$
\delta L_{GKF} / \delta \lambda
= 
(2 / c ) \nabla_i \widehat{A}^i
= 0
$. 
The gyrokinetic Poisson equation
is given by
\begin{equation}
\label{GKP2}
\frac{\delta L_{GKF}}{\delta \phi}
= 
- \rho_c + 
\frac{1}{4\pi} \frac{\partial}{\partial x^i} 
[\sqrt{g} (E_L)^i ]
= 0
, 
\end{equation}
where the charge density $\rho_c$ is written as 
\begin{equation}
\rho_c 
= -\frac{\delta L_{GK}}{\delta \phi}
= \rho^{(gc)} - \nabla_i P_G^i
, 
\end{equation}
with the generalized polarization vector density $P_G^i$ defined by
\begin{equation}
\label{PG}
 P_G^i
 \equiv   
 \sum_{n=0}^\infty  
(-1)^n
\nabla_{i_1} \cdots  \nabla_{i_n} 
Q^{i \, i_1 \cdots i_n }
. 
\end{equation}
Here, 
the multipole moments $Q^{i_1 \cdots i_m }$ 
are given by Eq.~(\ref{C7}) in Appendix~C. 
The gyrokinetic Amp\`{e}re's law is derived from 
the condition $\delta L_{GK}/\delta \widehat{A}_i =  0$ 
which is written as 
\begin{equation}
\label{GKA3}
\frac{\delta L_{GKF}}{\delta \widehat{A}_i}
= 
\frac{1}{c} j^i
- \frac{1}{4\pi} \varepsilon^{ijk} 
\frac{\partial ( B_k + \hat{B}_k) }{\partial x^j}
- \frac{\sqrt{g}}{4\pi c} g^{ij} 
\frac{\partial \lambda }{\partial x^j}
= 0
. 
\end{equation}
Here, the current density is written as 
\begin{equation}
j^l
= (j^{0})^l + c \nabla_k N^{kl}
, 
\end{equation}
where $(j^{0})^l$ is defined by Eq.~(\ref{D9}) in Appendix~D and 
$N^{kl}$ is given by 
\begin{equation}
 N^{kl}
\equiv 
\sum_{n=0}^\infty (-1)^{n+1} \nabla_{j_1}
\cdots \nabla_{j_n} R^{j_1 \cdots j_n k l}
, 
\end{equation}
with 
$R^{j_1 \cdots j_n k}$ defined in Eq.~(\ref{D5}).

\section{DERIVATION OF THE MOMENTUM BALANCE}

In this section, the invariance of the Lagrangian 
under arbitrary infinitesimal transformations of spatial coordinates 
is used to derive the local momentum balance equations for the single-particle-species 
system and for the whole system including particles of multiple species 
and electromagnetic fields. 

\subsection{Invariance of Lagrangians under 
infinitesimal transformations of spatial coordinates}

An arbitrary infinitesimal transformation of 
spatial coordinates from $x = (x^i)_{i=1,2,3}$ to $x' = (x'^i)_{i=1,2,3}$,  
is written as   
\begin{equation}
\label{xprime}
 x'^i 
=
x^i + \xi^i (x)
, 
\end{equation}
where the infinitesimal variation in the spatial coordinate $x^i$ 
is denoted by $\xi^i (x)$
which is an arbitrary function of 
only the spatial coordinates $x = (x^i)_{i=1,2,3}$ 
and independent of time $t$. 

The gyrokinetic Lagrangian $L_{GKa}$ is written as 
\begin{equation}
\label{LGKa}
L_{GKa}  
\equiv 
\int_V d^3 x  \, 
{\cal L}_{GKa}
\equiv 
\int_V d^3 x  
\int d^3 v\, 
F_a L_{GYa}  
\end{equation}
where $F_a$ and $L_{GYa}$ defined in Eq.~(\ref{LGYa}) behave as a scalar density field 
and a scalar field, respectively,  under the transformation of the spatial coordinates. 
The variation in $L_{GKa}$ under the infinitesimal spatial coordinate transformation
in Eq.~(\ref{xprime}) is written as 
\begin{equation}
\label{dLGKa}
\overline{\delta} L_{GKa}  
\equiv 
\int_V d^3 x  
\left(
\frac{\partial ( \xi^i {\cal L}_{GKa} )}{\partial x^i} 
+ 
\overline{\delta}
{\cal L}_{GKa}
\right)
\end{equation}
Here and hereafter,  the notation $\overline{\delta} \cdots$ 
represents the variation caused by  
the infinitesimal coordinate transformation in 
Eq.~(\ref{xprime}) and it
should be distinguished from 
the variation $\delta  \cdots$ due to the virtual 
displacement used in Secs.~III and IV C. 
%
The expression of the integral in Eq.~(\ref{dLGKa}) takes the form 
often found in conventional textbooks (see for example Ref.~\cite{Goldstein}) 
to give the change in the integral caused by the infinitesimal transformation. 
%
In the integrand in Eq.~(\ref{dLGKa}), the divergence term $\partial ( \xi^i {\cal L}_{GKa})/\partial x^i$ 
is obtained using Gauss's theorem for 
the difference between the domains of integrations in 
$x = (x^i)_{i=1,2,3}$ and $x' = (x'^i)_{i=1,2,3}$ 
while $\overline{\delta}{\cal L}_{GKa}$ is written using 
the Leibniz rule for the derivative operation by  
$\overline{\delta}$ as 
\begin{equation}
\label{dLGKa2}
\overline{\delta}  {\cal L}_{GKa}
=
 \int d^3 v \, \overline{\delta} (F_a  L_{GYa})
=
 \int d^3 v  \, ( \overline{\delta} F_a  \cdot L_{GYa} 
+ F_a \cdot \overline{\delta}   L_{GYa}
)
, 
\end{equation}
where 
$\overline{\delta} F_a$ and $\overline{\delta}   L_{GYa}$ 
represent
the variations in the spatial functional forms of 
$F_a$ and $L_{GYa}$ under the infinitesimal spatial 
coordinate transformation.  
Then, applying the chain rule formula for the derivative operation 
$\overline{\delta} L_{GYa}[u_{ax}^i, u_{a\vartheta}, 
\{ \partial_J A_i \}, \{ \partial_J \widehat{A}_i \}, \{ \partial_J \phi \}, \{ \partial_J g_{ij} \} ]$ 
yields 
\begin{eqnarray}
\label{dLGYa}
\overline{\delta} L_{GYa} 
 & = &
\frac{\partial L_{GYa}}{\partial u_{ax}^i} \overline{\delta} u_{ax}^i 
+ \frac{\partial L_{GYa}}{\partial u_{a\vartheta}} 
\overline{\delta} u_{a\vartheta}
+ \sum_J  \frac{\partial L_{GYa}}{\partial (\partial_J A_i)} 
\overline{\delta} (\partial_J A_i)
\nonumber \\ & & 
\mbox{} 
+ \sum_J  \frac{\partial L_{GYa}}{\partial  (\partial_J \widehat{A}_i)} 
\overline{\delta} (\partial_J \widehat{A}_i)
+ \sum_J  \frac{\partial L_{GYa}}{\partial (\partial_J \phi)} 
\overline{\delta} (\partial_J \phi)
\nonumber \\ & & 
\mbox{} 
+ \sum_J  \frac{\partial L_{GYa}}{\partial (\partial_J g_{ij})} 
\overline{\delta} (\partial_J g_{ij})
,
\end{eqnarray}
where 
$\partial L_{GYa}/\partial (\partial_J A_i) = 0$ 
when the order $n$ of $J=(j_1, \cdots, j_n)$ is greater than or equal to three [see Eq.~(\ref{LGYaf})]. 

As shown in Ref.~\cite{Sugama2021}, 
the variation in the functional form  under the infinitesimal spatial 
coordinate transformation can be represented by 
$\overline{\delta} = - L_\xi$, where 
$L_\xi$ is the Lie derivative~\cite{Marsden2} 
associated with the vector field given by $(\xi^i)_{i=1,2,3}$ 
and it acts on an arbitrary tensor field (as well as an arbitrary tensor field density). 
%
Using the fact that 
$L_\xi  {\cal L}_{GKa} =  
\partial ( \xi^i {\cal L}_{GKa} )/\partial x^i$ holds from the definition of the Lie derivative 
acting on a scalar density field, 
one finds that the integrand in Eq.~(\ref{dLGKa}) is written as 
$L_\xi  {\cal L}_{GKa} + \overline{\delta}{\cal L}_{GKa}$ which 
is found to vanish from $\overline{\delta} = - L_\xi$.
Then, the integral in Eq.~(\ref{dLGKa}) also vanishes and accordingly 
$\overline{\delta} L_{GKa} =0$, 
which means that 
$L_{GKa}$ is a scalar constant which is invariant under the coordinate transformation. 
Using Eqs.~(\ref{dLGKa})--(\ref{dLGYa}) and $\overline{\delta} = - L_\xi$,  
 $\overline{\delta} L_{GKa} =0$ can also be written as 
\begin{eqnarray}
\label{dLGKa3}
& &
\hspace*{-1mm}
\overline{\delta}
 L_{GKa}
  =
 \int_V d^3 x \int d^3 v \; 
 F_a \biggl( 
\xi^i \frac{\partial L_{GYa}}{\partial x^i} 
+ \overline{\delta} L_{GYa}
\biggr)
\nonumber \\ & & 
=
 \int_V d^3 x \int d^3 v \; 
 F_a \biggl( \xi^i \frac{\partial L_{GYa}}{\partial x^i}  + 
\frac{\partial L_{GYa}}{\partial u_{ax}^i} \overline{\delta} u_{ax}^i
+ \frac{\partial L_{GYa}}{\partial u_{a\vartheta}} 
\overline{\delta} u_{a\vartheta}
\nonumber \\ & & \mbox{} 
\hspace*{5mm}
+ \sum_J  \frac{\partial L_{GYa}}{\partial (\partial_J A_i)} 
\overline{\delta} (\partial_J A_i)
+ \sum_J  \frac{\partial L_{GYa}}{\partial (\partial_J \widehat{A}_i)} 
\overline{\delta} (\partial_J \widehat{A}_i)
\nonumber \\ & & \mbox{} 
\hspace*{5mm}
+ \sum_J  \frac{\partial L_{GYa}}{\partial (\partial_J \phi)} 
\overline{\delta} (\partial_J \phi)
+ \sum_J  \frac{\partial L_{GYa}}{\partial (\partial_J g_{ij})} 
\overline{\delta} (\partial_J g_{ij})
\biggr)
\nonumber \\
& &
=
0
.
\end{eqnarray}
%

Recall that $L_\xi$ annihilate any scalar constant. 
Therefore,  when $L_{GKa}$ is a scalar constant, 
one can naturally write 
$\overline{\delta} L_{GKa} = - L_\xi L_{GKa} (= 0)$. 
Thus, one can confirm 
that the relation $\overline{\delta}= - L_\xi$ is consistent and useful 
when treating all tensor variables (including scalar fields and scalar constants) 
and deriving the invariance properties associated with the coordinate transformation.  
This relation $\overline{\delta}= - L_\xi$ 
can be applied under the condition that all quantities, in which the variations due to the coordinate transformation are considered, can be written in terms of tensor fields (or tensor field densities) on which the operation of the Lie derivative can be defined. 
This condition means that integrals using such tensor fields yield scalar constants 
which represent geometric quantities and take invariant values 
independent of the choice of the spatial coordinates. 
It should be stressed that $\overline{\delta} L_{GKa} = - L_\xi L_{GKa} =0$ and 
Eq.~(\ref{dLGKa3}), which are derived only from 
the above-mentioned invariance property 
under the spatial coordinate transformation, 
are valid 
whether the gyrokinetic equation derived from the variational principle associated with 
the virtual variation in the phase-space trajectory in Secs.~II and III 
holds or not. 

In the same way as in Eq.~(\ref{dLGKa}),  
the invariance of the Lagrangian $L_{GKF}$ of the whole system 
under the infinitesimal spatial coordinate transformation 
can be written as 
\begin{eqnarray}
\label{dLGKF}
\overline{\delta} L_{GKF}
&  \equiv &
\int_V d^3 x  
\left(
\frac{\partial ( \xi^i {\cal L}_{GKF} )}{\partial x^i} 
+ 
\overline{\delta}
{\cal L}_{GKF}
\right)
\nonumber \\ 
& =  &
\sum_a
\overline{\delta}
 L_{GKa}
+
\int_V d^3 x  
\biggl(
\frac{\partial ( \xi^i {\cal L}_F )}{\partial x^i} 
+ 
\overline{\delta}
{\cal L}_F
\biggr)
\nonumber \\
& =  &
 0
,
\end{eqnarray}
where $L_{GKF}$ is defined by Eq.~(\ref{IGKF2}) with Eq.~(\ref{LGKF}). 
The variation $\overline{\delta} {\cal L}_F$ of the field Lagrangian 
density ${\cal L}_F$ defined in Eq.~(\ref{LGKF}) can be written as 
\begin{eqnarray}
\label{dLF}
& &
\hspace*{-3mm} 
\overline{\delta} {\cal L}_F
 = 
- \frac{\partial ( \xi^i {\cal L}_F)}{\partial x^i}
\nonumber \\
& & 
=
 \frac{\partial {\cal L}_F}{\partial (\partial_j A_i)} 
\overline{\delta} (\partial_j A_i)
+ \sum_J  \frac{\partial {\cal L}_F}{\partial (\partial_J \widehat{A}_i)} 
\overline{\delta} (\partial_J \widehat{A}_i)
\nonumber \\
&   & 
\hspace*{2mm}
+
\frac{\partial {\cal L}_F}{\partial (\partial_i \phi)} 
\overline{\delta} ( \partial_i \phi )
+  \sum_J \frac{\partial {\cal L}_F}{\partial ( \partial_J g_{ij} )} 
\overline{\delta} ( \partial_J g_{ij} )
+  \frac{\partial {\cal L}_F}{\partial \lambda} 
\overline{\delta} \lambda
,
\hspace*{7mm}
\end{eqnarray}
where $\overline{\delta} = - L_\xi$, 
$L_\xi {\cal L}_F =  \partial_i (\xi^i {\cal L}_F)$, 
 and the chain rule for 
$\overline{\delta}
{\cal L}_F [\{\partial_j A_i\}, \{\partial_J \widehat{A}_i\}, 
\{\partial_i \phi\}, 
\{\partial_J g_{ij}\}, \lambda ]$
are used. 
We now use Eqs.~(\ref{dLGKa3}),  and (\ref{dLF}) to rewrite 
Eq.~(\ref{dLGKF}) as 
\begin{eqnarray}
\label{dLGKF2}
& & 
\hspace*{-5mm} 
\overline{\delta}
L_{GKF}
= 
 \int_V d^3 x \biggl[ \sum_a \int d^3 v \,
 F_a \biggl( \xi^i \frac{\partial L_{GYa}}{\partial x^i}  + 
\frac{\partial L_{GYa}}{\partial u_{ax}^i} \overline{\delta} u_{ax}^i 
\nonumber \\ & & \mbox{} 
+ \frac{\partial L_{GYa}}{\partial u_{a\vartheta}} 
\overline{\delta} u_{a\vartheta}
+ \sum_J  \frac{\partial L_{GYa}}{\partial (\partial_J A_i)} 
\overline{\delta} (\partial_J A_i)
+ \sum_J  \frac{\partial L_{GYa}}{\partial (\partial_J \widehat{A}_i)} 
\overline{\delta} (\partial_J \widehat{A}_i)
\nonumber \\ & & \mbox{} 
+ \sum_J  \frac{\partial L_{GYa}}{\partial (\partial_J \phi)} 
\overline{\delta} (\partial_J \phi)
+ \sum_J  \frac{\partial L_{GYa}}{\partial (\partial_J g_{ij})} 
\overline{\delta} (\partial_J g_{ij})
\biggr)
+  \frac{\partial ( \xi^i {\cal L}_F)}{\partial x^i}
\nonumber \\ & & 
\mbox{} 
+  \frac{\partial {\cal L}_F}{\partial (\partial_j A_i)} 
\overline{\delta} (\partial_j A_i)
+ \sum_J  \frac{\partial {\cal L}_F}{\partial (\partial_J \widehat{A}_i)} 
\overline{\delta} (\partial_J \widehat{A}_i)
+
\frac{\partial {\cal L}_F}{\partial (\partial_i \phi)} 
\overline{\delta} ( \partial_i \phi )
\nonumber \\ 
&  & \mbox{} 
+  \sum_J \frac{\partial {\cal L}_F}{\partial ( \partial_J g_{ij} )} 
\overline{\delta} ( \partial_J g_{ij} )
+  \frac{\partial {\cal L}_F}{\partial \lambda} 
\overline{\delta} \lambda
\biggr]
\nonumber \\
&  & 
=  0
.
\end{eqnarray}

As seen from Eqs.~(\ref{dLGYa}), (\ref{dLGKa3}), (\ref{dLF}), 
and (\ref{dLGKF2}), 
the invariance of the scalar constants $L_{GKa}$ and $L_{GKF}$ 
under the infinitesimal spatial transformation can be confirmed using 
the chain rule formulas for the derivative operation 
$\overline{\delta} = - L_\xi$  acting on the scalar field 
$L_{GYa}[u_{ax}^i, u_{a\vartheta}, 
\{ \partial_J A_i \}, \{ \partial_J \widehat{A}_i \}, 
\{ \partial_J \phi \}, \{ \partial_J g_{ij} \} ]$ and the 
scalar field density 
${\cal L}_F[\{\partial_j A_i\}, \{\partial_J \widehat{A}_i\}, 
\{\partial_i \phi\}, 
\{\partial_J g_{ij}\}, \lambda ]$ which are given as composite functions. 
In Sec.V.B and Sec.~V.C, Eqs.~(\ref{dLGKa3}) and (\ref{dLGKF2}) are used to 
derive the local momentum balance equation for the single-particle-species 
system and that for the whole system consisting of particles of all species 
and electromagnetic fields.

\subsection{Momentum balance for a single particle species}

We now use $\overline{\delta} = - L_\xi$, 
$L_\xi u_{ax}^j = \xi^i \partial_i u_{ax}^j- u_{ax}^i \partial_i \xi^j$, 
$L_\xi u_{a\vartheta} = \xi^i \partial_i u_{a\vartheta}$, and 
the Euler-Lagrange equations for gyrocenter motion 
[Eq.~(\ref{dpidt}), 
$
( \partial L_{GYa}/\partial U )_u
= 0 
$, 
$
(\partial L_{GYa}/\partial \mu )_u
= 0 
$, 
and 
$
( d / d t )_a
( \partial L_{GYa} / \partial u_{a\vartheta} )
=
( \partial L_{GYa} / \partial \vartheta )_u
=
0  
$
]
to write the first three terms in the integrand on the right-hand side of 
Eq.~(\ref{dLGKa3}) as 
\begin{eqnarray}
\label{three_terms}
& & 
\hspace*{-6mm} 
 F \left(
\xi^i \frac{\partial L_{GYa}}{\partial x^i}
+\frac{\partial L_{GYa}}{\partial  u_{ax}^j} 
\overline{\delta} u_{ax}^j
+ \frac{\partial L_{GYa}}{\partial  u_{a \vartheta}} 
\overline{\delta}  u_{a\vartheta}
\right)
\nonumber \\ 
& & 
\hspace*{-3mm} 
=
\xi^i 
\Biggl[ 
\frac{\partial}{\partial t} 
\left( F_a \frac{\partial L_{GYa}}{\partial  u_{ax}^i} \right)+ \frac{\partial}{\partial U} 
\left( F_a u_{aU} \frac{\partial L_{GYa}}{\partial  u_{ax}^i} \right) 
\nonumber \\ 
& & \mbox{}
- \frac{\partial L_{GYa}}{\partial  u_{ax}^i} 
\biggl\{
\frac{\partial F_a}{\partial t} 
+ 
\frac{\partial}{\partial x^j} ( F_a u_{ax}^j)
+ 
\frac{\partial}{\partial U} ( F_a u_{aU})
\nonumber \\ 
& & \mbox{}
+ 
\frac{\partial}{\partial \vartheta} ( F_a u_{a\vartheta})
\biggr\}
\Biggr]
+ \frac{\partial}{\partial x^j} 
\biggl( \xi^i F_a u_{ax}^j 
\frac{\partial L_{GYa}}{\partial  u_{ax}^i} 
\biggr)
.
\end{eqnarray}
Then, substituting Eq.~(\ref{three_terms}) into Eq.~(\ref{dLGKa3}), 
using $\overline{\delta} \partial_J = \partial_J  \overline{\delta}$, 
and performing partial integrals yield
\begin{eqnarray}
\label{dLGKa4}
\overline{\delta}
L_{GKa}
& = & 
 \int_V d^3 x \, 
\left[ 
\xi^i \int d^3 v \left\{
\frac{\partial}{\partial t}
\left( 
 F_a 
\frac{\partial L_{GYa}}{\partial u_{ax}^i} 
\right)
\right.  \right. 
\nonumber \\ & & \mbox{} 
\left. 
-
D_t F_a
\frac{\partial L_{GYa}}{\partial u_{ax}^i} 
\right\}
+ \frac{\delta L_{GKa}}{\delta A_i} 
\overline{\delta} A_i
+ \frac{\delta L_{GKa}}{\delta \widehat{A}_i} 
\overline{\delta} \widehat{A}_i
\nonumber \\ & & \mbox{} 
\left. 
+ \frac{\delta L_{GKa}}{\delta \phi} 
\overline{\delta} \phi
+ \frac{\delta L_{GKa}}{\delta g_{ij}} 
\overline{\delta} g_{ij}
\right]
+  \mbox{B.T.}
\nonumber \\ 
& = & 0
, 
\end{eqnarray}
where 
\begin{eqnarray}
D_t F_a 
& \equiv & \frac{\partial F_a}{\partial t} 
+ \frac{\partial}{\partial x^j}
( F_a u_{ax}^j  )
+ \frac{\partial}{\partial U}
( F_a u_{a U}  )
+ \frac{\partial}{\partial \mu}
( F_a u_{a \mu} )
\nonumber \\ 
& & 
 \mbox{}
+ \frac{\partial}{\partial \vartheta}
( F_a u_{a \vartheta}  )
.
\end{eqnarray}
Furthermore, substituting 
$\overline{\delta} A_i = - \xi^j (\partial_j A_i) - (\partial_i \xi^j) A_j
= - \xi^j (\nabla_j A_i) - (\nabla_i \xi^j) A_j$, 
$\overline{\delta} \widehat{A}_i 
= - \xi^j (\partial_j \widehat{A}_i) - (\partial_i \xi^j) \widehat{A}_j
= - \xi^j (\nabla_j \widehat{A}_i) - (\nabla_i \xi^j) \widehat{A}_j$, 
$\overline{\delta} \phi = - \xi^j \partial_j \phi
= - \xi^j \nabla_j \phi$,  
and 
$\overline{\delta} g_{ij} = - \nabla_i \xi_j - \nabla_j \xi_i $ 
into Eq.~(\ref{dLGKa4})
and 
performing partial integrals, 
one obtains 
\begin{equation}
\label{dLGKa5}
\overline{\delta} L_{GKa}  
=
\int_V d^3 x \,
\xi^j  (J_{GKa})_j 
+ \mbox{B.T.} 
= 0
, 
\end{equation}
where
\begin{eqnarray}
\label{JGKaj}
\hspace*{-6mm}
(J_{GKa})_j
& \equiv & 
\frac{\partial}{\partial t}
\left( 
 \int d^3 v \,
 F_a 
\frac{\partial L_{GYa}}{\partial u_{ax}^j} 
\right)
- \int d^3 v \,
D_t F_a
\frac{\partial L_{GYa}}{\partial u_{ax}^j} 
\nonumber \\ & &
\mbox{} 
\hspace*{-3mm}
+ 2 \nabla_i  \left( g_{jk} \frac{\delta L_{GKa}}{\delta g_{ik}} \right)
- \frac{\delta L_{GKa}}{\delta \phi} 
\nabla_j \phi
- 
\frac{\delta L_{GKa}}{\delta A_k}
\nabla_j A_k 
\nonumber \\ & &
\mbox{} 
\hspace*{-3mm}
-
\frac{\delta L_{GKa}}{\delta \widehat{A}_k}
\nabla_j \widehat{A}_k
+
\nabla_k
\left(
\frac{\delta L_{GKa}}{\delta A_k}
A_j
+
\frac{\delta L_{GKa}}{\delta \widehat{A}_k}
\widehat{A}_j
\right)
.
\hspace*{3mm}
\end{eqnarray}
Here, it should be noted that 
\begin{equation}
\label{JGKaj0}
(J_{GKa})_j = 0
\end{equation}
holds because 
Eq.~(\ref{dLGKa5}) is valid for 
an arbitrary infinitesimal vector field  
represented by $\xi^j$ which vanishes on the boundary of $V$. 

Recalling that the canonical momentum of a single particle 
is given by  $p_{aj} \equiv \partial L_{GYa}/\partial u_{ax}^j$ , 
it is found that the first term on the right-hand side of 
Eq.~(\ref{JGKaj}) represents the rate of change in 
the momentum of particles per volume. 
The pressure tensor of the particle species species $a$ is given in terms of  
the variational derivative $\delta L_{GKa}/\delta g_{ij}$ as 
\begin{eqnarray}
\label{Paij}
P_a^{ij}
&  \equiv  & 
2 \frac{\delta L_{GKa}}{\delta g_{ij}}
\equiv
2 
\sum_J (-1)^{\#J}
\partial_J
\left(
\int d^3 v \; F_a \frac{\partial L_{GYa}}{\partial (\partial_J g_{ij})}
\right)
\nonumber \\ 
& = & 
P_{{\rm CGL}a}^{ij}
+ \pi_{\land a}^{ij}
+ \pi_{\parallel \Psi a}^{ij}
+ P_{\Psi a}^{ij}
, 
\end{eqnarray}
where $\#J = n$  represents the order of 
$J \equiv (j_1, j_2, \cdots, j_n )$, 
\begin{equation}
\label{PCGLa}
P_{{\rm CGL} a}^{ij}
=
\int d^3 v \, F_a
[ m_a U^2 b^i b^j + \mu B ( g^{ij} - b^i b^j ) ]
, 
\end{equation}
\begin{equation}
\label{piland}
\pi_{\land a}^{ij}
\equiv 
 \int d^3 v \, F_a
m_a U 
[ b^i ( u_{ax} )_\perp^j  + ( u_{ax} )_\perp^i b^j ]
, 
\end{equation}
\begin{equation}
\label{pipara}
\pi_{\parallel \Psi a}^{ij}
\equiv 
e_a  \int d^3 v \, F_a
U 
\frac{\partial \Psi_a}{\partial U}
\; b^i b^j 
, 
\end{equation}
and
\begin{equation}
\label{PPsia}
P_{\Psi a}^{ij}
\equiv 
- 2 e_a \frac{\delta}{\delta g_{ij}}
\left(
\int d^3 x \int d^3 v \; F_a \Psi_a 
\right)
.
\end{equation}
The pressure tensor 
$P_{{\rm CGL} a}^{ij}$ defined in Eq.~(\ref{PCGLa}) takes the 
Chew-Goldberger-Low (CGL) form~\cite{H&S} and it plays an 
essential role in the neoclassical transport. 
Effects of turbulent fluctuations on the momentum transport 
are included 
in $\pi_{\land a}^{ij}$, 
$\pi_{\parallel \Psi a}^{ij}$, 
and $P_{\Psi a}^{ij}$, 
which are detailedly investigated in Sec.~VIII. 
It is found from Eqs.~(\ref{Psi_a}), (\ref{spsia}), 
(\ref{vdef}), (\ref{vBa}),  and (\ref{pipara}) 
that $\pi_{\parallel \Psi a}^{ij}$ vanishes 
when $\widehat{\bf A} = 0$.

The particle density $N_a^{(p)}$ and the particle 
flux  $\Gamma_a^i$  of species $a$ are given from the 
functional derivatives
$\delta L_{GKa}/\delta \phi$ and 
$\delta L_{GKa}/\delta \widehat{A}_i$
by 
\begin{eqnarray}
\label{Nap}
e_a N_a^{(p)} 
&  \equiv & 
-
\frac{\delta L_{GKa}}{\delta \phi}
\equiv 
-
\sum_J (-1)^{\#J}
\partial_J
\left(
\int d^3 v \; F_a \frac{\partial L_{GYa}}{\partial (\partial_J \phi)} 
\right)
\nonumber  \\ 
& \equiv & 
-
\int d^3 v \; F_a \frac{\partial L_{GYa}}{\partial \phi} 
+ \sum_{n=1}^\infty (-1)^n
\nonumber \\ & & \mbox{}
\times
\partial_{j_1 \cdots j_{n-1}}
\left(
\int d^3 v \; F_a \frac{\partial L_{GYa}}{\partial 
(\partial_{j_1 \cdots j_{n-1}} (E_L)_{j_n})} 
\right)
,
\end{eqnarray}
and 
\begin{equation}
\label{Gamma_a}
\frac{e_a}{c} \Gamma_a^i
 \equiv 
 \frac{\delta L_{GKa}}{\delta \widehat{A}_i} 
\equiv
\sum_J (-1)^{\#J}
\partial_J
\left(
\frac{\partial {\cal L}_{GKa}}{\partial (\partial_J \widehat{A}_i)}  
\right) 
,
\end{equation}
respectively. 
The electric current density defined by 
$j^i \equiv \sum_a e_a \Gamma_a^i$
with the particle flux $\Gamma_a^i$ in Eq.~(\ref{Gamma_a})
enters the gyrokinetic Amp\`{e}re's law 
which is derived from the variational principle 
$\delta L_{GKF}/\delta \widehat{A}_i = 0$ in Eq.~(\ref{GKA3}). 
On the other hand, the variational derivative 
$\delta L_{GKa}/\delta A_i$ gives 
\begin{eqnarray}
\label{dLGKadA}
& & 
\hspace*{-3mm} 
\frac{e_a}{c} \Gamma_{\#a}^i
 \equiv 
 \frac{\delta L_{GKa}}{\delta A_i} 
\equiv
\sum_J (-1)^{\#J}
\partial_J
\left(
\frac{\partial {\cal L}_{GKa}}{\partial (\partial_J A_i)}  
\right) 
\nonumber \\ 
& & =
 \frac{\partial {\cal L}_{GKa}}{\partial A_i} 
- \frac{\partial}{\partial x^j}
\left(
\frac{\partial {\cal L}_{GKa}}{\partial (\partial_j A_i)}  
\right) 
+ \frac{\partial^2}{\partial x^j \partial x^k}
\left(
\frac{\partial {\cal L}_{GKa}}{\partial (\partial_j \partial_k A_i)}  
\right) 
,
\nonumber \\ & & 
\end{eqnarray}
from which another type of the particle flux $ \Gamma_{\#a}^i$ 
of species $a$ is derived as 
\begin{eqnarray}
\label{NV} 
 \Gamma_{\#a}^i
& \equiv & 
\int d^3 v \, F_a u_{ax}^i
+ \frac{c}{e_a} \epsilon^{ijk}
\frac{\partial }{\partial x^j}
\biggl(
\int d^3 v \, 
\frac{F_a}{\sqrt{g}}
\nonumber \\ & & 
\times \biggl[
- \mu b_k 
+\frac{m_a U}{B}
\left\{
(u_{ax})_k - (u_{ax})_l b^l b_k
\right\}
\nonumber \\ & & 
- e_a 
\biggl\{ 
\frac{\partial \Psi_a}{\partial B^k} 
-
\frac{1}{F_a}
\frac{\partial}{\partial x^l} 
\biggl( F_a
\frac{\partial \Psi_a}{\partial (\partial_l B^k)} 
\biggr)
\biggr\}
\biggr]
\biggr)
.
\hspace*{8mm}
\end{eqnarray}
Here, one note that 
neither of the last two terms in the last line 
of Eq.~(\ref{NV}) is seen to take the form of a vector 
component in the general spatial coordinate system. 
However, the sum of these two terms is shown to be 
rewritten by 
\begin{equation}
\label{PsiB}
\hspace*{-2mm} 
\frac{\partial \Psi_a}{\partial B^k} 
-
\frac{1}{F_a}
\frac{\partial}{\partial x^l} 
\biggl( F_a
\frac{\partial \Psi_a}{\partial (\partial_l B^k)} 
\biggr)
=
\left( 
\frac{\partial \Psi_a}{\partial B^k} 
\right)_{\nabla B}
-
\frac{1}{F_a}
\nabla_l
\biggl( F_a
\frac{\partial \Psi_a}{\partial (\nabla_l B^k)} 
\biggr)
.
\end{equation}
On the left-hand side of Eq.~(\ref{PsiB}), 
derivatives are performed with regarding $\Psi_a$ 
as taking a functional form of 
$\Psi_a (x^i, B^k (x^i), \partial_j B^k (x^i) )$
while, on the right-hand side, 
a functional form of 
$\Psi_a(x^i, B^k (x^i), \nabla_j B^k (x^i) )$ 
is considered. 
Then, each term of the right-hand side of Eq.~(\ref{PsiB}) 
is found to have the form of a vector component, 
and the sum of the two terms on the left-hand side 
also becomes a vector component. 
Therefore, it is understood that 
$\Gamma_{\#a}^i$ give by Eq.~(\ref{NV}) 
is transformed as a vector density under the general 
spatial coordinate transformation 
by noting that $F_a$ and $\epsilon^{ijk}$ are
a scalar density and a tensor density. 

The two types of particle fluxes 
$\Gamma_a^i = (c/e_a) \delta L_{GKa}/\delta \widehat{A}_i$ and 
$\Gamma_{\# a}^i = (c/e_a) \delta L_{GKa}/\delta A_i$ 
in Eqs.~(\ref{Gamma_a}) and (\ref{dLGKadA}) arise because 
the separation of the magnetic field into the average and fluctuating parts are done 
in the case of electromagnetic turbulence. 
As described in Ref.~\cite{Sugama2022}, 
it is $\Gamma_a^i$ that accurately represents 
both the average and fluctuating parts of the particle flux and is used to evaluate the current density 
in Amp\`{e}re's law as shown in Eqs.~(\ref{j}), (\ref{GKA3}), and (\ref{D1}). 
It is found from comparing 
 $\Gamma_{\# a}^i $ with
$\Gamma_a^i$
that the average part of $\Gamma_{\# a}^i$ equals that of $\Gamma_a^i$ to the lowest 
order in $\delta = \rho/L$
while their fluctuating parts differ from each other. 

It is emphasized here that 
$(J_{GKa})_j = 0$  in Eq.~(\ref{JGKaj0}) is valid for 
$(J_{GKa})_j $ defined in Eq.~(\ref{JGKaj}) 
where the gyrocenter distribution function $F_a$ can be arbitrarily chosen 
and it does not need to be a solution of the gyrokinetic 
Vlasov equation given by Eq.~(\ref{Vlasov2}) or Eq.~(\ref{GKE}). 
%
It is recalled that 
the variation associated with the spatial coordinate transformation 
should be clearly distinguished from the variation (or virtual displacement) 
used for deriving the gyrokinetic Vlasov equation; the fact that 
the the former variation of the Lagrangian 
vanishes can be used to derive the momentum balance equation even 
in a more general case where the governing kinetic equation differ from 
the gyrokinetic Vlasov equation derived using the latter variational principle. 
%
We now assume 
$F_a$ to satisfy not the gyrokinetic Vlasov equation, Eq.~(\ref{Vlasov2}), but 
a more general one, that is, the gyrokinetic Boltzmann equation 
given by 
\begin{eqnarray}
\label{GKB}
D_t F_a & \equiv & \frac{\partial F_a}{\partial t} 
+ \frac{\partial}{\partial x^j}
( F_a u_{ax}^j  )
+ \frac{\partial}{\partial U}
( F_a u_{a U}  )
+ \frac{\partial}{\partial \mu}
( F_a u_{a \mu} )
\nonumber \\ 
& & 
 \mbox{}
+ \frac{\partial}{\partial \vartheta}
( F_a u_{a \vartheta}  )
\nonumber \\ 
& = & {\cal K}_a
,
\end{eqnarray}
where  ${\cal K}_a$ represents the rate of temporal 
change in $F_a$ due to collisions and/or external sources for the species $a$. 
It is assumed in the present work that  
\begin{equation}
\label{eK}
\sum_a e_a \int d^3 v \,{\cal K}_a = 0
\end{equation}
is satisfied by ${\cal K}_a$.
Therefore, the charge density is not changed by ${\cal K}_a$. 
Using Eqs.~(\ref{JGKaj}), (\ref{Paij}),  (\ref{Nap})--(\ref{dLGKadA}), 
and (\ref{GKB}), 
Eq.~(\ref{JGKaj0}) implies that the solution of 
the gyrokinetic Boltzmann equation, Eq.~(\ref{GKB}), satisfies 
the canonical momentum balance equation, 
\begin{eqnarray}
\label{cmb1}
& & 
\hspace*{-3mm} 
\frac{\partial}{\partial t}
\left( 
 \int d^3 v \,
 F_a 
p_{aj}
\right)
- \int d^3 v \,
{\cal K}_a
p_{aj}
+ \nabla_i (P_a)^i_j
\nonumber \\ & &
\mbox{} 
=  
- e_a 
N_a^{(p)} 
\nabla_j \phi
+ 
\frac{e_a}{c} 
\Bigl[ \Gamma_{\#a}^k
\nabla_j A_k
+
\Gamma_a^k
\nabla_j \widehat{A}_k
\nonumber \\ & &
\mbox{} 
\hspace*{3mm}
- 
\nabla_k
\Bigl(
\Gamma_{\#a}^k
A_j
+
 \Gamma_a^k
\widehat{A}_j
\Bigr)
\Bigr]
, 
\end{eqnarray}
which can be written in the conventional dyadic notation representing 
vectors and tensors in terms of boldface letters 
as 
\begin{eqnarray}
\label{cmb2}
& & 
\hspace*{-3mm} 
\frac{\partial}{\partial t}
\left( 
 \int d^3 v \,
 F_a 
{\bf p}_a
\right)
- \int d^3 v \,
{\cal K}_a
{\bf p}_a
+ \nabla \cdot {\bf P}_a
\nonumber \\ & &
\mbox{} 
=  
- e_a 
N_a^{(p)} 
\nabla \phi
+ 
\frac{e_a}{c} 
\Bigl[ 
(\nabla  {\bf A} )
\cdot \boldsymbol{\Gamma}_{\#a}
+
(\nabla \widehat{\bf A} )
\cdot \boldsymbol{\Gamma}_a
\nonumber \\ & &
\mbox{} 
\hspace*{3mm}
- 
\nabla \cdot
\Bigl(
\boldsymbol{\Gamma}_{\#a}
{\bf A}
+
\boldsymbol{\Gamma}_a
\widehat{\bf A}
\Bigr)
\Bigr]
. 
\end{eqnarray}
Here, as expected from Noether's theorem, 
one can confirm that Eq.~(\ref{cmb2}) takes the conservation form of 
the canonical momentum in the direction of the 
constant vector ${\bf e}$
if the term including ${\cal K}_a$ vanishes and 
the electric and magnetic fields satisfy the symmetry conditions 
${\bf e} \cdot \nabla \phi = 0$ and 
${\bf e} \cdot \nabla {\bf A} = {\bf e} \cdot \nabla \widehat{\bf A} = 0$.
In a case where the electric and magnetic fields 
are axisymmetric, 
conservation of the toroidal angular momentum 
can also be derived from Eq.~(\ref{cmb2}) in the same manner
as shown in Sec.~VI. 

The canonical momentum balance equation, Eq.~(\ref{cmb2}), 
is also written as 
\begin{eqnarray}
\label{cmb3}
& & 
\hspace*{-3mm} 
\frac{\partial}{\partial t}
\left( 
 \int d^3 v \,
 F_a 
{\bf p}_a
\right)
- \int d^3 v \,
{\cal K}_a
{\bf p}_a
+ \nabla \cdot {\bf P}_a
\nonumber \\ & &
\mbox{} 
=  
- e_a 
N_a^{(p)} 
\nabla \phi
+ 
\frac{e_a}{c} 
\Bigl[
\Bigl(
\boldsymbol{\Gamma}_{\#a}
\times 
{\bf B} 
+
\boldsymbol{\Gamma}_a
\times
\widehat{\bf B}
\Bigr)
\nonumber \\ & &
\mbox{} 
\hspace*{3mm}
- 
{\bf A} 
(\nabla \cdot \boldsymbol{\Gamma}_{\#a} )
- 
\widehat{\bf A}
(\nabla \cdot \boldsymbol{\Gamma}_a )
\Bigr]
. 
\end{eqnarray}
Furthermore, Eq.~(\ref{cmb3}) is deformed to 
\begin{eqnarray}
\label{mombal}
& & 
\hspace*{-3mm} 
\frac{\partial}{\partial t} 
\left(  m_a N_a^{(g)} V_{a g \parallel} {\bf b} \right)
- \int d^3 v \, {\cal K}_a m_a U {\bf b}
+ \nabla \cdot {\bf P}_a
\nonumber \\
& & 
= 
e_a \left(
N_a^{(p)} {\bf E}_L
  + N_a^{(g)} {\bf E}_T 
  \right) 
\nonumber \\
& & 
\hspace*{3mm}
+ 
\frac{e_a}{c} 
\Bigl[
\Bigl(
\boldsymbol{\Gamma}_{\#a}
\times 
{\bf B} 
+
\boldsymbol{\Gamma}_a
\times
\widehat{\bf B}
\Bigr)
- 
\widehat{\bf A}
(\nabla \cdot \boldsymbol{\Gamma}_a )
\Bigr]
, 
\end{eqnarray}
where
${\bf E}_L = - \nabla \phi$ and 
${\bf E}_T = - c^{-1} \partial {\bf A}/\partial t$ are used.   
Here, 
\begin{equation}
\label{NVagpara}
 N_a^{(g)} 
\equiv 
\int d^3 v \, F_a 
\hspace{6mm}
\mbox{and} 
\hspace{6mm}
 N_a^{(g)} V_{ag \parallel}
\equiv 
\int d^3 v \, F_a U
\end{equation}
represent the density and the parallel flux of the gyrocenters, 
respectively. 
From Eqs.~(\ref{NV}), (\ref{GKB}), and (\ref{NVagpara}), 
one can obtain 
\begin{equation}
 \frac{\partial N_a^{(g)}}{\partial t}
+
\nabla \cdot \boldsymbol{\Gamma}_{\#a}
= \int d^3 v \; {\cal K}_a
,
\end{equation}
which is used to derive 
Eq.~(\ref{mombal}) from Eq.~(\ref{cmb3}). 
The first term on the left-hand side of Eq.~(\ref{mombal}) 
is the change rate of the density of the kinetic momentum 
which is obtained by extracting the vector potential term  
$
{\bf p}_a 
\equiv 
(e_a/c) {\bf A} + m_a U {\bf b}
$.
The second and third terms on the left-hand side of Eq.~(\ref{mombal}) 
represent the effects of collisions (or external sources) and 
the pressure tensor, respectively, 
while the right-hand side contains 
Lorentz forces due to the electric and magnetic fields. 
The last term on the right hand side appears due to 
the perturbed vector potential and it is carried over 
from Eq.~(\ref{cmb3}). 

%
It is noted that 
the time derivative terms in Eqs.~(\ref{cmb2})--(\ref{mombal}) are missing 
the perpendicular kinetic momentum part. 
This originates from the fact that 
the canonical momentum ${\bf p}_a$ associated with the gyrocenter Lagrangian 
does not include the perpendicular kinetic moment due to  
the perpendicular velocity ${\bf v}_\perp$ which depends on the gyrophase angle. 
The perpendicular part of the kinetic momentum density is given by 
$m_a \boldsymbol{\Gamma}_{a \perp}$ and its time derivative 
$m_a \partial \boldsymbol{\Gamma}_{a \perp}/\partial t$ 
is considered as neglected in the gyrokinetic 
momentum balance [Eqs.~(\ref{cmb2})--(\ref{mombal})]
which the gyrokinetic Boltzmann equation, Eq.~(\ref{GKB}), satisfies. 
The leading order of the magnitude of terms in the perpendicular part of 
Eqs.~(\ref{cmb2})--(\ref{mombal}) 
is given from the order of the Lorentz force terms and it is estimated to be 
${\cal O}(e_a | \boldsymbol{\Gamma}_{a \perp}| B/c) =
 {\cal O}(m_a \Omega_a |\boldsymbol{\Gamma}_{a \perp}|)$ 
where $|\boldsymbol{\Gamma}_{\# a \perp}| 
\sim |\boldsymbol{\Gamma}_{a \perp}| \sim N_a^{(g)} c |{\bf E}_L| / B \sim 
(\rho_a/L) N_a^{(g)} v_{Ta}$ 
is regarded as valid for both the average and fluctuating parts. 
Thus, the neglected term $m_a \partial \boldsymbol{\Gamma}_{a \perp}/\partial t$ 
in the perpendicular momentum balance is smaller than the leading-order terms 
by a factor ${\cal O}(\Omega_a^{-1} \partial/\partial t)$. 
Here, the transport time scale ordering gives
$\Omega_a^{-1} \partial/\partial t \sim (\rho_a/L)^3$ 
for the ensemble-averaged part 
while $\Omega_a^{-1} \partial/\partial t \sim \rho_a/L$ 
is obtained for the fluctuating part from the gyrokinetic ordering. 
Thus, neglecting $m_a \partial \boldsymbol{\Gamma}_{a \perp}/\partial t$ 
is not considered to give a significant influence on the perpendicular part of the local momentum balance 
although this higher-order term should be correctly included for accurately describing 
the flux-surface-averaged momentum balance along the symmetry direction in up-down symmetric tokamaks and stellarator-symmetric quasisymmetric 
stellarators~\cite{Sugama2011,Parra2012,Calvo2015}.

\subsection{Momentum balance for the whole system}

One can follow the same procedures as used 
in deriving Eq.~(\ref{dLGKa4}) 
to deform Eq.~(\ref{dLGKF2}) to
\begin{eqnarray}
\label{dLGKF3}
\overline{\delta}
L_{GKF}
& = & 
 \int_V d^3 x \, 
\left[ 
\xi^i  \sum_a \int d^3 v \left\{
\frac{\partial}{\partial t}
\left( 
 F_a 
\frac{\partial L_{GYa}}{\partial u_{ax}^i} 
\right)
\right.  \right. 
\nonumber \\ & & \mbox{} 
\left. 
-
D_t F_a
\frac{\partial L_{GYa}}{\partial u_{ax}^i} 
\right\}
+ \frac{\delta L_{GKF}}{\delta A_i} 
\overline{\delta} A_i
+ \frac{\delta L_{GKF}}{\delta \widehat{A}_i} 
\overline{\delta} \widehat{A}_i
\nonumber \\ & & \mbox{} 
\left. 
+ \frac{\delta L_{GKF}}{\delta \phi} 
\overline{\delta} \phi
+ \frac{\delta L_{GKF}}{\delta g_{ij}} 
\overline{\delta} g_{ij}
\right]
+  \mbox{B.T.}
\nonumber \\ & = & 0 
.
\end{eqnarray}
Substituting 
$\overline{\delta} A_i 
= - \xi^j (\nabla_j A_i) - (\nabla_i \xi^j) A_j$, 
$\overline{\delta} \widehat{A}_i 
= - \xi^j (\nabla_j \widehat{A}_i) - (\nabla_i \xi^j) \widehat{A}_j$, 
$\overline{\delta} \phi 
= - \xi^j \nabla_j \phi$
into Eq.~(\ref{dLGKF3}),  
and performing a partial integral, 
one obtains 
\begin{equation}
\label{dLGKF4}
\overline{\delta}
L_{GKF}
=
\int_V d^3 x \, 
\xi^j  (J_{GKF})_j 
+ \mbox{B.T.}
= 0
,
\end{equation}
where 
\begin{eqnarray}
\label{JGKF}
& & 
\hspace*{-5mm} 
(J_{GKF})_j 
 \equiv 
\frac{\partial}{\partial t}
\left( 
\sum_a \int d^3 v \,
 F_a 
\frac{\partial L_{GYa}}{\partial u_{ax}^i} 
\right)
- \sum_a \int d^3 v \,
D_t F_a
\frac{\partial L_{GYa}}{\partial u_{ax}^i} 
\nonumber \\ & &
\hspace*{5mm} 
\mbox{} 
- \frac{\delta L_{GKF}}{\delta \phi} 
\nabla_j \phi
- \frac{\delta L_{GKF}}{\delta A_i} 
\nabla_j A_i
- \frac{\delta L_{GKF}}{\delta \widehat{A}_i} 
\nabla_j \widehat{A}_i
\nonumber \\ & &
\hspace*{5mm} 
\mbox{} 
+ 
\nabla_i \left(
\frac{\delta L_{GKF}}{\delta A_i}
A_j
+\frac{\delta L_{GKF}}{\delta \widehat{A}_i} 
\widehat{A}_j
\right)
+ 2 \nabla_i 
\left( g_{jk} \frac{\delta L_{GKF}}{\delta g_{ik}}
\right)
. 
\nonumber \\ & &
\end{eqnarray}
Since Eq.~(\ref{dLGKF4}) is valid for any $\xi^j$ which vanishes on the boundary of $V$, 
\begin{equation}
\label{JGKFj0}
(J_{GKF})_j 
= 0
\end{equation}
holds for $(J_{GKF})_j$ defined in Eq.~(\ref{JGKF}) 
where $F_a$, $\phi$, $A_i$, and $\widehat{A}_i$ 
can be arbitrarily chosen and they do not need to 
be determined by any governing equations. 

When $F_a$, $\phi$, and $\widehat{A}_i$ satisfy the gyrokinetic 
Boltzmann equation shown in Eq.~(\ref{GKB}) and 
the gyrokinetic Poisson-Amp\`{e}re equations given by 
$\delta L_{GKF}/ \delta \phi = 0$ and  
$\delta L_{GKF}/ \delta \widehat{A}_i = 0$, 
Eqs.~(\ref{JGKF}) and (\ref{JGKFj0}) lead to 
the total canonical momentum balance equation, 
\begin{eqnarray}
\label{tcmbeq}
& & 
\hspace*{-3mm} 
\frac{\partial}{\partial t}
\left( 
\sum_a \int d^3 v \,
 F_a 
p_{aj}
\right)
- \sum_a \int d^3 v \,
{\cal K}_a
p_{aj}
+ \nabla_i \Theta^i_j
\nonumber \\ 
&  &
=
\frac{\delta L_{GKF}}{\delta A_i} 
\nabla_j A_i
- \nabla_i \Bigl(
\frac{\delta L_{GKF}}{\delta A_i} 
A_j
\Bigr)
, 
\end{eqnarray}
where 
$
p_{aj} \equiv \partial L_{GYa}/\partial u_{ax}^j
\equiv 
(e_a/c) A_j (x, t) + m_a U b_j (x, t)
$
and 
the total pressure tensor density 
$\Theta^{ij} \equiv \Theta^i_k g^{jk}$ is defined by 
\begin{equation}
\Theta^{ij}
\equiv 
2 \frac{\delta L_{GKF}}{\delta g_{ij}} 
\equiv 
\Theta_{GK}^{ij} + \Theta_F^{ij} 
. 
\end{equation}
The gyrokinetic part $\Theta_{GK}^{ij}$ of $\Theta^{ij}$ is 
written as  
\begin{equation}
\label{Theta}
 \Theta_{GK}^{ij}
\equiv 
2 \frac{\delta L_{GK}}{\delta g_{ij}} 
\equiv 
2 \sum_a
 \frac{\delta L_{GKa}}{\delta g_{ij}} 
=
P_{\rm CGL}^{ij}
+ \pi_\land^{ij} 
+ \pi_{\parallel \Psi}^{ij}
+ P_\Psi^{ij}
,
\end{equation}
where 
$P_{\rm CGL}^{ij}$, $\pi_\land^{ij}$, 
$\pi_{\parallel \Psi}^{ij}$, and $P_\Psi^{ij}$ are defined 
using Eqs.~(\ref{PCGLa})--(\ref{PPsia}) 
as 
\begin{equation}
\label{pppp}
\left[
P_{\rm CGL}^{ij}
,
\pi_\land^{ij} 
,
\pi_{\parallel \Psi}^{ij}
,
P_\Psi^{ij}
\right]
\equiv 
\sum_a
\left[
P_{\rm CGLa}^{ij}
,
\pi_{\land a}^{ij} 
,
\pi_{\parallel \Psi a}^{ij}
,
P_{\Psi a}^{ij}
\right]
.
\end{equation}
The field part 
$\Theta_F^{ij}$ 
of $\Theta^{ij}$
is given by  
\begin{eqnarray}
\label{Theta}
& &  \Theta_F^{ij}
\equiv 
2 \frac{\delta L_F}{\delta g_{ij}} 
\equiv 
2 \sum_J (-1)^{\#J}
\frac{\partial {\cal L}_F}{\partial (\partial_J g_{ij})} 
\nonumber \\
& & 
\hspace*{3mm}
=
\frac{\sqrt{g}}{4\pi} 
\left[ 
\frac{g^{ij} }{2} 
\left\{
(E_L)^k (E_L)_k + (B^k + \widehat{B}^k )(B_k+ \widehat{B}_k)
\right\}
\right.
\nonumber \\
& & 
\hspace*{8mm}
-
\left\{
(E_L)^i (E_L)^j+ (B^i + \widehat{B}^i ) (B^j + \widehat{B}^j )
\right\}
\nonumber \\ & & 
\hspace*{7mm}
\left. \mbox{}
+ \frac{1}{c}
\left\{
- g^{ij} ( \nabla_k \lambda ) \widehat{A}^k
+ ( \nabla^i \lambda ) \widehat{A}^j
+
( \nabla^j \lambda ) \widehat{A}^i
\right\}
\right]
,  
\hspace*{10mm}
\end{eqnarray}
which contains the well-known Maxwell stress tensor and 
the additional terms in the same form as 
found in the Vlasov-Darwin model.~\cite{Sugama2013}

Now, using Cartesian spatial coordinates and 
the conventional dyadic notation representing 
vectors and tensors in terms of boldface letters, 
Eq.~(\ref{tcmbeq}) is rewritten as 
\begin{eqnarray}
\label{tcmbeq2}
& & 
\frac{\partial}{\partial t}
\left( 
\sum_a \int d^3 v \,
 F_a 
{\bf p}_a
\right)
- \sum_a \int d^3 v \,
{\cal K}_a
{\bf p}_a
+ \nabla \cdot  \boldsymbol{\Theta}
\nonumber \\ 
&  &
\hspace*{5mm} 
=
( \nabla {\bf A} )
\cdot \frac{\delta L_{GKF}}{\delta {\bf A}} 
- \nabla \cdot \Bigl(
\frac{\delta L_{GKF}}{\delta{\bf A}} 
{\bf A}
\Bigr)
. 
\end{eqnarray}
Here,  $\delta L_{GKF}/\delta{\bf A}$ is given by 
\begin{eqnarray}
\label{dLGKFdA}
\frac{\delta L_{GKF}}{\delta {\bf A}} 
=
\frac{{\bf j}_\#}{c}
 -
 \frac{1}{4\pi} \nabla \times
({\bf B} + \widehat{\bf B} )
, 
\end{eqnarray}
where ${\bf j}_\#$ is 
defined from $\boldsymbol{\Gamma}_{\# a}$ in Eq.~(\ref{NV}) 
by 
\begin{equation}
\label{jmacro}
{\bf j}_\#
\equiv c \frac{\delta L_{GK}}{\delta {\bf A}} 
\equiv 
\sum_a e_a \boldsymbol{\Gamma}_{\# a}
\equiv 
{\bf j}^{(gc)} + c \nabla \times {\bf M}_\#
. 
\end{equation}
Here, ${\bf j}^{(gc)}$ is the gyrocenter current defined by 
Eq.~(\ref{C12}) in Appendix~C and 
${\bf M}_\#$ is given by 
\begin{eqnarray}
\label{Mmacro}
{\bf M}_\#
& \equiv & 
\sum_a e_a \int d^3 v \; F_a
\biggl[
- \mu {\bf b} 
+\frac{m_a U}{B}
({\bf u}_{ax})_\perp
\nonumber \\  & & \mbox{}
- e_a \biggl\{ 
\frac{\partial \Psi_a}{\partial {\bf B}} 
-
\frac{1}{F_a}
\frac{\partial}{\partial x^l} 
\biggl( F_a
\frac{\partial \Psi_a}{\partial (\partial_l {\bf B})} 
\biggr)
\biggr\}
\biggr]
. 
\end{eqnarray}
Now, one can confirm the validity of 
Noether's theorem again from Eq.~(\ref{tcmbeq2}) 
which takes the conservation form of 
the total canonical momentum in the direction 
specified by the constant vector ${\bf e}$
when the background magnetic field satisfies 
the symmetry condition 
${\bf e} \cdot \nabla {\bf A} = 0$
and 
$\sum_a \int d^3 v \; {\cal K}_a {\bf p}_a$ can be ignored. 
In Sec.~VI, toroidal angular momentum conservation is derived 
in the case of the axisymmetric background field. 
It should be noted 
that no specific conditions to determine 
${\bf A}$ are imposed from the variational principle in contrast 
to $\phi$ and $\widehat{\bf A}$ which are variationally determined. 
Thus, $\delta L_{GKF}/\delta{\bf A}$ given in Eq.~(\ref{dLGKFdA})
does not vanish generally. 

The total canonical momentum balance equation in Eq.~(\ref{tcmbeq2}) 
can be rewritten as 
\begin{eqnarray}
& & 
\frac{\partial}{\partial t}
\left( 
\sum_a \int d^3 v \,
 F_a 
{\bf p}_a
\right)
- \sum_a \int d^3 v \,
{\cal K}_a
{\bf p}_a
+ \nabla \cdot  \boldsymbol{\Theta}
\nonumber \\ & &
\hspace*{5mm} 
= 
\frac{\delta L_{GKF}}{\delta {\bf A}} 
\times {\bf B}
-  {\bf A} \; \nabla \cdot \Bigl(
\frac{\delta L_{GKF}}{\delta{\bf A}} 
\Bigr)
, 
\end{eqnarray}
which is also deformed to 
\begin{eqnarray}
\label{totmomb0}
& & 
\frac{\partial}{\partial t}
\left( 
\sum_a \int d^3 v \, F_a 
m_a U {\bf b}
\right) 
- \sum_a \int d^3 v \, {\cal K}_a m_a U 
{\bf b}
+ \nabla \cdot  \boldsymbol{\Theta}
\nonumber \\ 
& & 
\hspace*{5mm} 
= 
\rho_c^{(gc)}{\bf E}_T
+  
\frac{\delta L_{GKF}}{\delta {\bf A}} 
\times {\bf B}
, 
\end{eqnarray}
where $\rho_c^{(gc)} \equiv \sum_a  e_a N_a^{(g)}$ 
and 
${\bf E}_T \equiv - c^{-1} 
\partial {\bf A}/\partial t$. 
Equation~(\ref{totmomb0}) represents the total balance equation 
of the kinetic momentum instead of the canonical one. 
The effects of collisions (or external sources) and the total pressure tensor 
are shown on the left-hand side of Eq.~(\ref{totmomb0}) while 
the Lorentz forces due to the back ground inductive field and the 
background magnetic field appear on the right-hand side. 

Finally, Eq.~(\ref{totmomb0}) can be deformed through vector calculus to 
\begin{eqnarray}
& & 
\hspace*{-3mm} 
\frac{\partial}{\partial t}
\left( 
\sum_a \int d^3 v \, F_a 
m_a U {\bf b} 
+ \frac{1}{4\pi c} ( {\bf D}_L \times {\bf B} )
\right) 
\nonumber \\
& & \mbox{}
+
 \nabla \cdot  
\left( 
\boldsymbol{\Theta}
+ \frac{{\bf D}_L {\bf E}_T + {\bf E}_T {\bf D}_L}{4 \pi}
 \right)
+
\nabla \left( 
\frac{{\bf E}_T \cdot {\bf D}_L }{4 \pi}  
\right)
\nonumber \\ 
& & = 
\sum_a \int d^3 v \, {\cal K}_a m_a U {\bf b} 
+ 
\left(
\frac{\delta L_{GKF}}{\delta {\bf A}} 
\right)_T \times {\bf B}
,
\hspace*{6mm}
\end{eqnarray}
which is written in more detail as 
\begin{eqnarray}
\label{totmomb}
& & 
\hspace*{-3mm} 
\frac{\partial}{\partial t}
\left( 
\sum_a \int d^3 v \, F_a 
m_a U {\bf b} 
+ \frac{1}{4\pi c} ( {\bf D}_L \times {\bf B} )
\right) 
\nonumber \\ 
& & 
\mbox{} 
+
\nabla \cdot 
\left(
{\bf P}_{\rm CGL}
+ \boldsymbol{\pi}_\land
+ \boldsymbol{\pi}_{\parallel \Psi}
+ {\bf P}_\Psi
\right)
+
\nabla \left( 
\frac{|{\bf E}_L|^2}{8 \pi} + \frac{{\bf E}_T \cdot {\bf D}_L }{4 \pi}  
\right)
\nonumber \\ 
& & 
\mbox{} 
-  \nabla \cdot  \left( 
\frac{{\bf E}_L {\bf E}_L + {\bf D}_L {\bf E}_T + {\bf E}_T {\bf D}_L}{4 \pi}
\right) 
\nonumber \\ 
& & 
\mbox{} 
+
\nabla \left( 
\frac{|{\bf B}+\widehat{\bf B}|^2}{8 \pi}  
\right)
-  \nabla \cdot  
\left( \frac{ ({\bf B}+\widehat{\bf B})   ({\bf B}+\widehat{\bf B}) }{4 \pi} \right) 
\nonumber \\ 
& & 
\mbox{} 
- \nabla \left(
\frac{\nabla \lambda \cdot \widehat{\bf A} }{4\pi c}
\right)
+ \nabla \cdot
\left( \frac{ (\nabla \lambda )  \widehat{\bf A}
+ \widehat{\bf A} (\nabla \lambda ) }{4 \pi c} \right) 
\nonumber \\ 
& & = 
 \sum_a \int d^3 v \, {\cal K}_a m_a U {\bf b} 
+ 
\left(
\frac{({\bf j}_\#)_T }{c} 
- 
\frac{\nabla \times ({\bf B}+\widehat{\bf B})}{4\pi}  
\right) \times {\bf B}
.
\hspace*{8mm}
\end{eqnarray}
In Eq.~(\ref{totmomb}), ${\bf D}_L$ is the longitudinal part of the displacement vector 
defined by Eq.~(\ref{C4}) and ${\bf j}_\#$ is defined in Eq.~(\ref{jmacro}). 
The change rate 
of the kinetic momentum density plus the electromagnetic 
momentum density $({\bf D}_L \times {\bf B})/(4\pi c)$ 
is described by Eq.~(\ref{totmomb}). 
%
The left-hand side of Eq.~(\ref{totmomb}) shows all terms of momentum transport 
written as the divergence of pressure tensors due to particles' motion and Maxwell stresses 
including both average and fluctuating parts of the electromagnetic field. 
Except for the terms on the right-hand side, Eq.~(\ref{totmomb}) takes the conservation form 
similar to that of the total momentum conservation equation of the Vlasov-Darwin model 
derived in Refs.~\cite{Sugama2013}. 
Since ${\bf j}_\# = \sum_a e_a \boldsymbol{\Gamma}_{\# a}$ does not 
accurately represents the fluctuating part of the current density, 
$({\bf j}_\#)_T/c - (4\pi)^{-1}\nabla \times ({\bf B} + \widehat{\bf B})$ cannot 
be neglected as far as turbulent electromagnetic fields exist. 
In Sec.~VII, the self-consistency condition to determine  the average field ${\bf B}$ 
is considered to make the ensemble average of Eq.~(\ref{totmomb}) take
the conservation form. 
%

\section{MOMENTUM BALANCE IN TOROIDAL SYSTEMS}

In this section, we investigate the 
momentum balance in toroidal systems 
based on the results obtained in Sec.~V.
The background magnetic field 
${\bf B}$ with and without symmetry 
are considered and the momentum balance 
equations averaged over the ensemble and 
the flux surface are examined. 
Here, the background magnetic field ${\bf B}$ is assumed 
to satisfy the toroidal MHD equilibrium equation, 
\begin{equation}
\label{MHDeq}
\frac{1}{4 \pi}
( \nabla \times {\bf B} ) \times {\bf B}
= \nabla P_0
, 
\end{equation}
where $P_0$ is a magnetic flux surface function 
representing equilibrium pressure.

\subsection{Axisymmetric systems}

The axisymmetric toroidal background magnetic field is 
represented by 
\begin{equation}
\label{axB}
{\bf B} 
=
I  \nabla \zeta + 
 \nabla \zeta  \times \nabla \chi
,
\end{equation}
where $\zeta$ and 
$\chi$ represents the toroidal angle and the 
poloidal flux (divided by $2\pi$), respectively, 
and the covariant toroidal component 
$I$ is a flux surface function which 
is independent of the toroidal and poloidal 
angles. 
Denoting the major radius by $R$ and writing 
the contravariant basis vector ${\bf e}_\zeta$ 
in the toroidal direction as  
\begin{equation}
{\bf e}_\zeta  
\equiv 
R^2 \nabla \zeta
\equiv 
\frac{\partial {\bf x}}{\partial \zeta}
, 
\end{equation}
%
%
%
one obtains the following relation, 
\begin{equation}
\label{gez}
\nabla {\bf e}_\zeta  
=
R^{-1}  [
 ( \nabla R) {\bf e}_\zeta  
- {\bf e}_\zeta   ( \nabla R) ]
= {\bf n} \times {\bf I} 
= {\bf I} \times {\bf n} 
,  
\end{equation}
where ${\bf n} \equiv R^{-1} ({\bf e}_\zeta \times \nabla R)$ 
is the unit vector parallel to the direction of the major axis 
and ${\bf I}$ is the unit tensor. 
It is shown from Eq.~(\ref{gez}) that an arbitrary 
symmetric tensor ${\bf S}$ $(S^{ij}=S^{ji})$ 
satisfies 
\begin{equation}
\label{Sij}
(  \nabla \cdot {\bf S} ) \cdot {\bf e}_\zeta 
=
  \nabla \cdot  ({\bf S} \cdot {\bf e}_\zeta  )
. 
\end{equation}
It is also noted that $\nabla \cdot {\bf e}_\zeta = 0$ 
and 
\begin{equation}
\label{eS}
{\bf e}_\zeta \cdot \nabla S 
= \nabla  \cdot ( S {\bf e}_\zeta )
,  
\end{equation}
where $S$ is an arbitrary scalar function. 

In the axisymmetric background field ${\bf B}$, 
${\bf A}$ can also be given by 
the axisymmetric field which satisfies
\begin{equation}
\label{ezA}
{\bf e}_\zeta \cdot \nabla {\bf A}
= \frac{1}{R} {\bf A} \times {\bf n}
.  
\end{equation}
Then, the inner product of ${\bf e}_\zeta$ and 
Eq.~(\ref{tcmbeq}) is taken and Eqs.~(\ref{gez}), (\ref{Sij}), 
and (\ref{ezA}) are used to derive 
\begin{eqnarray}
\label{tcmbzeta}
& & 
\frac{\partial}{\partial t}
\left( 
\sum_a \int d^3 v \,
 F_a 
{\bf p}_a \cdot {\bf e}_\zeta
\right)
+ \nabla \cdot  
\Bigl[
\Bigl(
 \boldsymbol{\Theta} 
+ 
\frac{\delta L_{GKF}}{\delta{\bf A}} 
{\bf A} 
\Bigr)
\cdot {\bf e}_\zeta 
\Bigr]
\nonumber \\ 
&  &
\hspace*{5mm} 
=
\sum_a \int d^3 v \,
{\cal K}_a
{\bf p}_a \cdot {\bf e}_\zeta
. 
\end{eqnarray}
Except for the the right-hand side, 
Eq.~(\ref{tcmbzeta}) takes the conservation form of the canonical 
momentum conjugate to the toroidal angle as expected from 
Noether's theorem. 
It is found from the assumption given in Eq.~(\ref{eK}) 
that 
$
\sum_a \int d^3 v \,
{\cal K}_a
{\bf p}_a \cdot {\bf e}_\zeta
=
\sum_a \int d^3 v \, {\cal K}_a m_a U b_\zeta
$
where 
$b_\zeta = {\bf b} \cdot {\bf e}_\zeta$.
In the zero-gyroradius limit, when using a particle collision operator 
for ${\cal K}_a$, one finds that  
$\sum_a \int d^3 v \, {\cal K}_a m_a U b_\zeta$ 
to vanish because the momentum of particles is conserved in collisions. 
Furthermore, it can be shown that 
when ${\cal K}_a$ is given by 
the collision operator which appropriately includes the finite gyroradius 
effect,~\cite{Sugama2015,Brizard2004,Burby} 
$\sum_a \int d^3 v \, {\cal K}_a m_a U b_\zeta$  can be 
written as a divergence of the sum of classical momentum transport 
fluxes.
Therefore,
without external momentum sources,  
Eq.~(\ref{tcmbzeta}) keeps the conservation form 
even though the collision term is present. 
In addition to the case of axisymmetry, 
the canonical momentum conservation is confirmed 
in other cases of symmetry under continuous isometric transformations 
such as a translational symmetry and a helical (or screw) 
symmetry.~\cite{Sugama2021}

Next,  the toroidal component of the  momentum balance equation 
in Eq.~(\ref{totmomb}) 
is considered. 
The transverse part of ${\bf j}_\#$ on the right-hand side of 
Eq.~(\ref{totmomb})
can be written in terms of 
a certain field ${\bf B}_\#$ as 
\begin{equation}
({\bf j}_\#)_T 
=
\frac{c}{4\pi} \nabla \times {\bf B}_\#
\end{equation}
which is combined with  
${\bf B} \times {\bf e}_\zeta = \nabla \chi$
to derive
\begin{equation}
\label{jBz}
\bigl( ({\bf j}_\#)_T \times {\bf B} \bigr) 
\cdot {\bf e}_\zeta
=
({\bf j}_\#)_T 
\cdot \nabla \chi
=
\nabla \cdot
\left( 
\frac{c}{4\pi} {\bf B}_\# \times \nabla \chi 
\right). 
\end{equation}
One also obtains 
\begin{eqnarray}
\label{gBBz}
\hspace*{-3mm} 
\Bigl( \bigl( \nabla \times ( {\bf B} + \widehat{\bf B} ) \bigr) \times {\bf B} \Bigr) 
\cdot {\bf e}_\zeta
& = & 
\left( \nabla \times ( {\bf B} + \widehat{\bf B} ) \right)
\cdot \nabla \chi
\nonumber \\
& = &
\nabla \cdot
\left( 
( {\bf B} + \widehat{\bf B} ) \times \nabla \chi 
\right). 
\end{eqnarray}
Now, 
taking the inner product of Eq.~(\ref{totmomb}) and ${\bf e}_\zeta$, 
it is found that 
the toroidal angular momentum balance equation can be written in 
the following form,
\begin{eqnarray}
\label{ttmombal}
& & 
\hspace*{-3mm} 
\frac{\partial}{\partial t}
\left( 
\sum_a \int d^3 v \, F_a 
m_a U b_\zeta 
+ \frac{1}{4\pi c} {\bf D}_L \cdot \nabla \chi
\right) 
\nonumber \\ 
& & \mbox{}
+
 \nabla \cdot  
\Bigl[
\left( 
\boldsymbol{\Theta}
+ \frac{{\bf D}_L {\bf E}_T + {\bf E}_T {\bf D}_L}{4 \pi}
 \right)
\cdot {\bf e}_\zeta
+
\frac{{\bf E}_T \cdot {\bf D}_L }{4 \pi}  {\bf e}_\zeta
\Bigr]
\nonumber \\ 
& & \mbox{}
+\nabla \cdot
\Bigl[ \frac{1}{4\pi}
( {\bf B}_\# - 
{\bf B} - \widehat{\bf B} ) \times \nabla \chi 
\Bigr]
\nonumber \\ 
& & 
= \sum_a \int d^3 v \, {\cal K}_a m_a U b_\zeta
,
\end{eqnarray}
where 
$({\bf D}_L \times {\bf B}) \cdot {\bf e}_\zeta 
= {\bf D}_L \cdot \nabla \chi$
is used. 
The expression of the divergence term on the left-hand side 
of Eq.~(\ref{ttmombal}) 
is straightforwardly derived from Eq.~(\ref{totmomb}) 
using Eqs.~(\ref{Sij}), (\ref{eS}), (\ref{jBz}), and (\ref{gBBz}). 
Without external momentum sources,  
Eq.~(\ref{ttmombal}) keeps the conservation form 
in the same way as Eq.~(\ref{tcmbzeta}). 

It is shown above that the symmetry of the pressure tensor 
is essential in deriving the equation of 
the toroidal angular momentum conservation in an axisymmetric 
system. 
From this point of view, the derivation of the symmetric pressure 
tensor from the variational derivative of the Lagrangian with 
respect to the metric tensor is useful.

\subsection{Non-axisymmetric systems}

In non-axisymmetric toroidal systems, 
the background field ${\bf B}$ is expressed by 
\begin{equation}
{\bf B} = \nabla \psi (s) \times \nabla \theta 
+ \nabla \zeta \times \nabla \chi (s)
, 
\end{equation}
where 
$\theta$ and $\zeta$ are 
the poloidal and toroidal angles, respectively, and 
$s$ is an arbitrarily chosen flux-surface label. 
Here, we assume the background ${\bf B}$ to be stationary, 
$\partial {\bf B}/\partial t = 0$, for simplicity, and 
use the Hamada coordinates~\cite{Hamada} $(s, \theta, \zeta)$, in which 
the Jacobian $\sqrt{g} 
\equiv [\nabla s \cdot (\nabla \theta \times \zeta)]^{-1}$, 
the poloidal field 
$B^\theta \equiv {\bf B}\cdot \nabla \theta$, and 
the toroidal field 
$B^\zeta \equiv {\bf B}\cdot \nabla \zeta$ are 
flux-surface functions. 
Then,  
the contravariant basis vector ${\bf e}_\zeta$ 
in the toroidal direction is written as  
\begin{equation}
\label{ez2}
{\bf e}_\zeta  
\equiv 
\frac{\partial {\bf x}}{\partial \zeta}
\equiv 
\sqrt{g} \nabla s \times \nabla \theta
, 
\end{equation}
which satisfies $\nabla \cdot {\bf e}_\zeta = 0$
and ${\bf B} \times {\bf e}_\zeta = \nabla \chi$. 
Therefore, equations in the same form as
Eqs.~(\ref{eS}), (\ref{jBz}), and (\ref{gBBz}) hold while Eq.~(\ref{Sij}) 
does not because Eq.~(\ref{gez}) is not valid in the 
non-axisymmetric case. 
Now, taking the inner product of Eq.~(\ref{totmomb}) and ${\bf e}_\zeta$ 
gives 
the toroidal angular momentum balance equation in 
the following form,
\begin{eqnarray}
\label{tmombal_naxs}
& & 
\frac{\partial}{\partial t}
\left( 
\sum_a \int d^3 v \, F_a 
m_a U b_\zeta 
+ \frac{1}{4\pi c} {\bf D}_L \cdot \nabla \chi
\right) 
\nonumber \\ 
& & \mbox{}
\hspace*{5mm}
+
\nabla \cdot \bigl( \cdots \bigr)
+ {\bf e}_\zeta \cdot 
\bigl( \nabla \cdot ( {\bf P}_{\rm CGL} + \cdots ) \bigr)
\nonumber \\ 
& & 
\hspace*{3mm}
= 0 , 
\end{eqnarray}
where the effects of external momentum sources are ignored.
The conservation form is broken in Eq.~(\ref{tmombal_naxs}) 
because of 
${\bf e}_\zeta \cdot 
\bigl( \nabla \cdot ( {\bf P}_{\rm CGL} + \cdots ) \bigr)$
on the left-hand side of Eq.~(\ref{tmombal_naxs}). 
Here, the CGL and other pressure tensors and anisotropic 
Maxwell stress terms 
are included in 
${\bf e}_\zeta \cdot 
\bigl( \nabla \cdot ( {\bf P}_{\rm CGL} + \cdots ) \bigr)$ 
while, using  Eq.~(\ref{eS}), 
the tensors proportional to the unit tensor
can be transferred to the inside of 
the divergence term $\nabla \cdot \bigl( \cdots \bigr)$ 
in Eq.~(\ref{tmombal_naxs}). 
Using the flux-surface average 
defined by 
$\langle \cdots \rangle \equiv \oint d\theta \oint d\zeta \;
\sqrt{g} \cdots / V'(s)$ 
with 
$V'(s) \equiv \oint d\theta \oint d\zeta \;\sqrt{g}$, 
it is found that 
the divergence term $\nabla \cdot ( \cdots )$ in 
Eq.~(\ref{tmombal_naxs}) 
is annihilated by the flux-surface average 
because, as seen from Eqs.~(\ref{eS}), (\ref{jBz}), 
(\ref{gBBz}), and (\ref{ez2}), 
the inner products of the vectors in $( \cdots )$ 
and $\nabla s$ vanish and 
\begin{equation}
\langle \nabla \cdot {\bf T} \rangle 
= \frac{1}{V'(s)} \frac{d}{ds} 
\left( V'(s) \langle {\bf T}\cdot \nabla s \rangle \right)
\end{equation}
holds for any vector ${\bf T}$. 
Then, taking the flux-surface average, 
Eq.~(\ref{tmombal_naxs}) is reduced to the more compact form,
\begin{eqnarray}
\label{av_tmombal_naxs}
& & 
\frac{\partial}{\partial t}
\left\langle 
\sum_a \int d^3 v \, F_a 
m_a U b_\zeta 
+ \frac{1}{4\pi c} {\bf D}_L \cdot \nabla \chi
\right\rangle 
\nonumber \\ 
& & \mbox{}
\hspace*{5mm}
+ \left\langle  {\bf e}_\zeta \cdot 
\bigl( \nabla \cdot ( {\bf P}_{\rm CGL} + \cdots ) \bigr)
\right\rangle 
\nonumber \\ 
& & 
\hspace*{3mm}
= 0 . 
\end{eqnarray}
It is recalled that, 
in neoclassical theory for non-axisymmetric systems,~\cite{Wakatani} 
the lowest-order toroidal viscosity is given by 
$
\left\langle  {\bf e}_\zeta \cdot 
( \nabla \cdot {\bf P}_{\rm CGL} ) 
\right\rangle 
$. 
It is shown in Sec.~VII that, 
when using the ensemble average and the gyroradius expansion of  
Eq.~(\ref{av_tmombal_naxs}) in general non-axisymmetric toroidal systems,  
this neoclassical toroidal viscosity 
becomes a dominant term. 

\subsection{Quasi-axisymmetric systems}

In this subsection,  quasi-axisymmetric toroidal systems~\cite{QS3}
are considered using Eq.~(\ref{av_tmombal_naxs}) with 
the Hamada coordinates $(s, \theta, \zeta)$ 
to represent the equilibrium magnetic field ${\bf B}$. 
The quasi-axisymmetry is characterized by $B = |{\bf B}|$ 
being independent of the toroidal angle, 
$
\partial B / \partial \zeta = 0
$, 
which is equivalent to 
$\partial B/\partial \zeta_B = 0$ in the 
Boozer coordinates~\cite{Boozer} 
$(s, \theta_B, \zeta_B)$
as proved in Ref.~\cite{Sugama2002}. 
In the quasi-axisymmetric equilibrium field 
${\bf B}$,  
the CGL-type pressure tensor 
${\bf P}_{\rm CGL} \equiv
P_\parallel {\bf b} {\bf b}  + P_\perp ({\bf I} - {\bf b}  {\bf b} )$
is shown to satisfy 
\begin{equation}
\label{qstv}
\bigl\langle 
{\bf e}_\zeta \cdot ( \nabla \cdot {\bf P}_{\rm CGL} ) 
\bigr\rangle 
=
\left\langle 
( P_\perp -  P_\parallel) \frac{\partial \ln B}{\partial \zeta}
\right\rangle 
= 0
,
\end{equation}
which implies that  
the dominant neoclassical toroidal viscosity which exists 
in general non-axisymmetric systems vanishes as in 
the axisymmetric systems. 
Thus, one of the factors preventing conservation of the 
toroidal angular momentum in Eq.~(\ref{av_tmombal_naxs}) disappears even though 
perfect conservation is not allowed. 
Because of Eq.~(\ref{qstv}), 
the magnitude of the dominant terms in Eq.~(\ref{av_tmombal_naxs}) becomes of higher order. 
Then, as mentioned in Sec.~II, 
basic gyrokinetic equations including higher-order terms, 
which are not considered in the present study, 
are required to accurately describe 
the flux-surface-averaged momentum balance along the symmetry 
direction in stellarator-symmetric quasisymmetric stellarators as well as 
up-down symmetric tokamaks.~\cite{Sugama2011,Parra2012,Calvo2015} 

\section{ENSEMBLE-AVERAGED MOMENTUM BALANCE}

In this section,  
the momentum balance equation in Eq.~(\ref{totmomb}) is ensemble-averaged,  
by which all terms in the equation are smoothed to make 
their space-time scales of variations 
much larger than those of fluctuations.  
The ensemble average is used as the basic method of statistical mechanics 
to obtain the macroscopic mean values of physical valuables and 
it can also be considered to equal the local space–time average, 
the definition of which is described in detail in Ref.~\cite{Abel}.
The fluctuations are assumed to have wavelengths of the order of gyroradii 
in the directions perpendicular to the background magnetic field, and 
they are treated by the WKB representation in Sec.~VIII. 

Since the background magnetic field ${\bf B}$ is considered to include 
no fluctuations, 
one can write ${\bf B} = \langle {\bf B} \rangle_{\rm ens}$, 
where $\langle \cdots\rangle_{\rm ens}$ represents the 
ensemble average. 
It should also be noted here that no equation to determine the 
background magnetic field ${\bf B}$ is given from the variational principle 
while ${\bf B}$ is allowed to change with time in the present model. 
If the variational condition $\delta L_{GKF}/\delta {\bf A} = 0$ was employed, 
${\bf B}$ would include fluctuation components as seen from Eq.~(\ref{dLGKFdA}). 
Then, $\langle (\delta L_{GKF}/\delta {\bf A})_T \rangle_{\rm ens} = 0$ is assumed here 
instead of $\delta L_{GKF}/\delta {\bf A} = 0$ as the condition for determining 
${\bf B}$. 
Using Eq.~(\ref{dLGKFdA}), 
$\langle (\delta L_{GKF}/\delta {\bf A})_T  \rangle_{\rm ens} = 0$  is written as
\begin{equation}
\label{Bmacro}
\nabla \times {\bf B}
=
\frac{4\pi}{c} 
\langle ( {\bf j}_\#  )_T \rangle_{\rm ens}
, 
\end{equation}
which seems to be appropriate because 
the equilibrium part of 
${\bf j}_\# \equiv c \delta L_{GK}/\delta {\bf A}$ 
[see Eq.~(\ref{jmacro})] 
is equal to that of 
${\bf j} \equiv c \delta L_{GK}/\delta \widehat{\bf A}$ 
[see Eq.~(\ref{D1})] 
to the lowest order in the gyroradius expansion 
and it represents the current density consistent with the MHD equilibrium. 
%
In the neoclassical transport theory~\cite{H&S},  
the self-consistency condition for the background field ${\bf B}$ 
evolving in the transport time scale is given from the MHD equilibrium equation. 
In passing, the condition for ${\bf B}$ can be derived as the variational equation
in the drift kinetic model~\cite{Sugama2018,Brizard2023} and 
the variational derivation of the time-evolving axisymmetric background field is 
considered in the gyrokinetic model~\cite{Sugama2017,Sugama2014}.  

With the help of  Eq.~(\ref{jmacro}), Eq.~(\ref{Bmacro}) can be rewritten as 
\begin{equation}
\nabla \times \langle {\bf H}_\# \rangle_{\rm ens}
=
\frac{4\pi}{c} 
\langle ( {\bf j}^{(gc)})_T \rangle_{\rm ens}
, 
\end{equation}
Here, ${\bf j}^{(gc)}$ is the gyrocenter current given by Eq.~(\ref{C12})
and 
${\bf H}_\# $ is defined using ${\bf M}_\#$ in Eq.~(\ref{Mmacro}) as 
\begin{equation}
\label{Hmacro}
{\bf H}_\# \equiv {\bf B} + \widehat{\bf B} - 4 \pi {\bf M}_\#
, 
\end{equation}
from which one has
\begin{equation}
 \langle {\bf H}_\# \rangle_{\rm ens}
= {\bf B} - 4 \pi \langle  {\bf M}_\# \rangle_{\rm ens}
.
\end{equation}
Using Eq.~(\ref{Bmacro}), the ensemble average of Eq.~(\ref{totmomb}) 
is written as 
\begin{eqnarray}
\label{av-totmomb}
& & 
\hspace*{-5mm} 
\frac{\partial}{\partial t}
\left\langle
\sum_a \int d^3 v \, F_a 
m_a U {\bf b} 
+ \frac{1}{4\pi c} ( {\bf D}_L \times {\bf B} )
\right\rangle_{\rm ens}
\nonumber \\ 
& & 
\mbox{} 
\hspace*{-4mm} 
+
\nabla \cdot \left\langle
{\bf P}_{\rm CGL}
+ \boldsymbol{\pi}_\land
+ \boldsymbol{\pi}_{\parallel \Psi}
+ {\bf P}_\Psi
\right\rangle_{\rm ens}
+
\nabla \left\langle
\frac{|{\bf E}_L|^2}{8 \pi} + \frac{{\bf E}_T \cdot {\bf D}_L }{4 \pi}  
\right\rangle_{\rm ens}
\nonumber \\ 
& & 
\hspace*{-2mm} 
\mbox{} 
-  \nabla \cdot  \left\langle
\frac{{\bf E}_L {\bf E}_L + {\bf D}_L {\bf E}_T + {\bf E}_T {\bf D}_L}{4 \pi}
\right\rangle_{\rm ens}
\nonumber \\ 
& & 
\hspace*{-2mm} 
\mbox{} 
+
\nabla \left\langle
\frac{|{\bf B}+\widehat{\bf B}|^2}{8 \pi}  
\right\rangle_{\rm ens}
-  \nabla \cdot  
\left\langle
 \frac{ ({\bf B}+\widehat{\bf B})   ({\bf B}+\widehat{\bf B}) }{4 \pi} 
\right\rangle_{\rm ens}
\nonumber \\ 
& & 
\hspace*{-2mm} 
\mbox{} 
- \nabla \left\langle
\frac{\nabla \lambda \cdot \widehat{\bf A} }{4\pi c}
\right\rangle_{\rm ens}
+ \nabla \cdot
\left\langle
\frac{ (\nabla \lambda )  \widehat{\bf A}
+ \widehat{\bf A} (\nabla \lambda ) }{4 \pi c} 
\right\rangle_{\rm ens}
\nonumber \\ 
& & 
\hspace*{-3mm} 
= 
 \sum_a 
\left\langle
\int d^3 v \, {\cal K}_a m_a U {\bf b} 
\right\rangle_{\rm ens}
.
\hspace*{6mm}
\end{eqnarray}
The ensemble-averaged momentum balance equation, 
Eq.~(\ref{av-totmomb}), takes the conservation form 
when no external sources of momentum exist and 
the right-hand side is written 
as a divergence of the tensor representing 
classical momentum transport. 
It is emphasized here that the ensemble-averaged momentum 
conservation described above is satisfied even in non-axisymmetric 
systems when the background field ${\bf B}$ is determined by the equilibrium 
condition given in Eq.~(\ref{Bmacro}). 
It is interesting to compare Eq.~(\ref{av-totmomb}) with 
the momentum conservation law shown by Eqs.~(31)--(33) 
in Ref.~\cite{Sugama2013}
for the Vlasov-Poisson-Amp\`{e}re 
(or Vlasov-Darwin) system 
in which collisional effects are ignored and 
the magnetic field is not divided into background and turbulent parts.
One can see that kinetic and electromagnetic momenta, kinetic and electromagnetic  
pressure tensors, and longitudinal and transverse electric fields in 
the momentum conservation equation of the Vlasov-Darwin system appear 
in Eq.~(\ref{av-totmomb}) in a similar manner and 
that Eq.~(\ref{av-totmomb}) 
additionally includes 
polarization, magnetization, 
and other higher-order terms due to finite-gyroradius effects and electromagnetic 
microturbulence. 
The similarities and differences described above are regarded as natural results 
because the electromagnetic gyrokinetic systems are derived from 
the Vlasov-Darwin system through ordering assumptions regarding
gyroradius scales and fluctuation amplitudes. 

Hereafter in this section, Eq.~(\ref{av-totmomb}) is expanded using the ordering parameter 
given by the normalized gyroradius 
$\delta = \rho / L$ which is the ratio of the gyroradius $\rho$ to the equilibrium 
scale length $L$. 
%
%
The zeroth-order part $F_{a0}$ of the distribution function 
is assumed to be given by the local Maxwellian 
as 
\begin{eqnarray}
\label{FaM}
& & 
\hspace*{-5mm}
F_{a0} =
\langle F_{a0} \rangle_{\rm ens}
= 
F_{aM} \equiv D_{a0} f_{aM}
\nonumber \\
&  & 
\hspace*{-2mm}
\equiv
N_{a0} D_{a0}
\left( \frac{m_a}{2\pi T_{a0}} \right)^{3/2}
\exp \left[- \frac{1}{T_{a0}} 
\left( \frac{1}{2} m_a U^2 + \mu B \right) \right]
,
\hspace*{8mm}
\end{eqnarray}
where $N_{a0}$ and $T_{a0}$ are 
the background density and temperature of the particle species $a$, respectively, 
and $D_{a0} \equiv B/m_a$ is the zeroth-order part of 
$
D_a
\equiv
B^*_{a\parallel}/m_a
$. 
Here, the transport time scale ordering is used for the ensemble-averaged 
variables in Eq.~(\ref{av-totmomb}) which means that
$
\partial  \langle \cdots \rangle_{\rm ens} /  \partial t
 \sim \delta^2 ( v_T / L )
\langle \cdots \rangle_{\rm ens}
$
where $v_T$ is thermal velocity. 
The CGL pressure tensor defined in Eq.~(\ref{PCGLa}) 
is expanded in $\delta = \rho / L$ as  
\begin{equation}
{\bf P}_{{\rm CGL}a } = P_{a0} {\bf I} + ({\bf P}_{{\rm CGL}a})_1 +  {\cal O}(\delta^2)
, 
\end{equation}
where the zeroth-order part is isotropic and expressed by the scalar pressure, 
$P_{a0} \equiv N_{a0} T_{a0}$. 
Then, it is found that the zeroth-order part of the 
ensemble-averaged momentum conservation equation, Eq.~(\ref{av-totmomb}), 
is given by 
\begin{equation}
\label{P0B}
\nabla 
\left(
P_0 + \frac{B^2}{8\pi}
\right)
- \frac{1}{4 \pi} 
\nabla \cdot \left( {\bf B}   {\bf B} \right)
=
0
\end{equation}
where the equilibrium pressure is defined by 
$
P_0 \equiv  \sum_a P_{a0} 
\equiv \sum_a N_{a0} T_{a0}
$.
Equation~(\ref{P0B}) is easily confirmed to be equivalent to Eq.~(\ref{MHDeq}) 
representing the MHD equilibrium.


The first-order part of Eq.~(\ref{av-totmomb}) comes only from the CGL pressure tensor 
because the other pressure tensors and the turbulent Maxwell stress tensors 
are of the second order. 
Thus, one obtains 
\begin{equation}
\label{CGL1}
\nabla \cdot \langle ({\bf P}_{\rm CGL})_1 \rangle_{\rm ens}
=
0
. 
\end{equation}
The turbulent part $\widehat{F}_a$ of the distribution function $F_a$ 
has  
no contribution to $({\bf P}_{\rm CGL})_1$ and to Eq.~(\ref{CGL1}) because 
$\langle \widehat{F}_a \rangle_{\rm ens} = 0$. 
In neoclassical transport theory,~\cite{Hinton1976,H&S,Helander}  
the parallel component of 
$\nabla \cdot \langle ({\bf P}_{\rm CGL})_1 \rangle_{\rm ens}$ in 
Eq.~(\ref{CGL1}) automatically vanishes because 
of a quasineutrality condition and the momentum conservation property of the collision term. 
Also, 
the flux-surface average of the toroidal component of Eq.~(\ref{CGL1}), 
\begin{equation}
\label{CGL1z}
\langle \langle {\bf e}_\zeta \cdot
\left( \nabla \cdot  ({\bf P}_{\rm CGL})_1 \right) \rangle \rangle
= 0
,
\end{equation}
automatically holds in axisymmetric and quasi-axisymmetric toroidal systems 
as described in Sec.~VI. 
Here, $\langle \langle \cdots \rangle \rangle$ 
denotes the average over the flux surface and the ensemble. 
In general non-axisymmetric systems such as 
stellarator and heliotron plasmas, 
Eq.~(\ref{CGL1z}) is not automatically 
satisfied but it imposes an ambipolarity condition on 
neoclassical particle fluxes, 
from which the background radial electric field can be 
determined.~\cite{Wakatani} 


The effects of electromagnetic microturbulence with perpendicular 
wavelengths on the gyroradius scale appear on the second 
order in Eq.~(\ref{av-totmomb}). 
In Sec.~VIII, the turbulence contributions  
to the momentum transport are investigated in detail 
using the WKB representation for turbulent fluctuations.

\section{WKB REPRESENTATION}

The WKB (or ballooning) representation~\cite{WKB} 
is useful for treating turbulent fluctuations which 
have small wavelengths of the order of the gyroradius $\rho$ in the directions  
perpendicular to the background magnetic field. 
The WKB representation 
for the fluctuation part $\widehat{Q}$ of an arbitrary 
function $Q ({\bf x}, t)$ takes the form, 
\begin{equation}
\label{D-1}
\widehat{Q} ({\bf x}, t)
= 
\sum_{{\bf k}_\perp} \widehat{Q}_{{\bf k}_\perp} ({\bf x}, t)
\exp [ i S_{{\bf k}_\perp} ({\bf x}, t) ]
,
\end{equation}
where $\widehat{Q}_{{\bf k}_\perp}({\bf x}, t)$ has 
the equilibrium gradient scale length $L$
while the eikonal $S_{{\bf k}_\perp} ({\bf x}, t)$ represents 
a rapid variation with the wave number vector 
${\bf k}_\perp \equiv \nabla S_{{\bf k}_\perp} (\sim \rho^{-1})$ 
which satisfies ${\bf k}_\perp \cdot {\bf b} = 0$.

In the gyroradius expansion using $\delta = \rho/L \ll 1$, 
the zeroth-order gyrocenter distribution function is assumed to 
be given by the local Maxwellian, 
$F_{a0} = F_{aM} = D_{a0} f_{aM}$, as shown in Eq.~(\ref{FaM}). 
The fluctuation part of $F_a$ appears in the first order and 
it is given by the WKB representation as 
$
\widehat{F}_{a1}
= 
\sum_{{\bf k}_\perp} \widehat{F}_{a1{\bf k}_\perp}
\exp [ i S_{{\bf k}_\perp} ({\bf X}, t) ]
$,
where ${\bf X}$ is the gyrocenter position vector.  
As shown in Ref.~\cite{Sugama2022}, 
the ${\bf k}_\perp$-component $\widehat{F}_{a1{\bf k}_\perp}$ 
is written as 
$\widehat{F}_{a1{\bf k}_\perp} = D_{a0}\widehat{f}_{a1{\bf k}_\perp}$, 
where $\widehat{f}_{a1{\bf k}_\perp}$ is given by 
\begin{equation}
\label{ad-nad}
\hat{f}_{a1{\bf k}_\perp}
=
- f_{aM} \frac{e_a}{T_a}
\langle \widehat{\psi}_a \rangle_{\xi{\bf k}_\perp}
+ \widehat{h}_{a{\bf k}_\perp}
. 
\end{equation}
Here,  $\widehat{h}_{a{\bf k}_\perp}$
is the nonadiabatic part of the turbulent 
distribution function and 
the gyrophase-averaged potential 
$\langle \widehat{\psi}_a \rangle_{\xi{\bf k}_\perp}$ 
is defined by 
\begin{equation}
\hspace*{-2mm}
\langle \widehat{\psi}_a \rangle_{\xi{\bf k}_\perp}
=
J_0 \left( \frac{k_\perp v_\perp}{\Omega_a} \right)
\left(
\widehat{\phi}_{{\bf k}_\perp} 
- \frac{U}{c}  \widehat{A}_{\parallel{\bf k}_\perp}
\right)
+
J_1 \left( \frac{k_\perp v_\perp}{\Omega_a} \right)
\frac{v_\perp}{c}
\frac{\widehat{B}_{\parallel{\bf k}_\perp}}{k_\perp}
,
\end{equation}
where $J_0$ and $J_1$ are the Bessel functions. 
In addition, another kind of gyrophase-averaged potential 
is defined by 
\begin{eqnarray}
\langle \widehat{\chi}_a \rangle_{\xi{\bf k}_\perp}
 & = & 
- \frac{k_\perp v_\perp}{\Omega_a} J_1 \left( \frac{k_\perp v_\perp}{\Omega_a} \right)
\left(
\widehat{\phi}_{{\bf k}_\perp} 
- \frac{U}{c}  \widehat{A}_{\parallel{\bf k}_\perp}
\right)
\nonumber \\  & & \mbox{} 
+
\left[ 
\frac{k_\perp v_\perp}{\Omega_a}
J_0 \left( \frac{k_\perp v_\perp}{\Omega_a} \right)
- J_1 \left( \frac{k_\perp v_\perp}{\Omega_a} \right)
\right]
\frac{v_\perp}{c}
\frac{\widehat{B}_{\parallel{\bf k}_\perp}}{k_\perp}
\nonumber \\  & &
\end{eqnarray}
which is used later to express the turbulent pressure tensor in Eq.~(\ref{Psi_nad}). 

It is now recalled that the ensemble-averaged quantities are smooth spatial functions with 
the gradient scale length $L$. 
For arbitrary real-valued turbulent fluctuations $\hat{Q}$ and $\hat{Q}'$,  
$\langle \hat{Q}_{{\bf k}_\perp}^* 
\hat{Q}'_{{\bf k}'_\perp} \rangle_{\rm ens} = 0$ for 
${\bf k}_\perp \neq {\bf k}'_\perp$ and 
$\langle \hat{Q}  \hat{Q}' \rangle_{\rm ens}
= \sum_{{\bf k}_\perp}
\langle \hat{Q}_{{\bf k}_\perp}^* \hat{Q}'_{{\bf k}_\perp} \rangle_{\rm ens}$ 
hold. 
In the ensemble-averaged momentum balance equation given by Eq.~(\ref{av-totmomb}), 
the effects of the electromagnetic turbulence on the momentum 
transport enter  
$\langle \boldsymbol{\pi}_{\parallel \Psi} \rangle_{\rm ens}$
$\langle \boldsymbol{\pi}_\land \rangle_{\rm ens}$,
and $\langle {\bf P}_\Psi \rangle_{\rm ens}$ through 
the correlation between 
the turbulent distribution 
function and the turbulent potential. 
Using Eqs.~(\ref{piland}), (\ref{pipara}), and (\ref{pppp}), and 
neglecting terms of higher orders in $\delta =\rho/L$, 
one finds that 
\begin{eqnarray}
\label{piparapsi}
\langle 
\boldsymbol{\pi}_{\parallel \Psi}
\rangle_{\rm ens}
& = & 
- {\bf b}{\bf b} 
\sum_a 
\left(
\frac{n_{a0} e_a^2}{m_a c^2} \langle
(\widehat{A}_\parallel)^2 \rangle_{\rm ens}
\right.
\nonumber 
\\ 
& & 
\left.
\mbox{}
+ \frac{e_a}{c}
\int d^3 v \; 
U 
\langle
\widehat{h}_{a}
\langle
\widehat{A}_\parallel
\rangle_\vartheta
\rangle_{\rm ens}
\right) 
\nonumber
\\ 
 & = & 
- {\bf b}{\bf b} 
\left(
\frac{\omega_p^2}{4\pi c^2} \langle
(\widehat{A}_\parallel)^2 \rangle_{\rm ens}
+ \frac{1}{c}
\langle
\widehat{j}_\parallel
\widehat{A}_\parallel
\rangle_{\rm ens}
\right) 
\hspace*{5mm}
\end{eqnarray}
and 
\begin{eqnarray}
\langle 
\boldsymbol{\pi}_\land
\rangle_{\rm ens}
& = & 
\sum_a \int d^3 v \, \langle F_a \rangle_{\rm ens}
m_a U 
\bigl[ b^i \langle ( u_{ax} )_\perp^j \rangle_{\rm ens} 
\nonumber 
\\ & & \mbox{}
\hspace*{5mm}
+ \langle ( u_{ax} )_\perp^i \rangle_{\rm ens} b^j 
\bigr]
+ \langle 
\boldsymbol{\pi}_\land^{\rm turb}
\rangle_{\rm ens}
,
\end{eqnarray}
where the turbulent part 
$
\langle 
\boldsymbol{\pi}_\land^{\rm turb}
\rangle_{\rm ens}
$
is given by  
\begin{eqnarray}
\label{piens}
\langle 
\boldsymbol{\pi}_\land^{\rm turb}
\rangle_{\rm ens}
& = & 
\sum_a \int d^3 v \, \widehat{F}_a
m_a U 
[ b^i ( \widehat{u}_{ax} )_\perp^j  + ( \widehat{u}_{ax} )_\perp^i b^j ]
\nonumber \\
& = & 
\frac{c}{B}
\sum_{{\bf k}_\perp} 
 [ {\bf b} ( {\bf k}_\perp \times {\bf b}  ) 
 + ( {\bf k}_\perp \times {\bf b}  ) {\bf b} ]
\nonumber \\ & & 
\mbox{} \times
\sum_a 
\int d^3 v \; 
m_a U
{\rm Im} [
\langle 
\widehat{h}_{a{\bf k}_\perp}^* 
\langle \widehat{\psi}_a \rangle_{\vartheta{\bf k}_\perp}
\rangle_{\rm ens}
]
.
\hspace*{10mm}
\end{eqnarray}
Using Eqs.~(\ref{PPsia}) and (\ref{pppp}), 
$\langle {\bf P}_\Psi \rangle_{\rm ens}$ 
is written as 
\begin{equation}
\langle 
{\bf P}_\Psi
\rangle_{\rm ens}
=
\langle 
{\bf P}_{\langle \phi \rangle_{\rm ens}}
\rangle_{\rm ens}
+
\langle 
{\bf P}_\Psi^{\rm turb}
\rangle_{\rm ens}
\end{equation}
where the effects of the ensemble-averaged (or background) electric field 
$\langle {\bf E}_L \rangle_{\rm ens} \equiv 
- \nabla \langle \phi \rangle_{\rm ens}$ 
and the turbulent electromagnetic field are included in 
\begin{eqnarray}
\label{Pphi}
& & 
\hspace*{-5mm}
\langle 
{\bf P}_{\langle \phi \rangle_{\rm ens}}
\rangle_{\rm ens}
=
\frac{n_{a0} m_a c^2}{B^2}
( {\bf b} \times \nabla \langle \phi \rangle_{\rm ens}) 
( {\bf b} \times \nabla \langle \phi \rangle_{\rm ens}) 
- \frac{m_a c^2}{2 e_a B^2}
\nonumber \\ & & 
\hspace*{5mm}
\mbox{} 
\times 
\Bigl[
\nabla ( n_{a0} T_{a0} ) \nabla \langle \phi \rangle_{\rm ens}
+ \nabla \langle \phi \rangle_{\rm ens} \nabla ( n_{a0} T_{a0} ) 
\nonumber \\ & & 
\mbox{} 
\hspace*{10mm}
- ({\bf I} - {\bf b} {\bf b})
\nabla ( n_{a0} T_{a0} ) \cdot \nabla \langle \phi \rangle_{\rm ens}
\Bigr]
+ n_{a0} m_{a0} \frac{c^2 T_{a0}}{2 e_a B^2}
\nonumber \\ & & 
\mbox{} 
\hspace*{5mm}
\times 
\Bigl[
({\bf I} - 3 {\bf b} {\bf b}) {\bf b} {\bf b} : 
( \nabla \nabla \langle \phi \rangle_{\rm ens} )
+ {\bf b} ( {\bf b} \cdot \nabla \nabla \langle \phi \rangle_{\rm ens} )
\nonumber \\ & & 
\mbox{} \hspace{10mm}
+  ( {\bf b} \cdot \nabla \nabla \langle \phi \rangle_{\rm ens} ) {\bf b} 
+ 
\nabla \cdot ( {\bf b} {\bf b} )  \nabla \langle \phi \rangle_{\rm ens}
\nonumber \\ & & \hspace{10mm}
\mbox{} 
+  \nabla \langle \phi \rangle_{\rm ens} \nabla \cdot ( {\bf b} {\bf b} ) 
- \nabla \langle \phi \rangle_{\rm ens} \cdot \nabla  ( {\bf b} {\bf b} ) 
\nonumber \\ & & \hspace{10mm}
\mbox{} 
+ 2  \nabla_\perp \ln B   \;  \nabla \langle \phi \rangle_{\rm ens}
+ 2 \nabla \langle \phi \rangle_{\rm ens}  \; \nabla_\perp \ln B 
\nonumber \\ & & \hspace{10mm}
\mbox{} 
-  2  ( \nabla \langle \phi \rangle_{\rm ens} \cdot  \nabla \ln B )
( {\bf I} - {\bf b} {\bf b} ) 
\Bigr]
\end{eqnarray}
and 
\begin{equation}
\label{Psi_turb}
\langle 
{\bf P}_\Psi^{\rm turb}
\rangle_{\rm ens}
=
\langle 
{\bf P}_\Psi^{\rm ad}
\rangle_{\rm ens}
+
\langle 
{\bf P}_\Psi^{\rm nad}
\rangle_{\rm ens}
,
\end{equation}
respectively, 
where 
\begin{eqnarray}
\label{Psi_ad}
& & 
\hspace*{-5mm}
\langle 
{\bf P}_\Psi^{\rm ad}
\rangle_{\rm ens}
=
\bigl( {\bf I} - {\bf b} {\bf b} \bigr)
\;
\sum_a  \sum_{{\bf k}_\perp} 
\frac{n_{a0} e_a^2}{2T_a}
\Bigg[
\biggl\langle 
| \widehat{\phi}_{{\bf k}_\perp} |^2  
+ \frac{T_a}{m_a c^2}
| \widehat{A}_{\parallel {\bf k}_\perp} |^2  
\biggr\rangle_{\rm ens}
\nonumber \\ & & 
\hspace*{7mm}
\mbox{} 
\times
\bigl\{ 1 - \Gamma_0(b_{a{\bf k}_\perp} ) \bigr\}
+ \frac{T_a}{m_a c^2}
\biggl\langle  \frac{\widehat{B}_{\parallel {\bf k}_\perp} |^2  }{k_\perp^2}
\biggr\rangle_{\rm ens}
\bigl[ 1 - 2 b_{a{\bf k}_\perp}
\nonumber \\ & & 
\hspace*{7mm}
\mbox{} 
\times
 \bigl\{ \Gamma_0(b_{a{\bf k}_\perp}) 
- \Gamma_1(b_{a{\bf k}_\perp}) \bigr\} \bigr]
- \frac{v_{Ta}}{c} \;
{\rm Re} 
\biggl\langle  \frac{\widehat{\phi}_{{\bf k}_\perp}^* 
\widehat{B}_{\parallel {\bf k}_\perp}  }{k_\perp}
\biggr\rangle_{\rm ens}
\nonumber \\ & & 
\hspace*{7mm}
\mbox{} 
\nonumber \\ & & 
\hspace*{7mm}
\mbox{} \times 
( 2 b_{a{\bf k}_\perp} )^{1/2}
\bigl\{ \Gamma_0(b_{a{\bf k}_\perp}) 
- \Gamma_1(b_{a{\bf k}_\perp}) \bigr\}
\Biggr]
\end{eqnarray}
and 
\begin{eqnarray}
\label{Psi_nad}
\langle 
{\bf P}_\Psi^{\rm nad}
\rangle_{\rm ens}
& = & 
- \sum_a \sum_{{\bf k}_\perp} 
\left[
\frac{1}{2} 
({\bf I} - {\bf b} {\bf b})
-
\frac{{\bf k}_\perp {\bf k}_\perp }{k_\perp^2}
\right]
\nonumber \\ & & 
\mbox{} \times
e_a 
\int d^3 v \; 
{\rm Re} [
\langle 
\widehat{h}_{a{\bf k}_\perp}^* \langle \widehat{\chi}_a \rangle_{\vartheta{\bf k}_\perp}
\rangle_{\rm ens}
]
- \frac{1}{c}
\langle
\widehat{\bf j} \; \widehat{\bf A} 
\rangle_{\rm ens}
\nonumber \\ & & 
- \frac{1}{2c}
\langle
\widehat{\bf j}_\perp \cdot \widehat{\bf A}_\perp 
\rangle_{\rm ens}
({\bf I} - {\bf b} {\bf b})
\hspace*{10mm}
\end{eqnarray}
are derived from the adiabatic and nonadiabatic parts 
of the distribution function in Eq.~(\ref{ad-nad}), respectively. 
On the right-hand side of Eq.~(\ref{Psi_ad}), 
$b_{a{\bf k}_\perp}\equiv k_\perp^2 T_a /(m_a  \Omega_a^2)$ is used, and the 
functions $\Gamma_0$ and $\Gamma_1$ are defined by 
$\Gamma_0(b) \equiv I_0 (b) \exp (-b)$ and $\Gamma_1(b) \equiv I_1 (b) \exp (-b)$, 
respectively, where 
$I_0$ and $I_1$ are 
the modified Bessel functions. 

One finds that 
the pressure tensor terms consisting of only ${\bf b} {\bf b}$ and 
${\bf I}$ parts cannot produce transport of toroidal and poloidal momenta
across flux surfaces in toroidal magnetically confined systems. 
In the axisymmetric toroidal system,  
in which the magnetic field is given by Eq.~(\ref{axB}), 
one can use 
$
({\bf k}_\perp \times {\bf b} ) \cdot \nabla \chi
= - B R^2 \nabla \zeta \cdot {\bf k}_\perp 
$
and Eqs.~(\ref{piens}), (\ref{Psi_turb}), (\ref{Psi_ad}), and (\ref{Psi_nad}) 
to obtain  
the radial transport of the toroidal angular momentum 
due to the interaction of the nonadiabatic distribution function and the 
turbulent electromagnetic potential  as 
\begin{eqnarray}
\label{pi+PPsi}
& & 
\hspace*{-5mm}
\nabla \chi \cdot 
\langle 
\boldsymbol{\pi}_{\land}^{\rm turb}
+
{\bf P}_\Psi^{\rm turb}
\rangle_{\rm ens} 
\cdot R^2 \nabla \zeta
\nonumber \\  
&  & 
\hspace*{-3mm}= 
\sum_a 
\sum_{{\bf k}_\perp}
\int d^3 v \,
\left[
-
\frac{c I}{B} m_a U
{\rm Im} [
\langle 
\widehat{h}_{a{\bf k}_\perp}^* \langle \widehat{\psi}_a \rangle_{\vartheta{\bf k}_\perp}
\rangle_{\rm ens}
]
( {\bf k}_\perp \cdot R^2 \nabla \zeta ) 
\right.
\nonumber \\ & & 
\left. \mbox{} 
+
e_a
{\rm Re} [
\langle 
\widehat{h}_{a{\bf k}_\perp}^* \langle \widehat{\chi}_a \rangle_{\vartheta{\bf k}_\perp}
\rangle_{\rm ens} 
]
\frac{
( {\bf k}_\perp \cdot \nabla \chi )
( {\bf k}_\perp \cdot R^2 \nabla \zeta )
}{k_\perp^2}
\right]
.
\end{eqnarray}
The flux-surface average of 
Eq.~(\ref{pi+PPsi}) agrees with 
the low-flow ordering limit of the 
result given in Eq.~(53) of Ref.~\cite{Sugama1998} where the 
turbulent radial transport of the toroidal angular momentum 
double-averaged over the ensemble and the flux surface
is presented for the case of 
 high-flow ordering.~\cite{Sugama2017,Abel,Sugama1998} 
The radial turbulent transport of the toroidal angular momentum  in Eq.~(\ref{pi+PPsi}) is 
not the flux-surface average
but a spatially-local expression. 
It is shown in the case of the low-flow ordering~\cite{Sugama2011,Parra2011} 
that, in the axisymmetric configuration with 
up-down symmetry, the flux-surface average of Eq.~(\ref{pi+PPsi}) vanishes  
even though the local value of Eq.~(\ref{pi+PPsi}) does not.
It can also be shown from  Eq.~(\ref{Pphi}) that the flux-surface average of 
$\nabla \chi \cdot \langle 
{\bf P}_{\langle \phi \rangle_{\rm ens}}
\rangle_{\rm ens} \cdot R^2 \nabla \zeta$
vanishes as well.

Next, let us consider the Maxwell stress terms in the ensemble-averaged 
momentum balance equation in Eq.~(\ref{av-totmomb}). 
The Maxwell stress due to the electric field is dominantly given by 
${\bf E}_L$ because the magnitude of 
${\bf E}_T \equiv -c^{-1} \partial {\bf A}/\partial t$ 
is smaller than that of ${\bf E}_L$ by a factor of $\delta$. 
On the left-hand side of Eq.~(\ref{av-totmomb}), 
the turbulent magnetic pressure tensor also appears as 
\begin{equation}
\label{BB}
\frac{1}{8 \pi}  \left\langle |\widehat{\bf B}|^2\right\rangle_{\rm ens}
{\bf I}
-  
 \frac{1}{4 \pi} 
\left\langle
\widehat{\bf B} \widehat{\bf B}
\right\rangle_{\rm ens}
, 
\end{equation}
which has the opposite sign to 
that of the Maxwell stress tensor due to turbulent magnetic fields.

It is also noted that 
the terms proportional to $\nabla \lambda$ and $\widehat{\bf A}$ in 
Eq.~(\ref{av-totmomb}) are negligible compared with the other magnetic Maxwell 
stress terms because 
$
c^{-1}|\nabla \lambda| |\widehat{\bf A}| / |\widehat{\bf B}|^2
\sim |{\bf j}_L|/(  c k_\perp|\widehat{\bf B}| )
\sim (\partial {\bf E}_L / \partial t) / (  c k_\perp|\widehat{\bf B}| )
\sim (v_T/L) (cT/eB) (e\widehat{\phi}/T )/(c^2|\widehat{\bf B}|/B)
\sim (\rho/L) (v_T^2/c^2) \ll 1
$, 
where Eq.~(\ref{Lpart}), 
$
\nabla \cdot {\bf j}_L = - \partial \rho_c / \partial t
= - ( 4 \pi )^{-1} 
\partial ( \nabla \cdot {\bf E}_L ) / \partial t
$, 
$\partial/\partial t \sim v_T/L$, 
and
$
e \widehat{\phi}/T  \sim |\widehat{\bf B}|/B
\sim \rho/L 
$
are used.

One can also find that the nonadiabatic distribution function produces 
the turbulent current $\widehat{\bf j}$ which correlates with 
the turbulent vector potential $\widehat{\bf A}$ and produces 
the pressure tensor given by 
\begin{equation}
\label{jA}
- \frac{1}{c}
\langle
\widehat{j}_\parallel
\widehat{A}_\parallel
\rangle_{\rm ens}
{\bf b}{\bf b} 
- \frac{1}{c}
\langle
\widehat{\bf j} \; \widehat{\bf A} 
\rangle_{\rm ens}
- \frac{1}{2c}
\langle
\widehat{\bf j}_\perp \cdot \widehat{\bf A}_\perp 
\rangle_{\rm ens}
({\bf I} - {\bf b} {\bf b})
, 
\end{equation}
which are included in Eqs.~(\ref{piparapsi}) and (\ref{Psi_nad}). 
In the wavenumber representation, 
the turbulent magnetic field is given by 
\begin{equation}
\label{Bk}
\widehat{\bf B}_{{\bf k}_\perp} 
= 
i {\bf k}_\perp \times \widehat{\bf A}_{{\bf k}_\perp}
= 
\widehat{B}_{\parallel {\bf k}_\perp} {\bf b}
+ \widehat{\bf B}_{\parallel {\bf k}_\perp}
,
\end{equation}
where
$
\widehat{B}_{\parallel {\bf k}_\perp}
= 
i {\bf b} \cdot ({\bf k}_\perp \times \widehat{\bf A}_{{\bf k}_\perp} )
$
and
$
\widehat{\bf B}_{\perp {\bf k}_\perp}
= i  ({\bf k}_\perp \times {\bf b} ) \widehat{A}_{\parallel {\bf k}_\perp}
$. 
The turbulent vector potential is written by 
\begin{equation}
\widehat{\bf A}_{{\bf k}_\perp} = 
\widehat{A}_{\parallel {\bf k}_\perp} {\bf b}
+ \widehat{\bf A}_{\perp {\bf k}_\perp}
,  
\end{equation}
where
$
\widehat{A}_{\parallel {\bf k}_\perp}
=   {\bf b} \cdot \widehat{\bf A}_{{\bf k}_\perp} 
$
and 
$
\widehat{\bf A}_{\perp {\bf k}_\perp}
= 
k_\perp^{-2}({\bf b} \times {\bf k}_\perp) 
[({\bf b} \times {\bf k}_\perp )\cdot \widehat{\bf A}_{{\bf k}_\perp}]
$. 
Here, the Coulomb gauge condition 
$
{\bf k}_\perp \cdot \widehat{\bf A}_{{\bf k}_\perp} = 0
$
is used. 
Then, one has
\begin{equation}
| \widehat{\bf B}_{{\bf k}_\perp}|^2
=  | \widehat{B}_{\parallel {\bf k}_\perp}|^2
+ | \widehat{\bf B}_{\perp {\bf k}_\perp}|^2
=  k_\perp^2 
\left( 
| \widehat{\bf A}_{\perp {\bf k}_\perp}|^2
+ | \widehat{A}_{\parallel {\bf k}_\perp}|^2
\right)
, 
\end{equation}
and Amp\`{e}re's law is given by
\begin{equation}
\label{jk}
\widehat{\bf j}_{{\bf k}_\perp}
=  \frac{c}{4\pi} k_\perp^2 \widehat{\bf A}_{{\bf k}_\perp}
. 
\end{equation}
Using Eqs.~(\ref{Bk})--(\ref{jk}), one obtains 
\begin{eqnarray}
\label{BBjA}
& & 
\hspace*{-5mm}
- \frac{1}{4 \pi}
\langle \widehat{\bf B} \widehat{\bf B} \rangle_{\rm ens}  
- \frac{1}{c}
\langle \widehat{\bf j} \widehat{\bf A} \rangle_{\rm ens} 
= 
\frac{1}{4 \pi}
\sum_{{\bf k}_\perp}
\langle | \widehat{\bf A}_{{\bf k}_\perp}|^2 \rangle_{\rm ens} 
\left(
{\bf k}_\perp {\bf k}_\perp -k_\perp^2  {\bf I}
\right)
\nonumber \\
& & =
\frac{1}{4 \pi}
\sum_{{\bf k}_\perp}
\langle | \widehat{\bf B}_{{\bf k}_\perp}|^2 \rangle_{\rm ens} 
\left(
\frac{{\bf k}_\perp {\bf k}_\perp }{k_\perp^2} -  {\bf I}
\right)
,
\hspace*{10mm}
\end{eqnarray}
and the summation of Eqs.~(\ref{BB}) and (\ref{jA}) are written as 
\begin{eqnarray}
\label{BBpjA}
& & 
\hspace*{-3mm}
\mbox{Eq.~(\ref{BB})}
+ \mbox{Eq.~(\ref{jA})}
\nonumber \\ & & 
\hspace*{-2mm}
=
\frac{1}{8 \pi}
\sum_{{\bf k}_\perp}
\biggl[
- 
\langle
 | \widehat{\bf B}_{{\bf k}_\perp} |^2
    + 2 | \widehat{\bf B}_{\perp {\bf k}_\perp} |^2 
\rangle_{\rm ens} 
{\bf b} {\bf b} 
-
\langle
 | \widehat{\bf B}_{\perp {\bf k}_\perp} |^2 
\rangle_{\rm ens} 
\frac{{\bf k}_\perp {\bf k}_\perp }{k_\perp^2} 
\nonumber \\ & & 
\mbox{} \hspace*{5mm}
-
\langle
 | \widehat{\bf B}_{{\bf k}_\perp} |^2
    + | \widehat{\bf B}_{\parallel {\bf k}_\perp} |^2 
\rangle_{\rm ens} 
\frac{({\bf b} \times {\bf k}_\perp) ({\bf b} \times {\bf k}_\perp) }{k_\perp^2} 
\biggr]
,
\end{eqnarray}
which is given in terms of only the turbulent magnetic field. 

Again, in considering the axisymmetric toroidal system,   
the radial transport of the toroidal angular momentum 
due to the turbulent magnetic field is represented by 
a component of Eq.~(\ref{BBpjA}) obtained from a double-dot product 
with a dyad $(\nabla \chi) (R^2 \nabla \zeta)$,  
which is equivalent to that of Eq.~(\ref{BBjA}) and written as 
\begin{eqnarray}
\label{BBjAcz}
& & 
\hspace*{-3mm}
\nabla \chi \cdot
\biggl[
- \frac{1}{4 \pi}
\langle \widehat{\bf B} \widehat{\bf B} \rangle_{\rm ens}  
- \frac{1}{c}
\langle \widehat{\bf j} \widehat{\bf A} \rangle_{\rm ens} 
\biggr] \cdot R^2 \nabla \zeta
\nonumber \\
&  & 
= 
\frac{1}{4 \pi}
\sum_{{\bf k}_\perp}
\langle | \widehat{\bf A}_{{\bf k}_\perp}|^2 \rangle_{\rm ens} 
( {\bf k}_\perp \cdot \nabla \chi )
( {\bf k}_\perp \cdot R^2 \nabla \zeta )\nonumber \\
&  & 
= 
\frac{1}{4 \pi}
\sum_{{\bf k}_\perp}
\langle | \widehat{\bf B}_{{\bf k}_\perp}|^2 \rangle_{\rm ens} 
\frac{
( {\bf k}_\perp \cdot \nabla \chi )
( {\bf k}_\perp \cdot R^2 \nabla \zeta )
}{k_\perp^2}
.
\end{eqnarray}
Taking 
the flux-surface average of 
Eq.~(\ref{BBjAcz}) yields 
the same expression of the toroidal angular momentum transport 
across the flux surface
caused by the turbulent magnetic field 
as derived in Refs.~\cite{Sugama1998} and \cite{Abel}. 
However, in the same manner as in the case of Eq.~(\ref{pi+PPsi}), 
the flux-surface average of Eq.~(\ref{BBjAcz}) is shown to vanish 
in the axisymmetric configuration with 
up-down symmetry under the low-flow 
ordering~\cite{Sugama2011,Parra2011}.

It is also found from Eqs.~(\ref{piparapsi}) and (\ref{Psi_ad}) 
that the adiabatic part of the perturbed distribution function in Eq.~(\ref{ad-nad}) 
produces pressure tensor terms in 
$\langle 
\boldsymbol{\pi}_{\parallel \Psi}
\rangle_{\rm ens}$ 
and 
$\langle 
{\bf P}_\Psi^{\rm ad}
\rangle_{\rm ens}$
which are given in terms of 
turbulent electrostatic and vector potentials 
although these terms consist of only the ${\bf b} {\bf b}$ and 
${\bf I}$ parts so that 
they cannot produce toroidal and poloidal momentum transport 
across flux surfaces in toroidal plasmas.

\section{CONCLUSIONS}

In this paper, 
the Eulerian (or Euler-Poincar\'{e}) variational formulation is presented 
to obtain the governing equations of the electromagnetic turbulent gyrokinetic system, 
for which the local momentum balance equation is derived from the invariance of 
the Lagrangian of the system under an arbitrary spatial coordinate transformation. 
In addition, the effects of collisions and external sources are taken into account 
in the momentum balance equation.

Using the gyrokinetic Lagrangian which retains proper electromagnetic potential terms
and taking the variational derivatives of the Lagrangian with respect to the electrostatic 
and vector potentials of the perturbed magnetic field, 
one can obtain  
the gyrokinetic Poisson equation and Amp\`{e}re's law where the effects of 
the polarization and magnetization due to finite gyroradii and electromagnetic 
microturbulence are correctly included. 
Especially,  the derived gyrokinetic Amp\`{e}re's law can accurately express 
the current density from the microscopic gyroradius scale to the macroscopic equilibrium scale 
so that it is useful for long-time and global gyrokinetic turbulence simulations of high beta plasmas. 

The local momentum balance equation obtained in the present work contains 
the symmetric pressure tensor which is derived from the variational derivative of 
the Lagrangian with respect to the metric tensor. 
It is shown that
the pressure tensor obtained for the whole system consisting of all particles and fields 
involves the gyrokinetic and field parts; 
the neoclassical and turbulent momentum transport processes are described by 
the former part while the Maxwell stress is by the latter. 

One can confirm from the momentum balance equation that,  
when the background magnetic field has a symmetry such as a translational one and 
an axisymmetry, the canonical momentum conjugate to the coordinate in the symmetry 
direction 
is conserved as  predicted by Noether's theorem. 
The symmetry of the pressure tensor is found to be an important property 
for derivation of the momentum conservation in the symmetric background field. 
When the background field is assumed to satisfy the appropriate condition 
representing the macroscopic Amp\`{e}re's law, 
the ensemble-averaged total momentum balance equation 
is found to take
the conservation form even in the asymmetric background field. 
Thus, this condition can be conveniently applied to long-time gyrokinetic simulations in which the change in the background field occurs with the relaxation of high-beta plasmas. 
It is also shown that, 
in the toroidal systems with the quasi-axisymmetric background field, 
the toroidal angular momentum is not rigorously conserved although 
the flux-surface-averaged neoclassical toroidal viscosity, 
which is a dominant component for breaking the toroidal momentum conservation 
in general non-axisymmetric systems, vanishes. 

The WKB representation is employed to derive detailed expressions of the ensemble-averaged pressure tensor due to the electromagnetic microturbulence,
which provide a means for evaluating the local turbulent momentum transport by the local flux-tube gyrokinetic simulation.
The radial transport fluxes of the toroidal angular momentum caused by the nonadiabatic distribution function and the turbulent electromagnetic fields in the axisymmetric system are represented as a non-diagonal component of the pressure tensor, which are shown to agree with the results from the previous works based on the classical gyrokinetic formulation.
The local pressure tensor represented by a symmetric $3\times 3$ matrix contains further information on momentum transport which is useful for more detailed analyses of transport processes by gyrokinetic simulations.

\begin{acknowledgments}
This work is supported in part by 
the JSPS Grants-in-Aid for Scientific Research Grant No.~19H01879 
 and in part by the NIFS Collaborative Research Program NIFS23KIPT009. 
\end{acknowledgments}

\section*{AUTHOR DECLARATIONS}

\subsection*{Conflict of Interest}

The authors have no conflicts of interest to disclose.

\subsection*{Author Contributions}
\noindent
{\bf Hideo Sugama}: Conceptualization (lead); Formal analysis (lead); Funding acquisition (lead); Writing – original draft (lead).

\section*{DATA AVAILABILITY}
Data sharing is not applicable to this article as no new data were created or analyzed in this study.

\appendix

\section{DECOMPOSITION OF THE POTENTIAL FIELD}

The potential field defined in Eq.~(\ref{Psi_a}) is decomposed here
as 
\begin{eqnarray}
\label{Psia2}
\Psi_a ({\bf Z}, t)
& \equiv  & 
\phi ({\bf X}, t) + \Psi_{E1a} ({\bf X}, \mu, t) 
+ \Psi_{\widehat{A}1a} ({\bf X}, U,\mu, t)
\nonumber \\ & & \mbox{}
+ \Psi_{E2a} ({\bf X}, \mu, t)
+  \Psi_{E\widehat{A}a} ({\bf X}, U, \mu, t) 
\nonumber \\ & & \mbox{}
+ 
\Psi_{\widehat{A}2a} ({\bf X}, U,\mu, t)
, 
\end{eqnarray}
where 
\begin{equation}
\label{PsiE1a}
\Psi_{E1a} ({\bf X}, \mu, t) 
\equiv 
\langle \phi ({\bf X} + \boldsymbol{\rho}_a ,  t) \rangle_\vartheta
- \phi ({\bf X}, t) 
,
\end{equation}
\begin{equation}
\label{PsiA1a}
\Psi_{\widehat{A}1a} ({\bf X}, U,\mu, t)
 \equiv  
-  \frac{1}{c} 
\left[
\langle {\bf v} \cdot \widehat{\bf A} \rangle_\vartheta
+ {\bf v}_{Ba} \cdot  \langle \widehat{\bf A} \rangle_\vartheta
\right]
\end{equation}
\begin{equation}
\label{PsiE2a}
\Psi_{E2a} ({\bf X}, \mu, t)
 \equiv 
- \frac{e_a}{2 B}
\frac{\partial }{\partial \mu}
\bigl\langle 
(\widetilde{\phi} )^2 
\bigr\rangle_\vartheta
, 
\end{equation}
\begin{equation}
\label{PsiEAa}
 \Psi_{E\widehat{A}a} ({\bf X}, U, \mu, t) 
\equiv 
\frac{e_a}{c B}
\frac{\partial }{\partial \mu}
\langle \widetilde{\phi} 
({\bf v} \cdot \widehat{\bf A})
\rangle_\vartheta
,
\end{equation}
and
\begin{eqnarray}
\label{PsiA2a}
& & 
\hspace*{-3mm}
\Psi_{\widehat{A}2a} ({\bf X}, U,\mu, t)
\nonumber \\
&  &
\hspace*{-2mm}\equiv  
\frac{e_a}{2 m_a c^2}
\langle |\widehat{\bf A}|^2 \rangle_\vartheta
- \frac{e_a}{2c^2 B}
\frac{\partial }{\partial \mu}
\bigl\langle 
[({\bf v} \cdot \widehat{\bf A}) 
- \langle {\bf v} \cdot \widehat{\bf A}\rangle_\vartheta]^2
\bigr\rangle_\vartheta
.
\hspace*{5mm}
\end{eqnarray}

It should be noted that, in this Appendix, 
the Cartesian spatial coordinates and 
the conventional dyadic notation representing 
vectors and tensors in terms of boldface letters 
are used.
Then, the electrostatic potential 
$\phi ({\bf X} + \boldsymbol{\rho}_a )$ 
and the perturbed vector potential 
$\widehat{\bf A} ({\bf X} + \boldsymbol{\rho}_a)$ 
are Taylor expanded about 
the gyrocenter position ${\bf X}$ as 
\begin{equation}
\label{phixr}
\left[
\begin{array}{c}
\phi ({\bf X} + \boldsymbol{\rho}_a)
\\
\widehat{\bf A} ({\bf X} + \boldsymbol{\rho}_a)
\end{array}
\right]
= 
\sum_{n=0}^\infty  \frac{1}{n!}
\rho_a^{j_1} \cdots \rho_a^{j_n}
\partial_{j_1 \cdots j_n} 
\left[
\begin{array}{c}
\phi ({\bf X})
\\
\widehat{\bf A} ({\bf X})
\end{array}
\right]
,
\end{equation}
where $X^j$ and $\rho^j$ $(j=1,2,3)$ and 
the Cartesian spatial coordinates of the gyrocenter position vector  
${\bf X}$ and the gyroradius vector 
$\boldsymbol{\rho}$, 
respectively, and 
the partial derivatives are represented using the simplified notation, 
$
\partial_{j_1 \cdots j_n} 
\equiv
\partial^n/\partial X^{j_1} \cdots \partial X^{j_n}
$.
Here, for simplicity, we omit the $t$-dependence  of $\phi$  and 
employ the summation convention that the same symbol used 
for a pair of upper and lower indices indicates a summation over the range 
$\{ 1, 2, 3 \}$. 
Therefore, summation notations $\sum_{j_1 = 1}^3$ $\cdots$  $\sum_{j_n = 1}^3$ are 
dropped in Eq.~(\ref{phixr}).

Using Eqs.~(\ref{PsiE1a}) and (\ref{phixr}), 
one can write
the part of the potential function which linearly depends on 
the electric field  and its derivatives as 
\begin{eqnarray}
\label{PsiE1a2}
\Psi_{E1a} 
& = & 
\sum_{n=1}^\infty  
\frac{\alpha_a^{j_1 \cdots j_n} }{n!}
\partial_{j_1 \cdots j_n}  \phi ({\bf X})
\nonumber \\
& = &
- \sum_{n=1}^\infty  
\frac{\alpha_a^{j_1 \cdots j_n} }{n!}
\partial_{j_1 \cdots j_{n-1}} (E_L)_{j_n} ({\bf X})
, 
\end{eqnarray}
where the gyrophase average of a product of $n$ gyroradius 
vector components is denoted by 
\begin{equation}
\label{alpha}
\alpha_a^{j_1 \cdots j_n}
\equiv
\langle \rho_a^{j_1} \cdots \rho_a^{j_n}
\rangle_\vartheta
. 
\end{equation}
We see that $\alpha_a^{j_1 \cdots j_n}$ is symmetric with respect to 
arbitrary permutations of the indices $j_1, \cdots, j_n$. 
We also find that 
$
\alpha_a^{j_1 \cdots j_{n}}
= 0
$ 
 for odd $n$ 
and 
\begin{equation}
\label{alphaeven}
\alpha_a^{j_1 \cdots j_{2l}}
= 
\frac{1}{(2 l)!}
\sum_{\sigma \in \mathfrak{S}_{2l}}
\eta_a^{j_{\sigma (1)}  \cdots j_{\sigma (2l)}}
, 
\end{equation}
where 
$\mathfrak{S}_{2l}$ is the symmetric group of 
permutations of the set $\{1, 2, \cdots, 2l \}$ 
and 
$\eta_a^{j_1 \cdots j_{2l}}$ is defined by 
\begin{equation}
\label{eta}
\eta_a^{j_1 \cdots j_{2l}}
= 
\frac{(2l)!}{(l !)^2}
\left( \frac{\rho_a}{2} \right)^{2l}
h^{j_1 j_2} h^{j_3 j_4} \cdots h^{j_{2l-1} j_{2l}} 
,
\end{equation}
with 
$
\rho_a
\equiv 
(c /e_a)
\sqrt{2m_a\mu/B}
$
and 
$
h^{ij}
\equiv 
\delta^{ij} - b^i b^j
$.
Here, $b^i$ is the $i$th  component of 
${\bf b} \equiv {\bf B}/B$ and 
$\delta^{ij}$ represents the Kronecker delta; 
$\delta^{ij} = 1$ (for $i=j$), 0 (for $i\neq j$). 

Next, using Eqs.~(\ref{PsiA1a}), 
the part which linearly depends on  the vector potential 
and its derivatives is written as 
\begin{eqnarray}
\label{PsiA1a2}
\Psi_{\widehat{A}1a} 
& = &
 - \frac{1}{c} \bigl[
( U {\bf b} + {\bf v}_{Ba}) \cdot \langle \widehat{\bf A} ({\bf X} + \boldsymbol{\rho}_a) \rangle_\vartheta
+ \langle {\bf v}_\perp \cdot \widehat{\bf A} ({\bf X} + \boldsymbol{\rho}_a) \rangle_\vartheta
\bigr]
\nonumber \\
& = & 
 - \frac{1}{c} \bigl[
( U {\bf b} + {\bf v}_{Ba})  \cdot \langle \widehat{\bf A} ({\bf X} + \boldsymbol{\rho}_a)  \rangle_\vartheta 
\nonumber \\
&  & \mbox{} 
\hspace*{10mm}
+ \Omega_a  \langle ( \boldsymbol{\rho}_a \times  {\bf b} ) 
\cdot \widehat{\bf A} ({\bf X} + \boldsymbol{\rho}_a) \rangle_\vartheta
\bigr] 
\nonumber \\
& = & 
 - \frac{1}{c}
\sum_{n=0}^\infty
\frac{1}{n!}
\bigl[
\alpha_a^{j_1 \cdots j_n} \;  ( U b^i + v_{Ba}^i )
\nonumber \\
&  & \mbox{} 
\hspace*{10mm}
+ \Omega_a \; \varepsilon_{k l m} \; 
\alpha_a^{j_1 \cdots j_n k} \;  
b^l  \; \delta^{i m}
\bigr]
\partial_{j_1 \cdots j_n}  \widehat{A}_i ({\bf X})
,
\end{eqnarray}

Now, using Eqs.~(\ref{phixr}),  one has
\begin{equation}
\label{phitilde2}
\langle (\widetilde{\phi})^2 \rangle_\vartheta
 = 
\sum_{m=1}^\infty  \sum_{n=1}^\infty  
\frac{\beta^{i_1 \cdots i_m ;  j_1 \cdots j_n}}{m! \; n!}
\bigl(
\partial_{i_1 \cdots i_m} \phi ({\bf X})  
\bigr)
\bigl( 
\partial_{j_1 \cdots j_n} \phi ({\bf X})
\bigr)
.
\end{equation}
Here, $\beta^{i_1 \cdots i_m ;  j_1 \cdots j_n}$ is defined by 
\begin{equation}
\label{beta}
\beta^{i_1 \cdots i_m ;  j_1 \cdots j_n}
\equiv
\alpha_a^{i_1 \cdots i_m   j_1 \cdots j_n}
-
\alpha_a^{i_1 \cdots i_m}
\alpha_a^{j_1 \cdots j_n}
, 
\end{equation}
which satisfies
$
\beta^{i_1 \cdots i_m ;  j_1 \cdots j_n}
= 0
$
for odd $(m+n)$.
Substituting Eq.~(\ref{phitilde2}) into Eq.~(\ref{PsiE2a}), 
the part which is quadratically dependent 
on $\{ \partial_{i_1 \cdots i_{m-1}} (E_L)_{i_m} \}$ 
is derived as 
\begin{eqnarray}
\label{PsiE2a2}
\Psi_{E2a} ({\bf X}, \mu, t)
& = & 
\frac{1}{2}
 \sum_{m=1}^\infty  \sum_{n=1}^\infty
C_{E2a}^{i_1 \cdots i_m ; j_1 \cdots j_n}
\partial_{i_1 \cdots i_{m-1}} (E_L)_{i_m}
\nonumber \\  & & 
\mbox{} \times
\partial_{j_1 \cdots j_{n-1}} (E_L)_{j_n}
,
\end{eqnarray}
where $C_{E2a}^{i_1 \cdots i_m ; j_1 \cdots j_n}$
is given by 
\begin{equation}
\label{CE2a}
C_{E2a}^{i_1 \cdots i_m ; j_1 \cdots j_n}
 \equiv 
- \frac{e_a}{2\mu B}
\frac{(m+n) \beta^{i_1 \cdots i_m ;  j_1 \cdots j_n}}{m! \; n!}
.
\end{equation}
It is also found from Eqs.~(\ref{PsiEAa}) and (\ref{PsiA2a}) 
that
the remaining potential functions 
$\Psi_{E\widehat{A}a}$
and $\Psi_{ \widehat{A}2a}$ 
take bilinear and quadratic forms 
which are given by 
\begin{eqnarray}
\label{PsiEAa2}
\Psi_{E\widehat{A}a} ({\bf X}, \mu, t)
& = & 
 \sum_{m=1}^\infty  \sum_{n=1}^\infty
C_{E\widehat{A}a}^{i_1 \cdots i_m ; j_1 \cdots j_n}
\partial_{i_1 \cdots i_{m-1}} (E_L)_{i_m}
\nonumber \\  & & 
\mbox{} \times
\partial_{j_1 \cdots j_{n-1}}  \widehat{A}_{j_n}
, 
\end{eqnarray}
and 
\begin{eqnarray}
\label{PsiA2a2}
\Psi_{ \widehat{A}2a} ({\bf X}, \mu, t)
& = & 
\frac{1}{2}
 \sum_{m=1}^\infty  \sum_{n=1}^\infty
C_{\widehat{A}2a}^{i_1 \cdots i_m ; j_1 \cdots j_n}
\partial_{i_1 \cdots i_{m-1}}  \widehat{A}_{i_m}
\nonumber \\  & & 
\mbox{} \times
\partial_{j_1 \cdots j_{n-1}}  \widehat{A}_{j_n}
, 
\end{eqnarray}
respectively. 
One can use Eq.~(\ref{alpha}) 
to derive the coefficients 
$C_{E\widehat{A}a}^{i_1 \cdots i_m ; j_1 \cdots j_n}$ and 
$C_{\widehat{A}2a}^{i_1 \cdots i_m ; j_1 \cdots j_n}$ in 
Eqs.~(\ref{PsiEAa2}) and (\ref{PsiA2a2}) as functions of 
$\alpha_a^{j_1 \cdots j_n}$. 
The expressions given in 
Eqs.~(\ref{PsiE1a2}), (\ref{PsiA1a2}),  
(\ref{PsiE2a2}), (\ref{PsiEAa2}), and (\ref{PsiA2a2})
are valid in the 
Cartesian spatial coordinates although they can be easily transformed 
into those in general spatial coordinates as shown in Sec.~III.

\section{THE ELECTROMAGNETIC INTERACTION PART OF THE GYROKINETIC LAGRANGIAN DENSITY}

The gyrokinetic Lagrangian given by Eq.~(\ref{LGK}) in Sec.III is written as
$
L_{GK}  \equiv \int d^3 X \; {\cal L}_{GK}
$
where the gyrokinetic Lagrangian density ${\cal L}_{GK}$ is defined as a 
function of $({\bf X}, t)$ by
$
{\cal L}_{GK}
\equiv
\sum_a \int d^3 v \; F_a L_{GYa}
$.
The part of ${\cal L}_{GK}$ including 
the potential field $\Psi_a$ is represented by 
\begin{eqnarray}
\label{B1}
& & 
\hspace*{-5mm}
{\cal L}_\Psi 
\equiv
  \sum_a {\cal L}_{\Psi a}
= 
- \sum_a \int d^3 v \; F_a e_a \Psi_a
\nonumber \\ & & 
\hspace*{-5mm}
=
- \rho_c^{(g)} ({\bf X}, t)  \phi  ({\bf X}, t)
+ {\cal L}_ {E 1}   
+ {\cal L}_ {\widehat{A}1}  
+ {\cal L}_{E 2} + {\cal L}_ {E\widehat{A}}  
+ {\cal L}_ {\widehat{A}2}  
, 
\hspace*{6mm}
\end{eqnarray}
where Eq.~(\ref{Psia2}) is used. 
Equation~(\ref{B1}) describes the electromagnetic interaction of charged particles 
and is used to derive gyrokinetic expressions for polarization and magnetization 
as shown in Appendices C and D.
In Eq.~(\ref{B1}), 
the gyrocenter charge density $\rho_c^{(g)}({\bf X}, t)$ 
is given by 
$
\rho_c^{(g)} ({\bf X}, t) \equiv 
\sum_a e_a N_a^{(g)} ({\bf X}, t)
\equiv 
\sum_a e_a
\int d^3 v \, F_a ({\bf X}, U, \mu, t)
$
and the other components of the Lagrangian density are 
defined by 
\begin{eqnarray}
\label{B2}
& & 
\bigl[ {\cal L}_{E1},  {\cal L}_{\widehat{A} 1}, 
{\cal L}_{E 2},  
{\cal L}_{E \widehat{A}}, {\cal L}_{\widehat{A}  2} 
\bigr]
 \nonumber \\ 
&  & 
\equiv  
\sum_a
\bigl[ {\cal L}_{E 1 a},  {\cal L}_{\widehat{A} 1 a}, 
{\cal L}_{E 2 a},  
{\cal L}_{E \widehat{A} a}, {\cal L}_{\widehat{A}  2a} 
\bigr]
\nonumber \\ 
&  & 
\equiv 
-
\sum_a
\int d^3 v
\, F_a 
\,  e_a
\bigl[  \Psi_{E 1 a},  \Psi_{\widehat{A} 1a} , 
\Psi_{E 2 a} 
\Psi_{E \widehat{A} a}, \Psi_{\widehat{A} 2a} 
\bigr]
.
\hspace*{5mm}
\end{eqnarray}
Here, 
using Eqs.~(\ref{PsiE1a2}) and (\ref{B2}), 
${\cal L}_{E1a} $ can be represented in the linear form of 
${\bf E}_L$ and its spatial derivatives, 
\begin{eqnarray}
\label{B3}
 {\cal L}_{E1a} 
& = & 
\sum_{k = 1}^\infty 
Q_{0a}^{j_1 \cdots j_{2k}} 
\partial_{ j_1 \cdots  j_{2k-1}}
 (E_L)_{j_{2k}}
,
\end{eqnarray}
where 
$Q_{0a}^{j_1 \cdots j_{2k}}$ represents 
the multipole moment
of the electric charge distribution~\cite{Jackson} 
of species $a$ induced by finite gyroradius, 
\begin{equation}
\label{B4}
Q_{0a}^{j_1 \cdots j_{2k}} 
\equiv
e_a
\int d^3 v
\; F_a 
\frac{\alpha_a^{j_1 \cdots j_{2k}}}{(2k)!}
.
\end{equation}
Substituting Eq.~(\ref{PsiA1a2}) into Eq.~(\ref{B2}) yields 
the linear form of 
$\widehat{\bf A}$ and its spatial derivatives, 
\begin{eqnarray}
\label{B5}
 {\cal L}_{\widehat{A}1a} 
& = & 
\sum_{n=1}^\infty 
R_{0a}^{j_1 \cdots j_{n}} 
\partial_{j_1 \cdots  j_{n-1}} 
\widehat{A}_{j_{n}}
,
\end{eqnarray}
where
\begin{eqnarray}
\label{B6}
R_{0a}^{j_1 \cdots j_n i} 
& \equiv &
\frac{e_a}{c}
\int d^3 v
\; F_a 
\frac{1}{n!}
\bigl[
\alpha_a^{j_1 \cdots j_n} 
 ( U b^i + v_{Ba}^i )
\nonumber \\ & & \mbox{}
\hspace*{10mm}
+ \Omega_a \; 
\varepsilon_{k l m} \; \alpha_a^{j_1 \cdots j_n k} \; 
b^l \delta^{i m}
\bigr]
. 
\end{eqnarray}
Especially, in the cases of $n=0$ and 1, 
Eq.~(\ref{B6}) is written as 
\begin{eqnarray}
\label{B7}
R_{0a}^i 
& \equiv  & 
\frac{e_a}{c}
\int d^3 v
\; F_a  ( U b^i + v_{Ba}^i )
\nonumber \\
& = & 
\biggl[
\frac{e_a}{c}
N_a^{(g)} V_{ag \parallel} {\bf b}
+
\frac{\bf b}{B}  \times 
\left( 
P_{\parallel a} {\bf b} \cdot \nabla {\bf b} 
+ P_{\perp a} \nabla \ln B 
\right)
\biggr]^i
,
\hspace*{8mm}
\end{eqnarray}
and 
\begin{equation}
\label{B8}
R_{0a}^{j i} 
 \equiv 
\frac{e_a}{c}
\int d^3 v
\; F_a \; \Omega_a \; 
\varepsilon_{k l m} \; 
\alpha_a^{j  k} \; 
b^l \delta^{i m}
= 
\frac{P_{\perp a}}{B}
({\bf I} \times {\bf b})^{ji}
,
\end{equation}
respectively, 
where
$
[
P_{\parallel a}, 
\;
P_{\perp a}
]
\equiv
\int d^3 v
\; F_a 
\;
[
m_a U^2, 
\; 
\mu B
]
$.
From Eqs.~(\ref{PsiE2a2}), (\ref{PsiEAa2}), (\ref{PsiA2a2}), 
and (\ref{B2}), one obtains 
the quadratic forms, 
\begin{eqnarray}
\label{B9}
{\cal L}_{E2a} 
& =  & 
\frac{1}{2} \sum_{m=1}^\infty  \sum_{n=1}^\infty  
\chi_{Ea}^{i_1 \cdots i_m ;  j_1 \cdots j_n}
 \partial_{i_1 \cdots   i_{m-1}} (E_L)_{i_m} 
\nonumber \\ 
& & \mbox{}
\times 
\partial_{j_1 \cdots  j_{n-1}}(E_L)_{j_n}
,
\hspace*{8mm}
\end{eqnarray}
\begin{eqnarray}
\label{B10}
 {\cal L}_{E\widehat{A}a} 
& = & 
\sum_{n=1}^\infty 
Q_{\widehat{A}a}^{j_1 \cdots j_{n}} 
\partial_{j_1 \cdots  j_{n-1}} 
 (E_L)_{j_{n}}
\nonumber \\ 
& = & 
\sum_{n=1}^\infty 
R_{Ea}^{k_1 \cdots k_{n}} 
\partial_{k_1 \cdots  k_{n-1}} 
\widehat{A}_{k_{n}}
\nonumber \\ 
& = & 
\sum_{m=1}^\infty \sum_{n=1}^\infty 
\chi_{E\widehat{A}a}^{j_1 \cdots j_m ; k_1, \cdots k_n} 
\partial_{j_1 \cdots  j_{m-1}} 
 (E_L)_{j_m}
\nonumber \\ 
&  & \mbox{} \times
\partial_{k_1 \cdots  k_{n-1}} 
\widehat{A}_{k_n},
\end{eqnarray}
and 
\begin{eqnarray}
\label{B11}
 & & 
\hspace*{-12mm}
{\cal L}_{\widehat{A}2a} 
= 
\frac{1}{2}
\sum_{n=1}^\infty 
R_{\widehat{A}a}^{j_1 \cdots j_{n}} 
\partial_{j_1 \cdots  j_{n-1}} 
\widehat{A}_{j_{n}}
\nonumber \\ 
& &
\hspace*{-10mm}
=
\frac{1}{2}
\sum_{m=1}^\infty \sum_{n=1}^\infty 
\chi_{\widehat{A}a}^{j_1 \cdots j_m ; k_1, \cdots k_n} 
( \partial_{j_1 \cdots  j_{m-1}} 
\widehat{A}_{j_m})
(\partial_{k_1 \cdots  k_{n-1}} 
\widehat{A}_{k_n})
.
\end{eqnarray}
Here,  $\chi_{Ea}^{i_1 \cdots i_m ;  j_1 \cdots j_n}$, 
$\chi_{E\widehat{A}a}^{i_1 \cdots i_m ;  j_1 \cdots j_n}$, 
and 
$\chi_{\widehat{A}a}^{i_1 \cdots i_m ;  j_1 \cdots j_n}$ 
are regarded as the generalized electromagnetic susceptibilities 
which are defined by 
\begin{eqnarray}
\label{B12}
& & 
\hspace*{-2mm}
\left[
\chi_{Ea}^{i_1 \cdots i_m ;  j_1 \cdots j_n} 
,
\chi_{E\widehat{A}a}^{i_1 \cdots i_m ;  j_1 \cdots j_n} 
,
\chi_{\widehat{A}a}^{i_1 \cdots i_m ;  j_1 \cdots j_n} 
\right]
\nonumber \\
&  & 
\hspace*{-2mm}
\equiv
- e_a
\int d^3 v
\; F_a
\left[
C_{E2a}^{i_1 \cdots i_m ;  j_1 \cdots j_n} 
,
C_{E\widehat{A}a}^{i_1 \cdots i_m ;  j_1 \cdots j_n} 
,
C_{\widehat{A}2a}^{i_1 \cdots i_m ;  j_1 \cdots j_n} 
\right]
.
\hspace*{8mm}
\end{eqnarray}

\section{CHARGE DENSITY}

In this Appendix,  it is shown in detail that 
the charge density of particles consists of the charge 
density of gyrocenters and other components including 
multipole moments which appear due to finite gyroradii of 
charged particles and turbulent electromagnetic fields. 
The charge density $\rho_c$ which
appears in Poisson's equation, Eq.~(\ref{GKP}), 
is given by the variational derivative of the gyrokinetic Lagrangian 
$L_{GK}$
with respect to the electrostatic potential $\phi$ 
as shown in Eq.~(\ref{rhoc}) 
where the delta function 
$\delta^3 ({\bf X} + \boldsymbol{\rho}_a - {\bf x} )$
is used. 
Here, to obtain another expression of $\rho_c$, 
the variational derivative is written as 
\begin{eqnarray}
\label{C1}
& & 
\rho_c
=  - \frac{\delta L_{GK}}{\delta \phi} 
=
- \sum_{n = 0}^\infty  (-1)^n
\partial_{j_1 \cdots  j_n}
\left(
\frac{\partial {\cal L}_\Psi}{\partial (\partial_{j_1 \cdots  j_n} \phi)}  
\right) 
\nonumber \\ 
&  & \hspace*{3mm}
=
-
\frac{\partial {\cal L}_\Psi}{\partial \phi}  
+ \sum_{n = 1}^\infty 
(-1)^n
\partial_{j_1 \cdots j_n}
\frac{\partial {\cal L}_\Psi}{\partial (\partial_{j_1 \cdots  j_{n-1}} (E_L)_{j_n})}  
\nonumber \\ 
&  & \hspace*{3mm}
= 
\rho_c^{(gc)} 
-  \nabla \cdot {\bf P}_G
,
\end{eqnarray}
where 
$\rho_c^{(gc)}$ and  ${\bf P}_G$ are
 the gyrocenter charge density and  the polarization density vector, 
defined by 
\begin{equation}
\label{C2}
\rho_c^{(gc)} 
 \equiv 
-
\frac{\partial {\cal L}_\Psi}{\partial \phi}  
 \equiv 
\sum_a e_a \int d^3 v  \; F_a 
 \equiv  \sum_a e_a N_a^{(g)}
, 
\end{equation}
and
\begin{equation}
\label{C3}
{\bf P}_G
 \equiv 
\frac{\delta L_{GK}}{\delta {\bf E}_L}
\equiv
\sum_{n = 0}^\infty 
(-1)^n
\partial_{j_1 \cdots j_n}
\frac{\partial {\cal L}_\Psi}{\partial (\partial_{j_1 \cdots  j_n} {\bf E}_L)} 
,
\end{equation}
respectively. 
Then, the electric displacement field ${\bf D}$ is given by 
\begin{equation}
\label{C4}
{\bf D} 
\equiv 
{\bf E}
+
4 \pi {\bf P}_G
, 
\end{equation}
in terms of which the
gyrokinetic Poisson equation
is written as 
\begin{equation}
\label{C5}
\nabla \cdot {\bf D} 
=
4 \pi \rho_c^{(gc)} 
.
\end{equation}

From Eq.~(\ref{C3}), the $i$th component of 
the gyrokinetic polarization density vector 
is written as 
\begin{equation}
\label{C6}
P_G^i
=
\sum_{n=0}^\infty  
(-1)^{n}
\partial_{i_1 \cdots i_n} 
Q^{i \, i_1 \cdots i_{n} }
,
\end{equation}
where the multipole moments 
$Q^{i \, i_1 \cdots i_{n} }$ $(n=0, 1, 2, \cdots)$ 
are given using Eqs.~(\ref{B1}), (\ref{B2}), (\ref{B3}), (\ref{B9}), 
and (\ref{B10}) as 
\begin{equation}
\label{C7}
Q^{i \, i_1 \cdots i_{n} }
=
\frac{\partial {\cal L}_\Psi}{\partial (\partial_{i_1 \cdots  i_n} (E_L)_i )} 
= 
\sum_a
\bigl(
Q^{i \, i_1 \cdots i_{n} }_{0a}
+
Q^{i \, i_1 \cdots i_{n} }_{Ea}
+
Q^{i \, i_1 \cdots i_{n} }_{\widehat{A}a}
\bigr)
. 
\end{equation}
Here, $Q^{i \, i_1 \cdots i_{n} }_{0a}$ is defined in Eq.~(\ref{B4}).
The other multipole moments $Q_{Ea}^{i \, i_1 \cdots i_m}$ 
and $Q^{i \, i_1 \cdots i_{n} }_{\widehat{A}a}$
of 
the electric charge distribution of species $a$ 
are written 
in terms of in the linear forms   
with respect to  $(E_L)_i$,  $\widehat{A}_i$, and their spatial derivatives
as 
\begin{equation}
\label{C8}
 Q_{Ea}^{i \, i_1 \cdots i_m}
\equiv 
\sum_{n=1}^\infty
\chi_{Ea}^{i \, i_1 \cdots i_m ;  j_1 \cdots j_n}
\partial_{j_1 \cdots  j_{n-1}}  (E_L)_{j_n}
, 
\end{equation}
and 
\begin{equation}
\label{C9}
Q^{i \, i_1 \cdots i_{n} }_{\widehat{A}a}
= 
\sum_{m = 1}^\infty
\chi_{E\widehat{A}a}^{i \, i_1 \cdots i_{n} ; k_1 \cdots  k_m}
\partial_{k_1 \cdots  k_{m-1}} \widehat{A}_{k_m}
,
\end{equation}
respectively, 
where 
$\chi_{Ea}^{i \, i_1 \cdots i_{n} ; j \, j_1 \cdots  j_m}$ 
and 
$\chi_{E\widehat{A}a}^{i \, i_1 \cdots i_{n} ; k_1 \cdots  k_m}$ 
are defined in Eq.~(\ref{B12}). 
It is found from Eqs.~(\ref{beta}), (\ref{CE2a}), (\ref{B4}), and (\ref{B12}) 
that  $Q_{0a}^{i_1 \cdots i_m}$ and $Q_{Ea}^{i_1 \cdots i_m}$ 
are both symmetric with respect to arbitrary permutations of the 
indices $i_1,  \cdots,  i_m$ 
because $\alpha_a^{i_1 \cdots i_m}$ defined in Eq.~(\ref{alpha}) 
has the same symmetry.

When retaining only the $n=0$ term in Eq.~(\ref{C9})
and using the lowest order distribution function
 $F_{a0}$ given by the local Maxwellian, 
the polarization density vector is approximated by 
\begin{equation}
\label{C10}
P_G^i 
\simeq
\sum_a  Q_{Ea}^i
\simeq  
\sum_n \frac{n_{a0} m_a c^2}{B^2} {\bf E}_L
= \frac{c^2}{4\pi v_A^2} {\bf E}_L,
\end{equation}
where $n_{a0} \equiv \int d^3 v \; F_{a0}$ and 
$v_A \equiv B^2/(4\pi \sum_a n_{a0} m_a)$ represent 
the equilibrium density and the Alfv\'{e}n velocity, 
respectively. 
Equation~(\ref{C10}) presents a well-known expression of 
polarization.
It should be noted that Eq.~(\ref{C1}) with Eq.~(\ref{C3}) including
all multipole moments gives the charge density which 
is equivalent to that presented in Eq.~(\ref{rhoc}). 
Then, as shown in Ref.~\cite{Sugama2022}, 
the gyrokinetic Poisson equation
given by the classical gyrokinetic theory~\cite{Antonsen,CTB,F-C} based 
on the WKB formalism 
can be derived as well from the 
turbulent part of Eq.~(\ref{GKP}) using the charge 
density given by Eq.~(\ref{rhoc}) 
[or Eq.~(\ref{C1}) with Eqs.~(\ref{C2}) and (\ref{C3})]. 

It is remarked here that 
the gyrocenter charge density 
$\rho_c^{(gc)}$ defined in Eq.~(\ref{C2}) 
satisfies 
\begin{equation}
\label{C11}
\frac{\partial \rho_c^{(gc)}}{\partial t} 
+ \nabla \cdot {\bf j}^{(gc)}
= 0
, 
\end{equation}
where the gyrocenter current density 
${\bf j}^{(gc)}$ is given by 
\begin{equation}
\label{C12}
 {\bf j}^{(gc)} 
\equiv 
\sum_a e_a \boldsymbol{\Gamma}_a^{(gc)} 
\equiv 
\sum_a e_a \int d^3 v \; F_a 
 {\bf u}_{ax} 
. 
\end{equation}
Equation~(\ref{C11}) is derived by taking  the velocity-space integral 
and the species summation of Eq.~(\ref{GKB}) with the help of Eq.~(\ref{eK}). 
Combining Eqs.~(\ref{C5}) and (\ref{C11}), one obtains
\begin{equation}
\label{C13}
\nabla \cdot
\left( \frac{\partial {\bf D}}{\partial t} +
4 \pi \; 
{\bf j}^{(gc)}
\right)
= 0
,
\end{equation}
from which one can write
\begin{equation}
\label{C14}
4 \pi \; 
{\bf j}^{(gc)}_L
= -  \frac{\partial {\bf D}_L}{\partial t}  
=
\frac{\partial }{\partial t}
\left( \nabla 
 \phi_D
\right)
. 
\end{equation}
In Eq.~(\ref{C14}), the subscript $L$ denotes the longitudinal (or irrotational) vector part, 
and the potential $\phi_D$ for ${\bf D}_L$  is defined  
such that ${\bf D}_L = -\nabla \phi_D$ .

\section{CURRENT DENSITY}

In a similar way to Appendix~C, 
this Appendix shows how 
the current density of particles is expressed 
by the sum of the current density of gyrocenters and 
other components induced by finite gyroradii of 
charged particles and turbulent electromagnetic fields. 
The gyrokinetic
Amp\`{e}re's law is presented in Eq.~(\ref{GKA2}) which 
contains the transverse part ${\bf j}_T$ of the current 
density ${\bf j}$ given by the variational derivative 
of the gyrokinetic Lagrangian $L_{GK}$ with 
respect to the perturbed vector potential 
$\widehat{\bf A}$ as shown in Eq.~(\ref{j}). 
Here, another expression of ${\bf j}$ is obtained by 
writing the variational derivative as 
\begin{eqnarray}
\label{D1}
& & 
\hspace*{-3mm}
\frac{1}{c} {\bf j} 
= \frac{\delta L_{GK}}{\delta \widehat{\bf A}} 
=
\sum_{n=0}^\infty 
(-1)^n
\partial_{j_1 \cdots  j_n}
\left(
\frac{\partial {\cal L}_\Psi}{\partial 
(\partial_{j_1 \cdots  j_n}\widehat{\bf A})}  
\right) 
\nonumber \\ 
&  & 
=
 \sum_a e_a \int d^3 v  \; F_a 
\sum_{n=0}^\infty 
(-1)^ {n+1}
\partial_{j_1 \cdots  j_n}
\left(
\frac{\partial \Psi_a}{\partial (\partial_{j_1 \cdots  j_n}
 \widehat{\bf A})}  
\right) 
\nonumber \\ 
&  & 
= \sum_a \frac{e_a}{c} \boldsymbol{\Gamma}_a
.
\hspace*{8mm}
\end{eqnarray}
The particle flux $\boldsymbol{\Gamma}_a$ of species $a$ 
in Eq.~(\ref{D1}) is written as 
\begin{equation}
\label{D2}
\boldsymbol{\Gamma}_a
= \sum_{n=0}^\infty
\boldsymbol{\Gamma}_a^{(n)}
,
\end{equation}
where 
$\boldsymbol{\Gamma}_a^{(n)}$ 
is defined by 
\begin{equation}
\label{D3}
\boldsymbol{\Gamma}_a^{(n)}
= 
c
\int d^3 v  \; F_a 
(-1)^{n+1}
\partial_{j_1  \cdots  j_n}
\left(
\frac{\partial \Psi_a }{\partial (\partial_{j_1  \cdots  j_n} \widehat{\bf A})}
\right) 
.
\end{equation}
The zeroth-order flux 
$\boldsymbol{\Gamma}_a^{(0)}$ 
is written as 
\begin{eqnarray}
\label{D4}
\boldsymbol{\Gamma}_a^{(0)}
& \equiv & 
-
c
\int d^3 v  \; F_a 
\left(
\frac{\partial \Psi_a}{\partial  \widehat{\bf A}}  
\right) 
\nonumber \\
& = & 
\int d^6 Z
\; \delta^3 ({\bf X} - {\bf x} )
\biggl[
F_a ({\bf Z}, t) \biggl( {\bf v}  - \frac{e_a}{m_a c} \widehat{\bf A} 
+
{\bf v}_{Ba} \biggr)
\nonumber \\
& & 
+
\frac{e_a \widetilde{\psi}_a}{B} 
\frac{\partial F_a}{\partial \mu} {\bf v}
\biggr]
\nonumber \\
& = & 
\int d^3 v  \; F_a 
\biggl[
\left( 
U - \frac{e_a}{m_a c} \widehat{A}_\parallel \right) {\bf b}
+ {\bf v}_{Ba}
+ \frac{c}{B} {\bf b} \times \nabla \langle \psi_a \rangle_\vartheta 
\biggr]
\nonumber \\
&  & 
\mbox{} + {\cal O}(\delta^2)
,
\end{eqnarray}
which is equivalent to 
$\boldsymbol{\Gamma}_a^{(gc)}  \equiv \int d^3 v \; F_a {\bf u}_{ax}$ 
to the lowest order in $\delta$.  

Using Eqs.~(\ref{B1}), (\ref{B5}), (\ref{B10}), and (\ref{B11}),  
one can represent the derivatives 
$\partial {\cal L}_\Psi / \partial (\partial_{j_1 \cdots j_n} \widehat{A}_k)$ 
by 
\begin{eqnarray}
\label{D5}
R^{j_1 \cdots j_n k}
& \equiv &
\frac{\partial {\cal L}_\Psi}{\partial (\partial_{j_1 \cdots j_n} \widehat{A}_k)}  
=
 - \sum_a e_a \int d^3 v  \; F_a 
\frac{\partial \Psi_a}{\partial (\partial_{j_1 \cdots j_n} \widehat{A}_k)}  
\nonumber \\
& = & 
\sum_a 
\bigl(
R^{j_1 \cdots j_n k}_{0a}
+
R^{j_1 \cdots j_n k}_{Ea}
+
R^{j_1 \cdots j_n k}_{\widehat{A}a}
\bigr)
,
\end{eqnarray}
where 
\begin{equation}
\label{D6}
R^{j_1 \cdots j_n k}_{Ea}
=
\sum_{m=0}^\infty
\chi_{E\widehat{A}a}^{i_1 \cdots i_m i ;  j_1 \cdots j_n k}
\partial_{i_1 \cdots i_m} (E_L)_i
,
\end{equation}
\begin{equation}
\label{D7}
R^{j_1 \cdots j_n k}_{\widehat{A}a}
=
\sum_{m=0}^\infty
\chi_{\widehat{A}a}^{j_1 \cdots j_n k ; l_1 \cdots l_m l}
\partial_{l_1 \cdots l_m} \widehat{A}_l
,
\end{equation}
and $R^{j_1 \cdots j_n k}_{0a}$ is defined by 
Eq.~(\ref{B6}). 
The coefficients 
$\chi_{E\widehat{A}a}^{i_1 \cdots i_{m} i ;  j_1 \cdots  j_n k}$ 
and 
$\chi_{\widehat{A}a}^{j_1 \cdots j_n k ; l_1 \cdots l_m l}$ 
are given by  Eq.~(\ref{B12}). 

Now, it is found from Eqs.~(\ref{D1})--(\ref{D4}) 
that the $l$th component of ${\bf j}$ 
can be expressed as 
\begin{equation}
\label{D8}
j^l
=
(j^{(0)})^l + c \; \partial_k N^{kl}
,
\end{equation}
where 
\begin{equation}
\label{D9}
{\bf j}^{(0)}
\equiv
c \frac{\partial {\cal L}_\Psi}{\partial \widehat{\bf A}}  
=
 - c \sum_a e_a \int d^3 v  \; F_a 
\frac{\partial \Psi_a}{\partial \widehat{\bf A}}  
= \sum_a e_a \boldsymbol{\Gamma}_a^{(0)}
,
\end{equation}
 is regarded as the current of 
gyrocenters 
and 
\begin{equation}
\label{D10}
N^{k l}
\equiv 
\sum_{n=0}^\infty
(-1)^{n+1}
\partial_{j_1 \cdots j_n}
R^{j_1 \cdots j_n k l}
.
\end{equation}
Here, it should be recalled that 
$
{\bf j}^{(0)}
\equiv
\sum_a e_a \boldsymbol{\Gamma}_a^{(0)} 
$
defined above 
equals 
$
 {\bf j}^{(gc)} 
\equiv 
\sum_a e_a \boldsymbol{\Gamma}_a^{(gc)} 
\equiv 
\sum_a e_a \int d^3 v \; F_a 
 {\bf u}_{ax} 
$
to the lowest order in $\delta$ although 
the equality does not rigorously holds.
When assuming $|\boldsymbol{\rho_a} \cdot \nabla | < 1$ and retaining 
only the lowest order of $N^{kl}$ 
in the expansion with respect to $|\boldsymbol{\rho_a} \cdot \nabla | $, 
one just has the nonturbulent contribution to $N^{k l}$ as
\begin{equation}
\label{D11}
N^{k l}
\simeq
- 
R^{k l}
\simeq
-
\sum_a R_{0a}^{kl}
.
\end{equation}
Then, using Eqs.~(\ref{B8}) and (\ref{D11}) leads to 
\begin{equation}
\label{D12}
c \partial_k N^{k l}
\simeq
- 
\sum_a \left[ \nabla \times 
\left( \frac{c P_{\perp a}}{B} {\bf b} \right)
\right]^l
,
\end{equation}
where $P_{\perp a}$ is defined after Eq.~(\ref{B8}). 
Equation~(\ref{D12}) is a well-known expression of a magnetization current 
[see Ref.~\cite{magnetization_law}]. 
Then,  from using Eqs.~(\ref{D4}), (\ref{D8}), (\ref{D9}), and (\ref{D12})
with the lowest order distribution function 
$F_{a0}$ given by the local Maxwellian, 
the perpendicular component of the equilibrium current can be derived as
\begin{eqnarray}
\label{D13}
{\bf j}_\perp
& = &
\sum_a e_a 
\int d^3v \; F_{a0}
\left( {\bf v}_{Ba} 
+ \frac{c}{B} {\bf b} \times \nabla \langle \phi \rangle_{\rm ens}
\right)
\nonumber \\ & & 
\hspace*{3mm}
\mbox{}
- 
 \left[ 
\nabla \times 
\left( 
\frac{c P_0}{B}  
{\bf b} \right)
\right]_\perp
\nonumber \\
& = &
\frac{c}{B} {\bf b} \times \nabla P_0
,
\end{eqnarray}
where $P_0 \equiv \sum_a P_{a0}  \equiv \sum_a \int d^3 v \; F_{a0} \mu B$ 
denotes the equilibrium pressure 
and $\sum_a e_a \int d^3v \; F_{a0}= 0$ is used. 
Equation~(\ref{D13}) presents the magnetization law~\cite{magnetization_law} and 
one can see that, as pointed out in Ref.~\cite{Sugama2022},  
the diamagnetic current consistent with the MHD equilibrium 
$c^{-1}{\bf j}\times {\bf B} = \nabla P_0$ is correctly derived 
from the variational formulation with the ${\bf v}_{Ba}$ term retained 
in the potential part of the Hamiltonian given by Eq.~(\ref{HGYa}) with Eq.~(\ref{Psi_a}). 
It is also shown in Ref.~\cite{Sugama2022} 
that the gyrokinetic Amp\`{e}re's law 
given by the classical gyrokinetic 
theory~\cite{Antonsen,CTB,F-C} based on the WKB formalism 
can be derived from the 
turbulent part of Eq.~(\ref{GKA2}) using Eq.~(\ref{j}) 
which is equivalent to Eq.~(\ref{D1}) with 
all the gyroradius expansion terms retained.

Here, 
${\bf j}^{(0)}$ defined in Eq.~(\ref{D9}) 
is divided into the longitudinal and transverse parts as 
$
{\bf j}^{(0)}
=
{\bf j}^{(0)}_L + {\bf j}^{(0)}_T
$, 
where the longitudinal part can be written in terms of the scalar 
function  $\lambda^{(0)}$ as 
${\bf j}^{(0)}_L = \frac{1}{4\pi} \nabla \lambda^{(0)}$. 
Similarly,  $N^{k l}$ in Eq.~(\ref{D10})  is represented by the sum of two parts, 
\begin{equation}
\label{D14}
N^{k l}
=
N^{k l}_L + N^{k l}_T
, 
\end{equation}
where $N^{k l}_L$ and $N^{k l}_T$ are defined such that 
$\epsilon_{mnl} \partial^n N_L^{kl} = 0$ and 
and 
$\partial_l N_T^{kl} = 0$ 
are satisfied. 
Then, there exist $V^k$ and $ W^k_n$ in terms of which 
$N^{k l}_L$ and $N^{k l}_T$ are given by 
\begin{equation}
\label{D15}
N^{k l}_L = \partial^l V^k, 
\hspace*{5mm}
N^{k l}_T = \epsilon^{l m n} \partial_m W^k_n
.
\end{equation}
Using Eq.~(\ref{Lpart})  and 
${\bf j}^{(0)}_L \equiv \frac{1}{4\pi} \nabla \lambda^{(0)}$, 
one obtains 
\begin{equation}
\label{D16}
\frac{\lambda}{4\pi} = \frac{\lambda^{(0)}}{4\pi}
+ c \; \nabla \cdot {\bf V}
,
\end{equation}
and 
\begin{equation}
\label{D17}
{\bf j}_T  = {\bf j}^{(0)}_T + c \nabla \times {\bf M}_W
,
\end{equation}
where ${\bf V}$ is the vector with the components 
$V^k$ $(k=1,2,3)$,  and the $j$th component of the vector
${\bf M}_W$ is defined by 
\begin{equation}
\label{D18}
(M_W)_j = \partial_k W^k_j
. 
\end{equation}
It is found from Eq.~(\ref{D17}) that 
the magnetization field can be represented by 
${\bf M}_W$ 
up to  a gradient of an arbitrary scalar function. 
Using ${\bf M}_W$, 
 the magnetic intensity field ${\bf H}$ is defined by 
\begin{equation}
\label{D19}
{\bf H} = {\bf B} + \widehat{\bf B} - 4 \pi {\bf M}_W
\end{equation}
which is distinguished  from ${\bf H}_\#$ given in Eq.~(\ref{Hmacro}). 
Then, using Eq.~(\ref{D19}), 
the gyrokinetic Amp\`{e}re's law, Eq.~(\ref{GKA2}) 
can be rewritten as 
\begin{equation}
\label{D20}
\nabla \times {\bf H}
= 
\frac{4\pi}{c} {\bf j}^{(0)}_T
.
\end{equation}

\section{ENERGY BALANCE IN ELECTROMAGNETIC GYROKINETIC TURBULENCE}

This Appendix presents energy balance equations in electromagnetic 
gyrokinetic turbulence. 
Here, the Cartesian coordinate system is used and three-dimensional
vectors are written in terms of boldface letters. 
The energy of a single charged particle of species $a$ is denoted by 
${\cal E}_a$ which is equal to the gyrocenter Hamiltonian $H_{GYa}$
in Eq.~(\ref{HGYa}) 
and written as 
\begin{equation}
\label{E1}
{\cal E}_a 
 \equiv 
\frac{1}{2} m_a U^2 
+ \mu B + e_a \Psi_a
\equiv 
H_{GYa} 
 =   
\frac{\partial L_{GYa} }{\partial {\bf u}_{aZ}} 
\cdot {\bf u}_{aZ} 
 -  L_{GYa}
. 
\end{equation}
It can be shown from Eqs.~(\ref{ddt}), (\ref{ELeq2}), 
and (\ref{E1}) that the total derivative of 
$H_{GYa}$ is written as
\begin{eqnarray}
\label{E2}
& & 
\dot{\cal E}_a 
  \equiv 
\left( \frac{d}{dt} \right)_a H_{GYa} 
\equiv
\left(
\frac{\partial }{\partial t} 
+ {\bf u}_{aZ} \cdot \frac{\partial}{\partial  {\bf Z}}  
\right) 
H_{GYa}
\nonumber \\ 
& & 
\hspace*{3mm}
=
- \left( \frac{\partial L_{GYa}}{\partial t} \right)_u
=
 e \frac{\partial \Psi_a}{\partial t} 
+ \mu \frac{\partial B}{\partial t} 
- \frac{e_a}{c} 
{\bf u}_{ax}  \cdot  \frac{\partial {\bf A}_a^* }{\partial t} 
,
\hspace*{5mm}
\end{eqnarray}
where $( \partial L_{GYa}/\partial t)_u$ denotes  
the time derivatives of $L_{GYa}$ 
with ${\bf u}_{aZ} $ kept fixed in $L_{GYa}$. 
Multiplying Eq.~(\ref{E2}) with $F_a$ and taking its velocity-space integral, 
the local energy balance equation for the system of the single particle species 
is obtained as 
\begin{eqnarray}
\label{E3}
& & 
\hspace*{-3mm}
\frac{\partial }{\partial t} 
\left(
\int d^3 v \, F_a {\cal E}_a 
\right)
+ \nabla
\cdot 
\left(
\int d^3 v \, 
F_a {\cal E}_a {\bf u}_{ax}
 \right)
\nonumber \\
& & 
=
\int d^3 v \, \left(  F_a   \dot{\cal E}_a 
+ {\cal K}_a {\cal E}_a 
\right) 
,
\end{eqnarray}
where the gyrokinetic Boltzmann equation shown in 
Eq.~(\ref{GKB}) is used 
and 
${\bf u}_{ax}= (d/dt)_a {\bf X}$ represents the gyrocenter velocity defined at 
the right-hand side of Eq.~(\ref{dXdt}).

Next, the energy balance in the whole system including particles of all species 
and the turbulent electromagnetic fields is considered. 
From 
Eq.~(\ref{LGKF}), one finds 
\begin{eqnarray}
\label{E4}
& & 
\hspace*{-3mm}
\sum_a  \int d^3 v \, F_a 
\frac{\partial L_{GYa}}{\partial t}
+ \frac{\partial {\cal L}_F }{\partial t}
\nonumber \\ 
&  & =
\sum_J
\biggl\{
\frac{\partial {\cal L}_{GKF} }{\partial (\partial_J \phi)} 
\frac{\partial (\partial_J \phi)}{\partial t} 
+  \frac{\partial {\cal L}_{GKF} }{\partial  (\partial_J {\bf A}) } 
\cdot \frac{\partial (\partial_J {\bf A}) }{\partial t}
\nonumber \\ 
& & \hspace*{8mm}
\mbox{}
+  \frac{\partial {\cal L}_{GKF} }{\partial (\partial_J \widehat{\bf A}) } 
\cdot \frac{\partial (\partial_J  \widehat{\bf A}) }{\partial t}
\biggr\}
+ \frac{\partial {\cal L}_{GKF} }{\partial \lambda } 
\frac{\partial \lambda }{\partial t}
. 
\end{eqnarray}
Then, taking the species summation of Eq.~(\ref{E3}) 
and using Eqs.~(\ref{E1}), (\ref{E2}), and (\ref{E4}), 
they yield
\begin{eqnarray}
\label{E5}
& & 
\hspace*{-3mm}
\frac{\partial}{\partial t} 
\left( 
\sum_a \int d^3 v \; F_a \; H_{GYa} - {\cal L}_F
\right) 
\nonumber \\ & & \mbox{} 
+ \nabla 
\cdot 
\left(
\sum_a \int d^3 v \; F_a \; H_{GYa} \; {\bf u}_{ax} 
\right)
+ \sum_J
\biggl\{
\frac{\partial {\cal L}_{GKF} }{\partial (\partial_J \phi)} 
\frac{\partial (\partial_J \phi)}{\partial t} 
\nonumber \\ 
&  & 
\mbox{}
+  \frac{\partial {\cal L}_{GKF} }{\partial  (\partial_J {\bf A}) } 
\cdot \frac{\partial (\partial_J {\bf A}) }{\partial t}
+  \frac{\partial {\cal L}_{GKF} }{\partial (\partial_J \widehat{\bf A}) } 
\cdot \frac{\partial (\partial_J  \widehat{\bf A}) }{\partial t}
\biggr\}
+ \frac{\partial {\cal L}_{GKF} }{\partial \lambda } 
\frac{\partial \lambda }{\partial t}
\nonumber \\ 
&  & 
=
\sum_a \int d^3 v \;  {\cal K}_a  \;
H_{GYa}
.
\end{eqnarray}
Now,   the
variational equations, 
$\delta L_{GK}/\delta \phi = 0$, 
$\delta L_{GK}/\delta \widehat{\bf A} = 0$, 
and $\delta L_{GK}/\delta \lambda = 0$, 
which are equivalent to the gyrokinetic Poisson equation, 
Amp\`{e}re's law, and the Coulomb gauge condition, 
are used to rewrite  Eq.~(\ref{E5}) as
\begin{eqnarray}
\label{E6}
& & 
\hspace*{-8mm}
\frac{\partial}{\partial t} 
\left( 
\sum_a \int d^3 v \; F_a \; H_{GYa} - {\cal L}_F
\right) 
\nonumber \\ & & \mbox{} 
\hspace*{-3mm}
+ \nabla 
\cdot 
\left(
\sum_a \int d^3 v \; F_a \; H_{GYa} \; {\bf u}_{ax} 
\right)
+ 
\sum_{n=1}^\infty
\sum_{k=1}^n (-1)^{k-1}
\nonumber \\ & & \mbox{} 
\times
\partial_{j_k}
\biggl\{
\partial_{j_1 \cdots j_{k-1}}
\biggl(
\frac{\partial {\cal L}_{GKF} }{\partial 
( \partial_{j_1  \cdots  j_n} \phi ) }
\biggr)
\frac{\partial
(
\partial_{j_{k+1} \cdots  j_n} \phi 
)
}{\partial t}
\nonumber \\ & & \mbox{} 
\hspace*{3mm}
+
\partial_{j_1 \cdots j_{k-1}}
\biggl(
\frac{\partial {\cal L}_{GKF} }{\partial 
( \partial_{j_1  \cdots  j_n} \widehat{A}_l ) }
\biggr)
\frac{\partial
(
\partial_{j_{k+1} \cdots  j_n} \widehat{A}_l  
)
}{\partial t}
\biggr\}
\nonumber \\ & & \mbox{} 
\hspace*{-3mm}
+ \nabla 
\cdot 
\left[
\frac{c}{4\pi} {\bf E}_T \times  ( {\bf B} + \widehat{\bf B} - 4\pi {\bf M}_\# ) 
\right]
+ \partial_i 
\left[
\Xi^i_j \frac{\partial B^j}{\partial t}
\right]
\nonumber \\ & & 
\hspace*{-8mm}
= 
\left[ 
{\bf j}_\#
- \frac{c}{4 \pi} \nabla \times
( {\bf B} + \widehat{\bf B} )
\right]
\cdot {\bf E}_T 
+
\sum_a \int d^3 v \;  {\cal K}_a  \;
H_{GYa}
, 
\hspace*{2mm}
\end{eqnarray}
where $\Xi^i_j$ on the left-hand side is defined by
\begin{equation}
\label{E7}
\Xi^i_j 
= \frac{\partial {\cal L}_{GK}}{\partial (\partial_i B^j)}
= \sum_a \int d^3 v \; F_a 
\frac{\partial L_{GYa}}{\partial (\partial_i B^j)}
.
\end{equation}
Equation~(\ref{E6}) can be further deformed to 
obtain the local energy balance equation of the whole system as 
\begin{eqnarray}
\label{E8}
& & 
\frac{\partial}{\partial t} 
\biggl[
\sum_a \int d^3 v \, F_a 
\biggl(
\frac{1}{2} m_a \Bigl|
{\bf v}  - \frac{e_a}{m_a c} \widehat{\bf A}
\Bigr|^2
- \frac{e_a}{c} {\bf v}_{Ba} \cdot  \widehat{\bf A}
\nonumber \\ & & \mbox{} \hspace*{3mm}
- \frac{e_a^2}{2 c^2 B} \frac{\partial}{\partial \mu}
\Bigl[
\bigl(
\widetilde{{\bf v} \cdot \widehat{\bf A}}
\bigr)^2
\Bigr]
\biggr)
+ \frac{|{\bf E}_L|^2}{8\pi}
+ \frac{{\bf E}_L \cdot {\bf P}_D}{2} 
\nonumber \\ & & \mbox{} \hspace*{3mm}
+
\frac{1}{2}\sum_{n=1}^\infty
( \partial_{j_1 \cdots  j_n} (E_L)_i  )
Q_E^{i j_1 \cdots j_n}
+ \frac{{\bf E}_T \cdot {\bf D}_L}{4 \pi} 
+ \frac{|{\bf B} +  \widehat{\bf B} |^2}{8\pi}
\biggr]
\nonumber \\ & & \mbox{} \hspace*{3mm}
+ \nabla
\cdot 
\biggl[
\sum_a \int d^3 v \, F_a 
\biggl\{ \frac{1}{2} m_a U^2 + \mu B 
+ e_a (\Psi_a -\phi (x) ) \biggr\} {\bf u}_{ax} 
\nonumber \\ & & \mbox{} \hspace*{5mm}
+ 
\frac{c}{4\pi} {\bf E} \times {\bf H}_\#
+ \frac{1}{4\pi} \phi_D \frac{\partial {\bf E}_T}{\partial t} 
+ \phi \frac{\partial ({\bf P}_G)_T}{\partial t}
\biggr]
+ \partial_i 
\left[
\Xi^i_j \frac{\partial B^j}{\partial t}
\right]
\nonumber \\ & & \mbox{} \hspace*{3mm}
+ 
\partial_i
\biggl[
\sum_{n=1}^\infty
\sum_{k=0}^{n-1} (-1)^{n-k}
\biggl\{
(\partial_{j_{n-k} \cdots j_{n-1}} (E_L)_{j_n} )
\nonumber \\ & & \mbox{} \hspace*{8mm}
\times
\partial_{j_1 \cdots j_{n-k-1}}
\left(
\frac{\partial Q^{i j_1 \cdots j_n}}{\partial t}
\right)
+ c (\partial_{j_{n-k} \cdots j_{n}} (\widehat{E}_T)_l )
\nonumber \\ & & \mbox{} \hspace*{8mm}
\times
\partial_{j_1 \cdots j_{n-k-1}} 
R^{i j_1 \cdots j_n l}
\biggr\}
\biggr]
\nonumber \\ & & \mbox{} \hspace*{3mm}
+ \nabla \cdot 
\left[ \frac{c}{4 \pi} \widehat{\bf E}_T \times ( {\bf B} + \widehat{\bf B} )
- \frac{\lambda}{4 \pi} \widehat{\bf E}_T
\right]
+ \partial_i \left[
c  N^{ij} (\widehat{E}_T)_j
\right]
\nonumber \\ & & 
= 
\left[ 
({\bf j}_\#)_T
- \frac{c}{4 \pi} \nabla \times
( {\bf B} + \widehat{\bf B} )
\right]
\cdot 
\left( {\bf E}_T + {\bf E}_L \right)
\nonumber \\ & & \mbox{}
\hspace*{6mm} 
+
\sum_a \int d^3 v \,  {\cal K}_a 
\biggl\{ \frac{1}{2} m_a U^2 + \mu B 
+ e_a (\Psi_a -\phi (x) ) \biggr\} 
, 
\end{eqnarray}
where ${\bf H}_\# $ is defined in Eq.~(\ref{Hmacro}). 
The rate of change in the sum of the kinetic and electromagnetic energy densities 
is described by 
Eq.~(\ref{E8}). 
There appear 
the effects of polarization and magnetization including all 
multipole moments. 
The energy flux on the left-hand side of Eq.~(\ref{E8}) 
contains the kinetic energy flow due to the gyrocenter motion,   
the Poynting vector, and the extra contributions due to the 
electromagnetic microturbulence. 
The last terms on the left-hand side of Eq.~(\ref{E8}) can be deformed into 
\begin{eqnarray}
\label{E9}
& &
\hspace*{-3mm}
\nabla \cdot 
\left[ \frac{c}{4 \pi} \widehat{\bf E}_T \times ( {\bf B} + \widehat{\bf B} )
- \frac{\lambda}{4 \pi} \widehat{\bf E}_T
\right]
+ \partial_i \left[
c  N^{ij} (\widehat{E}_T)_j
\right]
\nonumber \\ 
&  & =
\nabla \cdot 
\biggl[ \frac{c}{4 \pi} \widehat{\bf E}_T \times 
{\bf H}
- \frac{\lambda^{(0)}}{4 \pi} \widehat{\bf E}_T
+ c  {\bf V} \cdot \nabla \widehat{\bf E}_T 
\biggr] 
\nonumber \\ & & \mbox{} \hspace*{3mm}
+ \partial_i 
\left[
c \epsilon^{ijk} W^l_j \partial_l (\widehat{E}_T)_k
+ W^i_j \frac{\partial \widehat{B}^j}{\partial t}
\right]
\end{eqnarray}
where ${\bf H}$ is given in Eq.~(\ref{D19}). 
It is seen from Eqs.~(\ref{E8}) and (\ref{E9}) that the energy balance equation 
two types of the Poynting vector, 
$(c/4\pi) {\bf E} \times {\bf H}_\#$ and 
$(c/4\pi)\widehat{\bf E}_T \times {\bf H}$, 
where ${\bf E} \equiv {\bf E}_L + {\bf E}_T$, 
${\bf E}_L = - \nabla \phi$, 
${\bf E}_T = - c^{-1} \partial {\bf A}/\partial t$ and 
$\widehat{\bf E}_T = - c^{-1} \partial \widehat{\bf A}/\partial t$. 
On the right-hand side of Eqs.~(\ref{E6}) and (\ref{E8}), 
the effects of collisions and/or external sources 
are represented by terms including ${\cal K}_a$. 
It is noted that
these terms can be written as the divergence of classical energy transport flux 
when ${\cal K}_a$ is given by the collision operator including 
the finite gyroradius effect. 
In addition, terms including ${\bf E}_L \equiv -\nabla \phi$ on the right-hand 
side of Eq.~(\ref{E8}) are given in the divergence form as 
\begin{equation}
\label{E10}
\left[ 
({\bf j}_\#)_T
- \frac{c}{4 \pi} \nabla \times
( {\bf B} + \widehat{\bf B} )
\right]
\cdot 
{\bf E}_L 
 = 
\nabla \cdot 
\left[ \frac{c}{4\pi} \nabla \phi
\times 
(
{\bf B}_\# 
-
{\bf B} - \widehat{\bf B}
)
\right]
. 
\end{equation}
Therefore, 
in the stationary background magnetic field where 
${\bf E}_T \equiv - c^{-1} \partial {\bf A}/\partial t = 0$ and with no
external energy sources, 
Eqs.~(\ref{E6}) and (\ref{E8}) take the conservation form. 
Furthermore, even when ${\bf E}_T \equiv - c^{-1} \partial {\bf A}/\partial t \neq 0$, 
the ensemble average of Eq.~(\ref{E6}) and (\ref{E8}) can take the form of 
total energy conservation on macroscopic spatiotemporal scales 
in the background field determined by the condition in Eq.~(\ref{Bmacro}). 
It can also be confirmed from comparison 
with Eq.~(22) in Ref.~\cite{Sugama2013}
that the kinetic and electromagnetic energies, the kinetic energy flux, 
the Poynting vector, and the longitudinal and transverse electric fields in the 
the energy conservation equation of the Vlasov-Darwin system  
are retained in Eq.~(\ref{E8}). 
There, additional terms 
due to finite-gyroradius effects and electromagnetic 
microturbulence are included as well.

\end{document}